\documentclass[%
 reprint,
 amsmath,amssymb,
 aps,
]{revtex4-2}
\usepackage{siunitx}
\usepackage{graphicx}% Include figure files
\usepackage{dcolumn}% Align table columns on decimal point
\usepackage{bm}% bold math
\usepackage{hyperref}
\usepackage{url}
\usepackage[dvipsnames]{xcolor}
\hypersetup{hypertexnames=True, colorlinks = true, linkcolor=blue, citecolor=blue, urlcolor=blue }
\usepackage{multirow}
\usepackage{booktabs}
\usepackage{longtable}
\usepackage{amsmath}
\usepackage[version=3]{mhchem} % Formula subscripts 
\usepackage{adjustbox}
\begin{document}

\preprint{APS/123-QED}

\title{Prediction of atomic H adsorption energies in metalloid doped MSSe (M = Mo/W) Janus layers: A combined DFT and machine learning study}% Force line breaks with \\
%\thanks{A footnote to the article title}%

\author{G. Tejaswini$^{1}$}
\author{Anjana E Sudheer$^{1}$}
\author{Amrendra Kumar$^{2,3}$}
\author{M. Vallinayagam$^{4,5}$}
\author{Pavan Kumar Perepu$^{6}$}  
\author{Attila Cangi$^{7,8}$}
\author{Mani Lokamani$^{8}$}
\author{M. Posselt$^{8}$}
\author{M. Zschornak$^{4,5}$}
\author{C. Kamal$^{2,3}$}
\author{D. Amaranatha Reddy$^{1}$}
\author{D. Murali$^{1}$}%
\email{dmurali@iiitk.ac.in}
\affiliation{$^{1}$Indian Institute of Information Technology Design and Manufacturing, Kurnool, India}
\affiliation{$^{4}$IEP,TU Bergakademie Freiberg, 09599 Freiberg, Germany}
\affiliation{$^{5}$Fakultät Maschinenbau/Energietechnik/Physik, Hochschule für Technik und Wirtschaft, Friedrich-List-Platz 1, 01069 Dresden, Germany}
\affiliation{$^{2}$Theory and Simulations Laboratory, Theoretical and Computational Physics Section, Raja
Ramanna Centre for Advanced Technology, Indore 452013, India}
\affiliation{$^{3}$Homi Bhabha National Institute, Training School Complex, Mumbai 400094, India}
\affiliation{$^{6}$Indian Institute of Information Technology, Sri City, Andhra Pradesh, India}
\affiliation{$^{7}$Center for Advanced Systems Understanding (CASUS), D-02826 Görlitz, Germany}
\affiliation{$^{8}$Helmholtz-Zentrum Dresden-Rossendorf, D-01328 Dresden, Germany}
%\affiliation{$^{9}$Helmholtz-Zentrum Dresden-Rossendorf, Institute of Ion Beam Physics and Materials Research,01328 Dresden, Germany}

\date{\today}% It is always \today, today,
             %  but any date may be explicitly specified

\begin{abstract}
Janus derivatives of 2H MX$_2$ (M = Mo/W; X = S/Se), namely MSSe, have already been experimentally realized and explored for applications in photocatalysis, photovoltaics, and optoelectronics. Focusing on the photocatalytic properties of these layers, we investigate the adsorption of atomic hydrogen on the MSSe layers in the presence of metalloid dopants B, Si, and Ge. The layers in their pristine form exhibit positive adsorption energies (E$_{ad}$), indicating an endothermic nature. Substitution of a dopant in the pristine MSSe layers alters the local symmetry, bonding character, and charge distribution, thereby increasing the number of active sites for hosting H adsorption and reducing the adsorption energy. The doping concentration and the hydrogen coverage are taken into account by considering supercells of different sizes. We select distinct sites, both atomic and interstitial, for the substitution of dopants. The spin-polarized electronic structure calculations in doped MSSe reveal a significant magnetic moment in addition to the metallic and half-metallic nature. The energetics of the H atom at various sites is studied to find the most favorable active site on the MSSe Janus layers. Our results based on density functional theory (DFT) calculations show that the adsorption process becomes spontaneous and less attractive in the presence of atomic site dopant substitution, whereas the interstitial site results in an endothermic behavior. Moreover, having the data from DFT, we develop a supervised machine learning (ML) model for predicting the hydrogen adsorption energy, E$_{ad}$. We propose an ML model based solely on elemental features for predicting E$_{ad}$ in doped monolayers. For this purpose, we utilize 23 elemental features of the atoms involved in the structure, thereby eliminating the need for DFT calculations in feature design. The dimensionality reduction technique, principal component analysis (PCA), is employed to reduce the dimensionality of the feature space, yielding independent features that are mutually orthogonal. The model is implemented as a multi-layer perceptron regressor with two hidden layers. The data augmentation technique is employed to artificially expand the dataset size, thereby enhancing the accuracy of the neural network model by 0.90\% on the testing data.   
\end{abstract}

%\keywords{Suggested keywords}%Use showkeys class option if keyword
                              %display desired
\maketitle

%\tableofcontents

\section{\label{sec:intro}Introduction}

In the post-graphene era, transition metal dichalcogenides (TMDs), characterized by the MX$_2$ (M- transition metal, X - chalcogen) stoichiometry and semiconducting property~\cite{Ruijie-Angew.Chem.Int.Ed.2023, Manzeli-Nat.Rev.Mater2017} have emerged as a new class of prospective 2D materials, since the exfoliation of graphene layers from graphite~\cite{Novoselov-Science2004}. Unlike graphene, which does not possess a band gap, TMDs are characterized by band gaps, which can be tuned by increasing or decreasing the number of layers, making them suitable for various optoelectronic applications such as photo detectors~\cite{Xiangfeng-Chem.Reviews2024}, light-emitting diodes~\cite{Wang-Nanoscale.Adv.2020}, solar cells~\cite{Zhou-Adv.Funct.Mater.2024}, and water splitting driven by visible light~\cite{Priyakshi-Mater.Today.Sus2024}.
The materials studied most extensively in the TMD class are MX$_2$, where M = Mo/W; X = S/Se~\cite{Eftekahri-JMCA2017}. The most stable phase of MX$_2$ TMDs is the trigonal prismatic phase, with single M and six X atoms located at the center and corners of the prism. Recently, polar TMDs known as Janus layers (JLs), which are derived from traditional TMDs, have gained a lot of attention due to the asymmetry-induced enhanced properties, such as Rashba spin splitting~\cite{Tao-PRB2018, Ayushi-JAP2024}, second harmonic generation \cite{Yadong-PCCP2019, Shi-Qi-Adv.Optical.Mater.2022}, skyrmionics~\cite{Gorkan-PRM2023, Weiy-PRB2024}, excitonic~\cite{Ting-NanoLetters2021, Sudheer-PCCP2024, Sudheer-Comput.Mater.Sci2024}, thermoelectric~\cite{Abhishek-ACS2020,VM2020, Ozbey-PRB2024,VM2025} and 2D magnetic ordering~\cite{Yagmurcukardes-Appl.Phys.Rev.2020, Zhang-2DMaterials2023}. JLs can be formed by substituting one of the two X atomic layers in MX$_2$ structures with the other chalcogen atomic layer of different electronegativity. Due to the difference in the chemical environment on either side of the central metal atomic layer, the out-of-plane mirror symmetry is broken, resulting in an intrinsic dipole moment along the direction perpendicular to the layer~\cite{Ju-JPM2020}. Such out-of-plane polarization in the JLs facilitates the spatial separation of the photogenerated electron-hole pairs, thereby minimizing their recombination. Thus, JLs are well explored for photocatalytic water splitting and photovoltaic applications~\cite{Vallinayagam-JPCC2023, Ju-ACS2020, Jamdagni-PCCP2022, Zhou-PRM2021, Huiqin-Chem.Phys2022}. The first and widely explored layers of this kind are transition metal-based JLs with the MXY stoichiometry, where M = Mo, W; X/Y = O, S, Se, Te~\cite{Varjovi-PRB2021, Haman-IJHE2024, Bikerouin-Appl.Surf.Sci.2022, Rajneesh-Adv.Theory.Simul.2025}. Given the experimental challenges in synthesizing JLs, theoretical investigations serve as a powerful approach to explore and predict their fundamental properties and potential functionalities. To date, the following three JLs MoSSe~\cite{Lu-Nature.Nanotech2017, Zhang-ACSNano2017} WSSe~\cite{Lin-ACSNano2022} and PtSSe~\cite{Sant-npj.2D.Mater.Appl2020} are synthesized.

In the current scenario, the world economy is lagging in the production of green hydrogen to achieve net zero emissions goals~\cite{Bidattul-RSER2024}. Photocatalytic water splitting is a method that requires only sunlight and a semiconducting photocatalyst. Understanding the adsorption energetics of the intermediates involved in photocatalytic water splitting is crucial for increasing the efficiency and yield of the photocatalyst. Herein, we focus on the adsorption of H atoms, a crucial step in the sub-reaction known as the hydrogen evolution reaction (HER) in photocatalytic water splitting. The H adsorption process can be characterized by calculating the adsorption energy (E$_{ad}$), which can be defined as E$_{ad}$ = E$_{H*}$ - E$_{*}$ - 0.5 E$_{H_2}$~\cite{Norskov-JPCB2004}. Here, E$_{H*}$ and E$_{*}$ are the total energies with and without adsorbed H atom at the active site (*) of the JL, and E$_{H_2}$ is the total energy of the H$_2$ molecule. The atomic H adsorption reaction is reported to be endothermic in both the regular MX$_2$ and Janus MXY layers~\cite{Rahman-ACSOmega2022}, which limits the efficiency of solar-driven hydrogen production~\cite{Eugene-IJHE2012}. To improve the performance, various structural modification techniques can be adopted, such as heterostructure construction~\cite{Tejaswini-PCCP2024, Karthikeyan-JMCC2025, Ma-JMCA2025, Haoyang-MaterialsTodayChemistry2023}, strain modulation~\cite{Chen-ChemPhotoChem2023, Yu-Small2024}, vacancy creation~\cite{Ouahrani-PRM2023, Mehdipour-PRB2022, Maarisetty-JMCA2020}, and doping of a metallic or non-metallic element~\cite{Hou-PRM2021, Obeid-PCCP2020}. Employing these structural modification techniques not only increases the number of active sites for adsorption but also improves the straddling band alignment and enhances the efficiency of light absorption.

In this study, we focus on the substitutional doping at various atomic and interstitial sites of the MSSe JLs. The breaking of the local surface symmetry by introducing a dopant will highly impact the electronic, optical, and adsorption properties~\cite{Thajitr_2022, Zhang_2020}. We consider the metalloid elements B, Si, and Ge as dopants, which are non-toxic. The screening of efficient photocatalysts via an experimental synthesis approach is a time-consuming and resource-intensive process. In general, DFT-based calculations can predict several physical properties of new materials and provide a microscopic understanding of observed physical properties. However, one has to work with larger supercells to reduce finite-size effects~\cite{Komsa-PRL2013}, which renders the simulations computationally demanding. Thus, emerging machine learning methods have become indispensable tools to speed up the screening process, e.g., by employing models tuned to predict desired properties, which are trained on large datasets ~\cite{Jonathan-npjComput.Mater.2019}

In recent years, numerous machine learning (ML) techniques have been developed for predicting various atomic and molecular adsorption energies.
For example, the molecular CO$_2$ and CO adsorption energies in binary alloys have been studied with the physicochemical properties of the material using a neural network based on multi-layer perceptrons, decision tree regression, support vector regression, and extreme gradient boosting~\cite{Xiaofeng-RSC.adv2024}.
The atomic H adsorption energies have been studied on Pt nanoclusters with the help of local structural features and with Gaussian Process Regression, and Random Forest Regression methods~\cite{Zhiheng-CATC2024}.
By taking the electronic density of states (DOS) of the surface atoms as input, the adsorption energies of monoatomic elements H, C, N, O, S, and their hydrogenated counterparts, CH, CH$_2$, CH$_3$, NH, OH, SH on the bimetallic alloy surfaces have been predicted accurately using the convolution neural networks~\cite{Victor-Nat.Commun2021}.
The adsorption of various molecules, including H$_2$, CH$_4$, CO$_2$, and biogas in metal organic frameworks has been studied using ML models~\cite{Nokubonga-JES-2025}. The adsorption energy predictions of various transition metal atoms on the MX$_2$, where M = Mo/W; X = S/Se transition metal dichalcogenide monolayers in the presence of X point defects, have been done using the Sure
Independence Screening and Sparsifying Operator (SISSO) generated features~\cite{Brian-npj2DMaterAppl2025}.
Using the same SISSO technique, the molecular adsorption energies of various molecules, including H$_2$O, CH$_3$, CO, N, NH, NH$_2$, NO, O, and OH on various transition metal surfaces are studied, and the importance of the d-band center combined with the surface energy is explored for better adsorption energy predictions~\cite{Wen-PRB2025}. Adsorbate Chemical Environment-based Graph
Convolution Neural Networks (ACE-GCN) have been introduced for the accurate prediction of the adsorption energies for various adsorbates on the Pt$_3$Sn alloy and Pt facet~\cite{Ghanekar-NatCommun2022}. Using 15 chemical features and the ML models Gradient Boosted Regression, Random Forest Regression, K-nearest neighbours, Adaptive Boosting, and support vector regression, the Gibbs free adsorption energy is predicted in the transition metal doped WSn$_2$N$_4$ 2D layers~\cite{Wang-IJHE2024}. 
\begin{figure*}[ht]
     \centering
     \includegraphics[width=\textwidth]{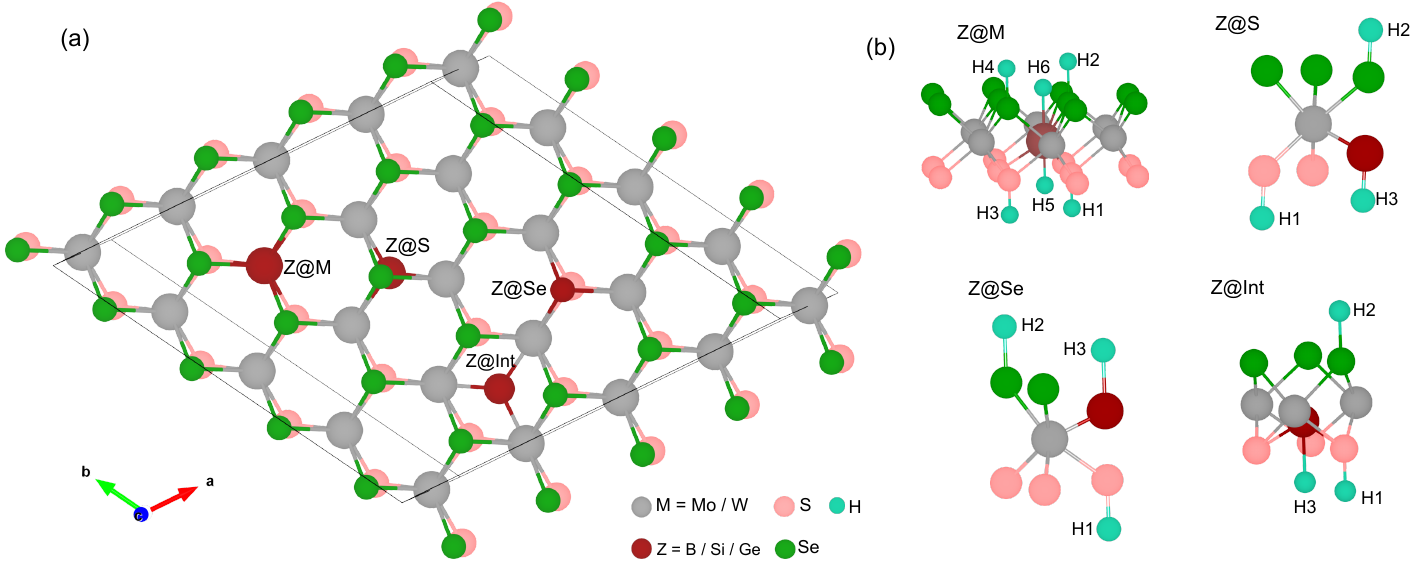}
     \caption{Schematic illustration of (a) Z doping at various atomic sites in the MSSe JL and (b) H adsorption sites in Z-doped MSSe JLs. By convention, H1 and H2 denote H atoms adsorbed atop S and Se sites, respectively. H3 and H4 represent adsorption at the top of the dopant Z, with H approaching through the S and Se planes, respectively. For the Z@Mo configuration, two additional adsorption sites, located on the S and Se planes away from the dopant, are also considered. The same set of adsorption sites is used for both the $3\times3\times1$ and $4\times4\times1$ supercells.}
     \label{fig:ballstick}
\end{figure*}

In the present work, extending the work of Vallinayagam et al.~\cite{Vallinayagam-JMCA2024}, where they studied the atomic H adsorption in the doped MoSSe layer in the 4$\times$4$\times$1, we predict the H adsorption energy at various sites using ML techniques on the MSSe JLs, in the presence of three dopants, B, Si, and Ge, substituted at distinct sites. We consider two different doping concentrations 2.1\% and 3.8\%, and study the effect on H adsorption. To date, there are no reports available that predict the H adsorption energies using ML in the doped MSSe JLs, including the dependence on the doping concentration. As descriptors for the ML model, we utilize the chemical features of the elements, which require no computational effort. The present work is structured as follows. 
Initially, we discuss our DFT results, which include the binding energy, charge transfer, spin polarization, density of states, and the H atom adsorption energies in doped MSSe layers. Followed by this, discuss the feature selection and the physical importance of features in predicting the H adsorption energy. We discuss our dimensionality reduction technique, based on principal component analysis (PCA), to reduce the dimensionality of the original feature space. We present the implementation of the neural network model and its prediction performance. Finally, we discuss expanding the dataset using the synthetic minority over-sampling technique for regression (SMOTER), and the performance of the neural network model on the combined dataset of original and synthetic data points has been tested.

\section{\label{sec:methodology}Computational methods}
Spin-polarized electronic structure calculations based on DFT have been carried out using the Vienna ab-initio simulation package~\cite{kresse1996}. The electron-ion interaction is treated within the projector-augmented wave (PAW) method~\cite{Blochl1994}, where plane-waves are generated with the cutoff energy of \SI{600}{\electronvolt}. For exchange-correlation functional, we employ the Perdew-Burke-Ernzerhof (PBE) functional within the generalized gradient approximation (GGA)~\cite{pbe} scheme. As a common condition, the forces on atoms are relaxed down to the accuracy level of \SI{1E-3}{\electronvolt\per\angstrom} and the energy convergence criteria is set as \SI{1E-6}{\electronvolt}. To determine the H adsorption energy at different H concentration and doping concentration levels, supercell sizes of 3×3×1 and 4×4×1 were considered for the calculations with a vacuum of 15 \AA~ in the direction normal to the layer to avoid the interactions with repeated images. The integration in the reciprocal space is carried out on 3${\times}$3${\times}$1 and 5${\times}$5${\times}$1 k-point grid (adapted depending on simulation cell size) generated within Monkhorst–Pack scheme~\cite{monkhorst1976}. The dopant is introduced in these cells by replacing one of M/S/Se, see Fig.~\ref{fig:ballstick}(a), by the dopant $Z$, thereby yielding the doping concentration of 2.1\% and 3.8\%. Different H adsorption sites on and near the dopant, Fig.~\ref{fig:ballstick}(b), are considered to generate the required dataset for further ML modeling. The $E^H_a$ is calculated for all H adsorption sites in two doping concentrations to infer the effect of doping concentration in predicting the $E^H_a$.

\begin{figure*}[tb]
     \centering
     \includegraphics[width=\textwidth]{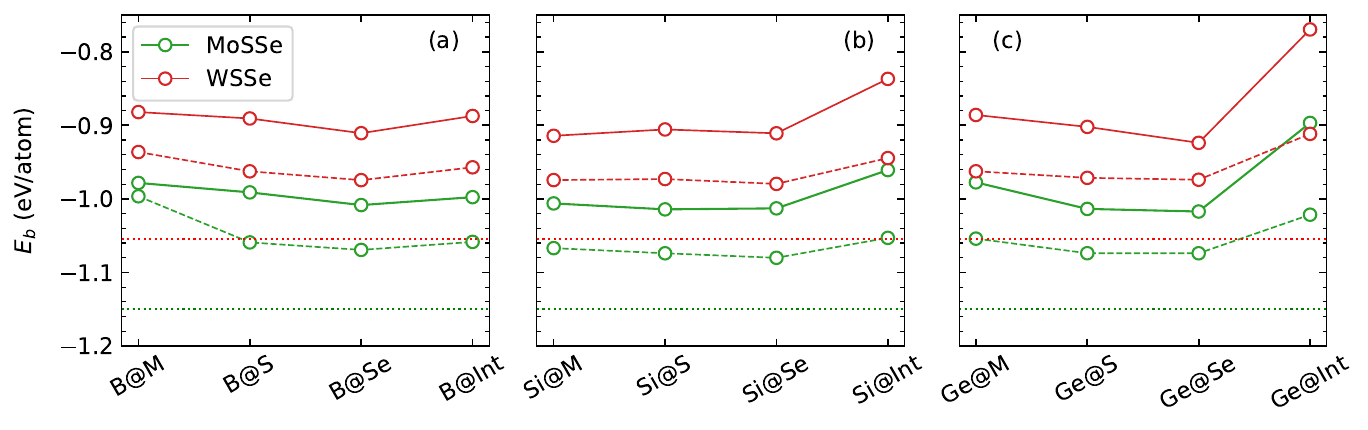}
     \caption{Calculated binding energy $E_b$ of (a) B, (b) Si, and (c) Ge doping in MoSSe (green lines) and WSSe (red lines) host layers. Driven by the atomic size effect, the binding strength in Z@Int doping is less than that in the remaining doping sites, except for the B dopant. Overall, the chalcogen sites are the energetically preferred doping sites. The Z@M notation denotes doping at Mo in MoSSe and W in WSSe atomic sites. The continuous and dashed lines represent the $E_b$ calculated with $3\times3\times1$ and $4\times4\times1$ cells, respectively. The dotted red and green lines represent the binding energy of pristine MSSe layers, respectively.}
     \label{fig:ebind}
\end{figure*}

\section{\label{sec:results}Results and discussion}
\subsection{\label{sec:binding}Dopant binding in MSSe JLs}

The stability and interaction of the dopant in MSSe JLs are analyzed by the binding energy per atom $E_b$ of Z with the host layers. The $E_b$ is calculated by the expression,
\begin{equation}
E_b = \frac{1}{N}(E_\text{Z-MSSe} - n_\text{M} \cdot E_\text{M} - n_\text{S} \cdot E_\text{S} - n_\text{Se} \cdot E_\text{Se} - n_\text{Z} \cdot E_\text{Z})
\label{eq:eb}
\end{equation}
where $E_\text{Z-MSSe}$ is the ground state energy of $Z$ doped MSSe layers. The ground state energies $E_\text{M}$, $E_\text{S}$, $E_\text{Se}$, and $E_\text{Z}$ of bulk crystals or dimers of M, S, Se, and $Z$ elements, respectively. The total number of atoms $N$ taken in the simulation is equal to the sum of the number of individual atomic species $n_\text{M}$, $n_\text{S}$, $n_\text{Se}$, and $n_\text{Z}$. The bulk crystals of Mo (space group: 229; symmetry: \texttt{I$\Bar{m}$3m}), W (space group: 229; symmetry: \texttt{I$\Bar{m}$3m}), B (space group: 166; symmetry: \texttt{R$\Bar{3}$m}), Si (space group: 227; symmetry: \texttt{Fd$\Bar{3}$m}), and Ge (space group: 227; symmetry: \texttt{Fd$\Bar{3}$m}) are used to get their ground state energies. For chalcogens, S$_2$ and Se$_2$ dimers are considered. The calculated $E_b$ is shown in Fig.~\ref{fig:ebind}. According to the definition in Eq.~\ref{eq:eb}, a negative (positive) $E_b$ denotes an attractive (repulsive) interaction between the dopant and host layer, indicating a stable (unstable) doping scenario. As shown in Fig.~\ref{fig:ebind}, all doping scenarios yield negative $E_b$, exposing a stable and energetically feasible dopants and doping configurations for MSSe JLs. Particularly, the dopants are more stable at chalcogen sites. In comparison, the $4\times4\times1$ supercell yields a better $E_b$ prediction than the $3\times3\times1$ supercell. For reference, we also plot the binding energy of pristine MSSe with the green and red dotted lines in Fig.~\ref{fig:ebind}. As expected, the pristine MSSe layers show larger binding energies compared to the doped layers.

The atomic size of the dopant strongly influences the $E_b$. For instance, B shifts toward the S atomic plane due to its small size in B@Mo doped layer. Also, B intercalates favorably at the interstitial site, turning the $E_b$ comparable to that of B@S doping. In contrast, larger dopants such as Si and Ge are energetically less stable at interstitial positions, see Fig.~\ref{fig:ebind}. Furthermore, dopants induce charge redistribution, activating the nearest-neighbour atoms, while atoms beyond the first coordination shell recover their pristine charge state, as in undoped MoSSe. This behavior is particularly pronounced for Z@Mo doping \cite{Vallinayagam-JMCA2024} and is expected to occur similarly in $Z$-doped WSSe layers.

\begin{figure*}[tb]
    \centering
    \includegraphics[width=\textwidth]{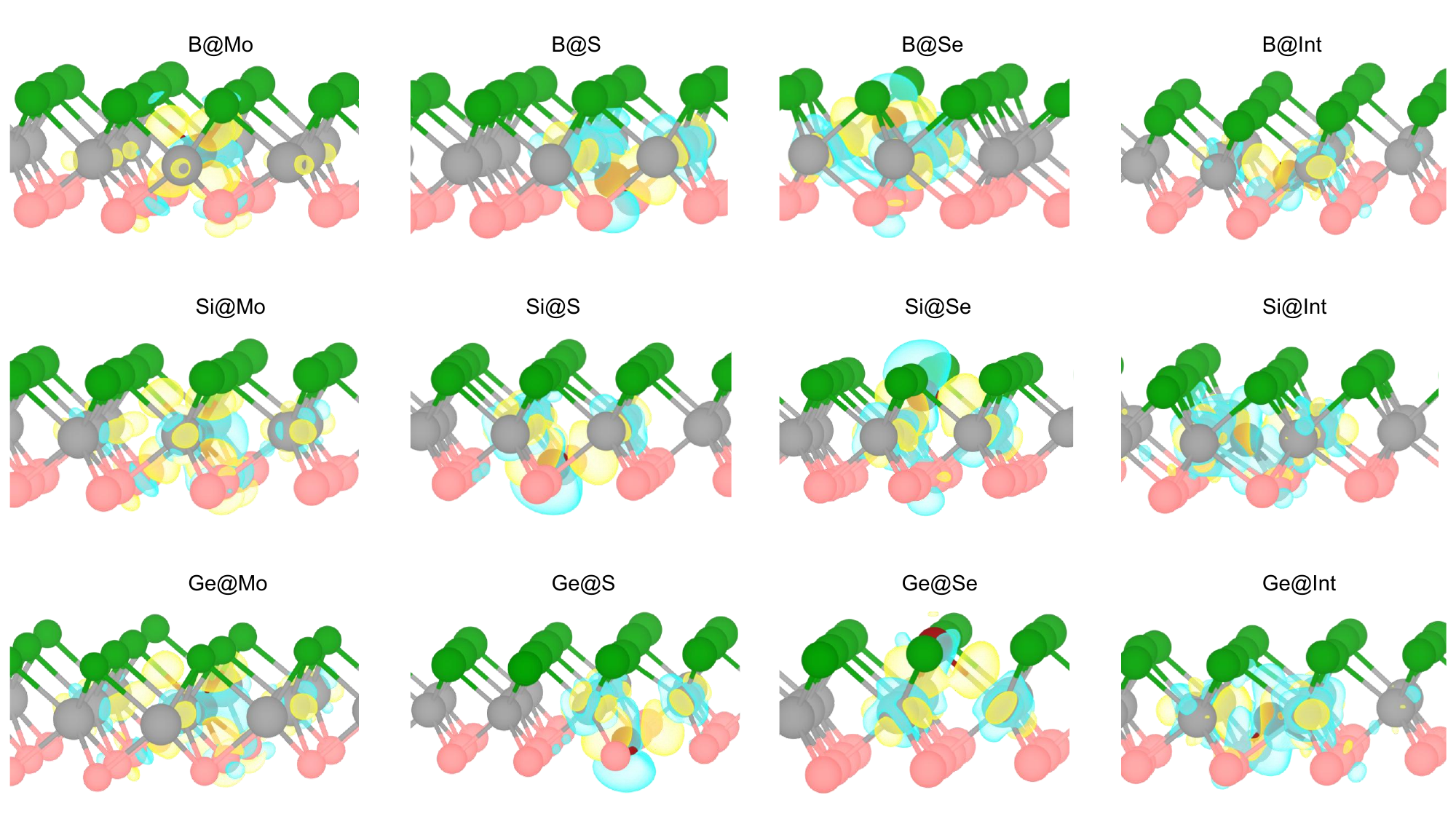}
     \caption{Calculated differential charge density for B, Si, and Ge doped MoSSe JL, respectively. The differential charge density is defined as $\rho(Z+MoSSe)-\rho(MoSSe)-\rho(Z)$. Here $\rho(Z+MoSSe)$ and $\rho(Z)$ are the charge densities of doped MoSSe and dopant, respectively. The term $\rho(MoSSe)$ represents the charge density of dopant removed MoSSe JL.}
     \label{fig:cdd}
\end{figure*}

\subsection{\label{sec:DCD}Charge density in doped MSSe}

To understand charge redistribution when a dopant atom is introduced at distinct sites of the MSSe Janus layers, we calculate the differential charge density (DCD). As a representative case, we present the DCD in the MoSSe when dopants B, Si, and Ge are introduced at the Mo, S, Se, and interstitial sites, which is shown in Fig.~\ref{fig:cdd}. From the valence electron number of B (3: 2s$^2$2p$^1$), Si(4: 3s$^2$3p$^2$), and Ge(4: 4s$^2$4p$^2$), we expect that the doping in MSSe leads to p-type doping. But, on the contrary, depending on the doping site, the properties of the dopants result in both p-type and n-type doping. It is clear from Fig.~\ref{fig:cdd} that the charge redistribution is confined only around the dopant, and the atoms that are far from the dopant have not been affected. The presence of dopants in the MSSe layers redistributes the charges in the nearby atomic sites, tuning them as activated sites~\cite{Vallinayagam-JMCA2024}. Since the HER is an electron-demanding reaction, the n-type doping in MSSe layers is favorable.

In the case of Z@Mo (Mo: 4d$^5$5s$^1$), a similar trend of charge distribution is seen for B, Si, and Ge. The charge depletion is found near the dopant atom and is being transferred to the S and Se atomic planes. However, due to the large electronegativity of S atoms, more charge accumulation is seen towards the S atomic plane compared to that of Se. From the observed charge distribution, we can infer that doping the Z@M site of MSSe leads to n-type doping. The pronounced interaction between the dopant and the first nearest Mo and S atoms surpasses the Z-Se interaction. Specifically, the charge distribution is markedly concentrated along the Z-S bond. Conversely, the Z-Se interaction is scarcely perceptible, reinforcing the assertion that the Z-S interaction prevails over the Z-Se interaction. Consequently, the S sites emerge as more optimal for the HER compared to Se. Therefore, the electronegativity and atomic size differences between the dopant and host atoms alter the charge states of nearby chalcogens, making them a favorable site for HER activity.  

When B is introduced in the S(3s$^2$3p$^4$) atomic plane, the electron accumulation is seen near the B atom (-0.33 e), which is confirmed by Bader analysis. Thus, doping B at the S site of MoSSe leads to p-type doping. In contrast to the B, the doping of Si and Ge at S shows electron depletion near the dopant and accumulation on the first nearest neighbour three Mo atoms. Therefore, the dopants Si and Ge act as donors when introduced at the S atomic plane. From Fig.~\ref{fig:cdd} it is clearly evident that the introduction of the dopant in the S atomic plane does not influence the charge distribution on the Se(4s$^2$4p$^4$) atomic plane. A similar distribution of charge is observed for the doping of Z in the Se atomic plane as that of the Z doping in the S atomic plane. The dopant B, by accepting the charge, leads to p-type doping, while Si and Ge, by depleting the charge, result in n-type doping. When the doping at the interstitial site is concerned, the three dopants, B, Si, and Ge, donate charge, which is distributed across the interstitial region and has not transferred to either of the S or Se atomic planes as observed from Fig.~\ref{fig:cdd}. The presence of the dopant at distinct atomic and interstitial sites alters the local bonding character and charge distribution within the structure, which in turn induces spin polarization and significantly affects the electronic properties of the structure.

\subsection{Magnetic moment and electronic structure of doped MSSe}

\begin{figure*}[htp]
     \centering
     \includegraphics[width=\textwidth]{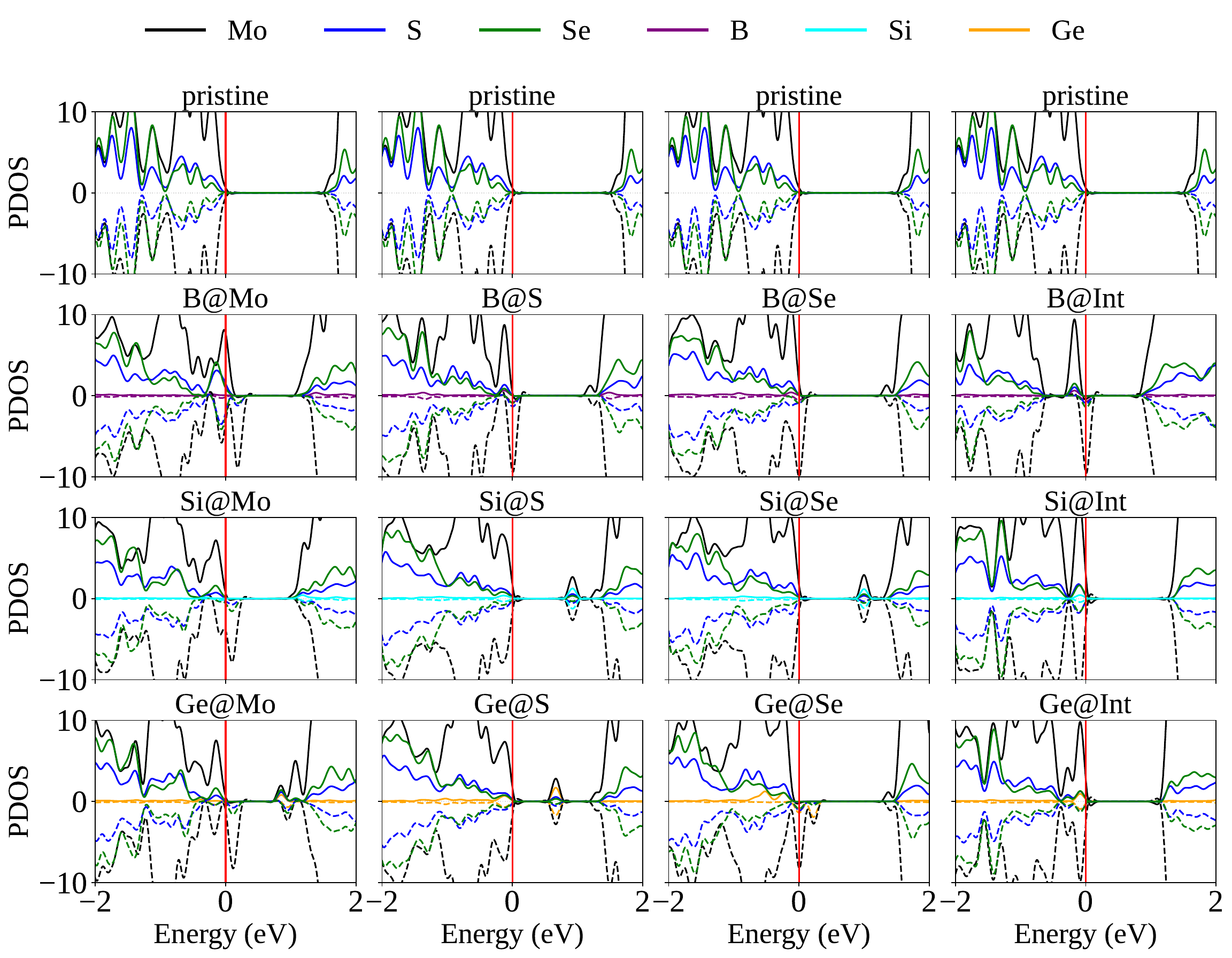} %16cm
     \caption{Atom resolved density of states of pristine and B, Si, and Ge doped MoSSe layer at a doping concentration of 2.1\% (4$\times$4$\times$1 supercell). The solid red line represents the Fermi-level. }
     \label{fig:PDOS}
\end{figure*}

The spin-polarized electronic structure calculations reveal the presence of defect states within the band gap for all doped systems. The observed GGA band gaps of pristine MSSe are 1.56 eV and 1.69 eV, respectively. From the electronic DOS (see Fig.~\ref{fig:PDOS} and SFig. 1-3), it is clear that the impurity states are present near the Fermi level, which reduces the original band gap of MSSe. With respect to the size of the chosen supercell, the DOS of doped MSSe shows dissimilar trends. However, for a particular size of the supercell, the MSSe DOS show similar behaviour. As a representative case in Fig.~\ref{fig:PDOS}, we present the atom-resolved DOS for the Z doping in a 4$\times$4$\times$1 supercell of MoSSe, representing the doping concentration of 2.1\%. From Fig.~\ref{fig:PDOS}, it is observed that the doping of B at all the considered four distinct sites shows a shift between the spin-up and spin-down states, resulting in a net magnetic moment. The observed magnetic moments are given in Table~\ref{tab:mag}. In the case of B doping, the highest magnetic moment is observed for B@Mo doping, with a value of 1.64 $\mu_B$. The major contribution to this moment comes from the Mo-d (1.10 $\mu_B$) orbitals with a small contribution from the S-p and Se-p orbitals (0.24 $\mu_B$), as also evident from the spin density given in SFig. 6. The moments are mainly localized around six Mo atoms, which are the second-nearest neighbours to the dopant B. In the case of B@S, B@Se, and B@Int as well, the magnetic moment is mainly induced due to the Mo-d orbitals. Overall, doping B in MoSSe at four distinct sites has resulted in a metallic behavior due to the crossing of Mo-d states above the Fermi level. Spin-up carriers dominate the conducting channel in B@Mo, while the down-spin carriers provide channels for B@S, B@Se, and B@Int.

Belonging to the same group in the periodic table, having the same number of valence electrons, the doping of Si and Ge shows a similar trend in the magnetic moment and electronic structure for doping at distinct sites in MoSSe (as well as in WSSe), except for the doping in the Se atomic plane. In the case of Si(Ge) doping at the Mo site, MoSSe shows half-metallic behavior due to the Fermi crossing of the spin-down Mo-d state and has resulted in a net magnetic moment of 2.02 (2.03) $\mu_B$. The corresponding spin density is shown in SFig. 6, which indicates an intense spin polarization on the first nearest three Mo atoms. The doping of Si at S, Se, and interstitial sites has resulted in zero magnetization, and MoSSe exhibits semiconducting behavior with band gaps of 0.93 eV, 1.00 eV, and 1.38 eV, respectively. The doping of Ge@S and Interstitial sites shows zero magnetic moment, and the band gaps are 0.67 eV and 1.25 eV, respectively. However, in the case of Ge@Se, we observe a magnetic moment of 2.00 $\mu_B$ and half-metallic behaviour with the Fermi crossing of the Mo-d states and Ge-p states. The observed magnetic moments reflect this with net magnetization of the MoSSe having main contributions from the Mo-d (0.90 $\mu_B$) and Ge-p (0.54 $\mu_B$) orbitals.

A similar trend to that of the 4$\times$4$\times$1 supercell of MoSSe in terms of magnetic moment and band gaps is observed in the 4$\times$4$\times$1 WSSe Janus layer as well, which are provided in SFig. 3 and SFig. 7. However, the largest magnetic moment of 3.03 $\mu_B$ is found in WSSe in the case of B doping at the W atomic site, which is mainly coming from the W-d (1.49 $\mu_B$) and S-p and Se-p (0.56 $\mu_B$) states. The observed spin densities and the electronic DOS for the 3$\times$3$\times$1 supercells of MSSe are given in SFig. 1-2 and SFig. 4-5.

\begin{table*}[htp]
\centering
\caption{Magnetic moment $m$ and band gap E$_g$ information for the MSSe JLs with the dopant substitution at distinct sites and concentration levels. }
\label{tab:mag}
\begin{ruledtabular}
\begin{tabular}{cccccccccc}
S.No & Structure & Dopant & \textbf{$m$ ($\mu_B$)} & \textbf{$E_g$ $(eV)$} &
S.No & Structure & Dopant & \textbf{$m$ ($\mu_B$)} & \textbf{$E_g$ $(eV)$} \\
\midrule
1 & \multirow{12}{*}{MoSSe} & B@Mo & 0.99 & 0.00 & 
25 & \multirow{12}{*}{MoSSe} & B@Mo & 1.64 & 0.00 \\

2 &  & B@S & 0.00 & 0.01 &
26 &  & B@S & 1.10 & 0.00 \\

3 &  & B@Se & 0.97 & 0.00 &
27 &  & B@Se & 1.03 & 0.00 \\

4 &  & B@Int & 0.92 & 0.00 &
28 &  & B@Int & 1.06 & 0.00 \\

5 & $3\times3\times1$ & Si@Mo & 1.90 & 0.03 &
29 & $4\times4\times1$ & Si@Mo & 2.02 & 0.00 \\

6 &  & Si@S & 0.00 & 0.77 &
30 &  & Si@S & 0.00 & 0.93 \\

7 &  & Si@Se & 2.07 & 0.00 &
31 &  & Si@Se & 0.00 & 1.00 \\

8 &  & Si@Int & 0.00 & 1.37 &
32 &  & Si@Int & 0.00 & 1.38 \\

9 &  & Ge@Mo & 1.87 & 0.00 &
33 &  & Ge@Mo & 2.03 & 0.00 \\

10 &  & Ge@S & 0.00 & 0.49 &
34 &  & Ge@S & 0.00 & 0.67 \\

11 &  & Ge@Se & 2.02 & 0.00 &
35 &  & Ge@Se & 2.03 & 0.00 \\

12 &  & Ge@Int & 0.00 & 1.18 &
36 &  & Ge@Int & 0.00 & 1.25 \\
%%%%%%%%%%%%%%%%%%%%%%%%%%%%%%%%%%%%%%%%%%%%%%%%%%%%%%%%%
\hline
13 & \multirow{12}{*}{WSSe} & B@W & 0.92 & 0.00 & 
37 & \multirow{12}{*}{WSSe} & B@W & 3.03 & 0.17 \\

14 &  & B@S & 0.00 & 0.01 &
38 &  & B@S & 1.01 & 0.00 \\

15 &  & B@Se & 0.69 & 0.00 &
39 &  & B@Se & 0.07 & 0.00 \\

16 &  & B@Int & 0.87 & 0.00 &
40 &  & B@Int & 1.10 & 0.00 \\

17 & $3\times3\times1$ & Si@W & 1.81 & 0.00 &
41 & $4\times4\times1$ & Si@W & 1.75 & 0.00 \\

18 &  & Si@S & 0.00 & 0.63 &
42 &  & Si@S & 0.00 & 0.82 \\

19 &  & Si@Se & 1.90 & 0.00 &
43 &  & Si@Se & 0.00 & 0.95 \\

20 &  & Si@Int & 0.00 & 1.49 &
44 &  & Si@Int & 0.00 & 1.49 \\

21 &  & Ge@W & 1.85 & 0.00 &
45 &  & Ge@W & 1.92 & 0.07 \\

22 &  & Ge@S & 0.00 & 0.38 &
46 &  & Ge@S & 0.00 & 0.58 \\

23 &  & Ge@Se & 2.05 & 0.15 &
47 &  & Ge@Se & 2.00 & 0.00 \\

24 &  & Ge@Int & 0.00 & 1.28 &
48 &  & Ge@Int & 0.00 & 1.34 \\

\end{tabular}
\end{ruledtabular}
\end{table*}
\begin{figure*}[htp]
     \centering
     \includegraphics[width=\textwidth]{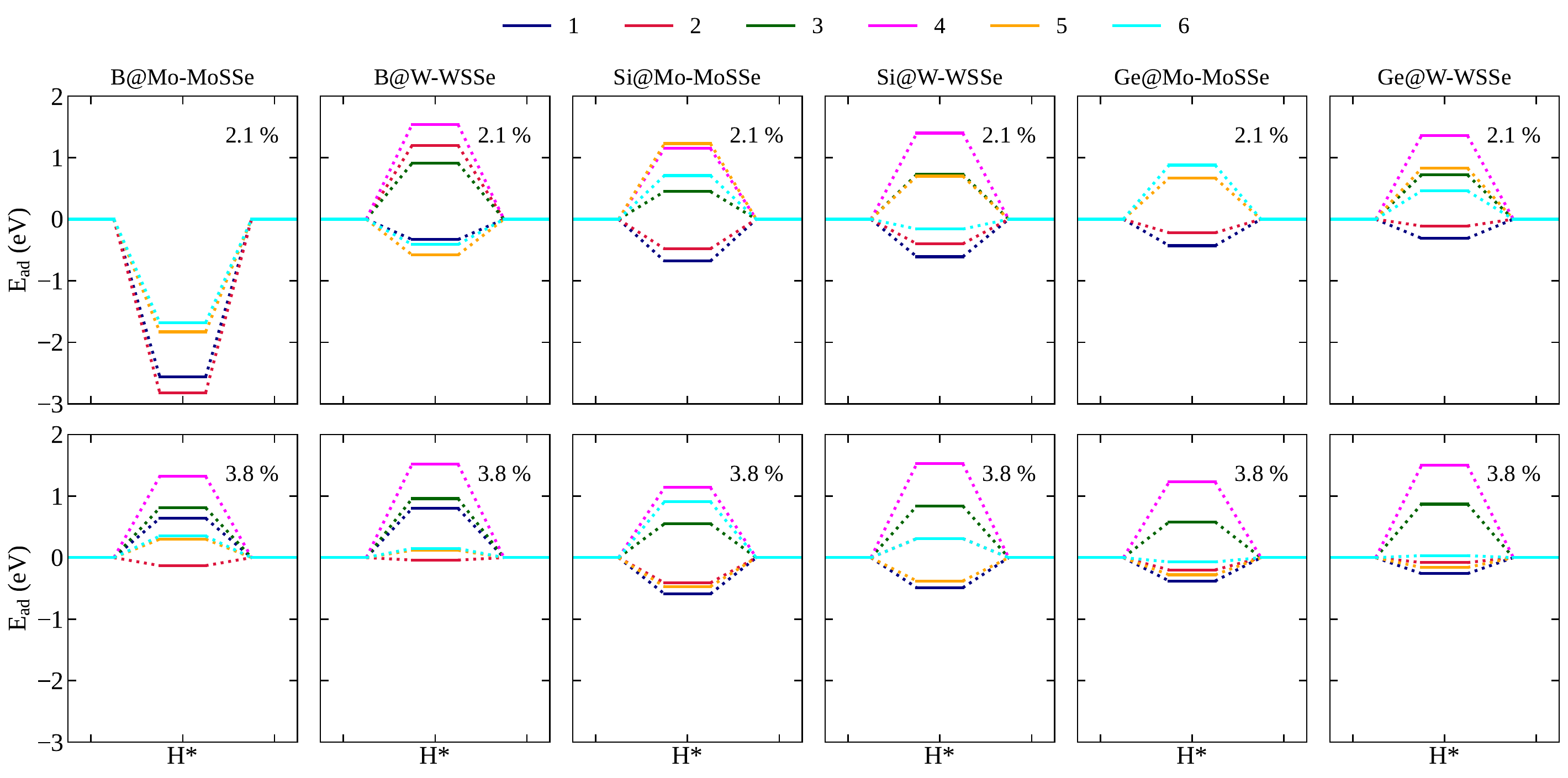} %16cm
     \caption{Atomic H adsorption energies (E$_{ad}$) when the metalloid elements B, Si, and Ge are on a M site of MSSe JLs under the doping concentrations (DCs) 2.1\% and 3.8\%.}
     \label{fig:xatm}
\end{figure*}

\subsection{\label{sec:Hadsorb}H atom adsorption energies in MSSe JLs}
In the ref.~\cite{Vallinayagam-JMCA2024}, we have discussed the behavior of the H atom in the MoSSe JL in the presence of the metalloids B, Si, and Ge, doped at distinct atomic and interstitial sites at a doping concentration of 2.1\%. In this study, a similar schematic is applied to investigate the effect of dopant concentration on H adsorption in MSSe JLs (M = Mo, W). The dopants are substituted in MSSe at four different sites, as mentioned in Fig.~\ref{fig:ballstick} (a).
By analyzing the charge distribution in $Z$ doped JLs, various feasible H adsorption sites are selected and the selection is depicted in Fig.~\ref{fig:ballstick}(b).  Furthermore, these simulations explore the effects of doping concentration and H coverage on the adsorption energy of H, which are internally controlled by cell size. The doping concentrations of 2.1\% and 3.8\% are finally achieved, and the H adsorption behavior is analyzed for each doping scenario. The interaction of atomic H with distinct sites on the MSSe Janus layers is probed by calculating the adsorption energy, which can be defined as 
\begin{equation}
    E_{ad} = E_{Z@MSSe+H} - E_{Z@MSSe} - \frac{1}{2} E_{H_2}
\end{equation}

Where $E_{Z@MSSe+H}$, $E_{Z@MSSe}$ are the total energies of the Z-doped MSSe layer energies with and without H atom adsorption, and $E_{H_2}$ is the total energy of the H$_2$ molecule. The positive and negative values of E$_{ad}$ indicate endothermic and exothermic behavior of H on the MSSe layers. For the optimal HER, we require such scenarios where the E$_{ad}$ is only slightly negative. Too large a negative E$_{ad}$ also indicates the strong binding of H towards the surface, which is not favorable for HER, since the second step of HER, known to be the H$_2$ releasing step, will become challenging. In pristine MSSe, the values of E$_{ad}$ for the H adsorption on top of the S (Se) atom are found to be 1.67 eV (2.19 eV) and 1.90 eV (2.12 eV), respectively, in a 4$\times$4$\times$1 supercell~\cite{Vallinayagam-JMCA2024, Tejaswini-JMCA2025}. In the case of a 3$\times$3$\times$1 supercell of pristine MSSe layers, the E$_{ad}$ values for the adsorption on top of the S (Se) atom are 1.61 eV (1.88 eV) and 1.89 eV (2.03 eV), respectively~\cite{Tejaswini-JMCA2025}. The E$_{ad}$ values indicate an extensive repulsive interaction of H in pristine MSSe under the H coverage of 2.08\% (4$\times$4$\times$1 supercell) and 3.70\% (3$\times$3$\times$1). Herein, we discuss the trends in E$_{ad}$ with respect to the site of the dopant. 

\subsubsection{Doping at M site}

In the $Z$@M scenario, the selected dopants are substituted at the M site in MSSe JLs, irrespective of differences in the atomic size of the dopant and host atoms. The $Z$@M doping alters the charge distribution and coordination with nearest neighbours due to the size effect. For example, B@Mo in MoSSe JL reduces the effective coordination due to the shift of B towards the S-atomic plane~\cite{Vallinayagam-JMCA2024}. These changes in dopant sites alter the charges on constituent atoms, producing activated sites. The six distinct sites considered for the H adsorption are shown in Fig.~\ref{fig:ballstick}(b). The H adsorption on the dopant has two possibilities, where H can interact through either the S(5) or the Se(6) plane.  The calculated E$_{ad}$ for the six distinct H adsorption sites in the Z@M doping case are given in Fig.~\ref{fig:xatm}. In the MoSSe JL, at 2.1\% doping concentration (DC), the adsorption energy for sites 3 and 4 is the same as that for 5 and 6 when the doping atom is B or Ge. But this is not the case with Si, where all six sites differ in E$_{ad}$. However, E$_{ad}$ differs for all six sites in the WSSe JL for both DCs. For B@Mo doped MoSSe JL, the chemisorption dominates at a DC of 2.1\% for all four sites with a large negative E$_{ad}$, which indicates the strong binding of H with the layer and is not favorable for HER. For the same B@Mo case, at 3.8\% doping, the mentioned six distinct sites show different E$_{ad}$. For site 2, the E$_{ad}$ is slightly negative, indicating spontaneous H adsorption and a less negative value, which favors HER. For B@W of WSSe JL, the adsorption of H is endothermic for sites 2, 3, and 4, whereas it is exothermic for sites 1, 5, and 6 at a DC of 2.1\%. The sites 2, 3, and 4, which represent the H adsorption on the immediate Se neighbour and the next appearing S and Se neighbours to the dopant, possess the same charge as that in the case of pristine WSSe. This is due to the charge delivered by the B atom being mainly transferred to the first nearest, highly electronegative three S atoms, as is evident from the Bader analysis. On top of the dopant atom from either of the sides, S  and Se E$_{ad}$ values are found to be negative, which indicates that the B-H bond is stronger than that of the S-H and Se-H bonds. The H adsorption on S is favourable only when the dopant is acting as a donor in the MSSe layers. At a doping concentration of 3.8\%, all sites show positive E$_{ad}$ with sites 2 and 5 having E$_{ad}$ close to zero, and the E$_{ad}$ for site 2 is found to be slightly negative (-0.04 eV). At 3.8\% DC of B@Mo of MoSSe and B@W of WSSe, the E$_{ad}$ shows exactly similar behavior for both MSSe for all six sites. This is not the same for the DC 2.1\%.

In the case of Si@Mo in MoSSe, for the two DC levels, the E$_{ad}$ shows a similar behavior except for the case of site 5. For site 5, i.e., on top of the Si atom facing the S-atomic plane, the E$_{ad}$ is 1.23 eV for the 2.1\% DC, and it is -0.47 eV for 3.8\% DC. For the Si@W in WSSe, the H adsorption is endothermic for sites 3, 4, and 5, whereas it is exothermic for sites 1, 2, and 6. The E$_{ad}$ for sites 3 (0.73 eV) and 4 (1.40 eV) at 2.1\% DC is increased by 0.10 and 0.13 eV for the DC of 3.8\%. Similarly, for sites 1 and 2, E$_{ad}$ increased by an amount of 0.11 and 0.09 eV, respectively, from the DC 2.1\% to 3.8\%. At the DC of 2.1\%, E$_{ad}$ is negative, indicating the exothermic reaction for sites 2 and 6, and these reactions became endothermic at 3.8\% DC. In the case of 2.1\%, the E$_{ad}$ of MSSe JLs, for sites 1 and 2, the E$_{ad}$ is nearly equal (differs only by 0.08 eV). At a DC of 3.8\%, the E$_{ad}$  for sites 1 and 5 differ only by 0.10 eV and 0.09 eV, respectively.

For 2.1\% Ge doping, at Mo of MoSSe, sites 3 and 4 have shown the same adsorption behavior as that of 1 and 2, which is a similar scenario as that of the  B@Mo case. The H adsorption is found to be endothermic for sites 5 and 6, whereas it is exothermic for sites 1 and 2. When the doping concentration is increased, all four sites, 1, 2, 5, and 6, show exothermic behavior. In contrast to the DC of 2.1\%, sites 3 and 4 show dissimilar behavior compared to sites 1 and 2 at DC 3.8\%, which are found to be highly endothermic. In the case of Ge@W of WSSe, at the 3.8\% DC, the H adsorption behavior follows the same trend as that of MoSSe with slightly higher E$_{ad}$. At 2.1\% DC for the sites, 1 and 2, the H adsorption is exothermic, as that of MoSSe, and only slightly higher. Unlike in the case of MoSSe, site 6 has a lower E$_{ad}$ compared to site 5. It means the E$_{ad}$ is lower on the Ge when H is adsorbed on the S(Se) atomic side of MoSSe (WSSe). Similar to that at the 3.8\% DC, the sites 3 and 4 show a positive E$_{ad}$. 

%Overall, in the majority of the cases, it is evident that the E$_{ad}$ increases with an increase in DC. 
Overall, for sites 1--4, the E$_{ad}$ on MoSSe is lower compared to that of the WSSe for both DCs. This result aligns with our analysis of H adsorption energy trends in MXY (M = Mo/W; X/Y = S, Se, Te) layers, which was recently reported~\cite {Tejaswini-JMCA2025}. The reason for this is attributed to the greater electronegativity (EN) of W compared to Mo. Also, in the pristine form of the MSSe JLs, the valence band maximum of WSSe lies at a higher energy than that of the MoSSe. However, this trend is not observed for sites 5 and 6, i.e., when the H atom is adsorbed on the dopant atom. 

\subsubsection{Doping at S site}

For doping at the S site, we consider three distinct adsorption sites for H: on top of S (1), on top of Se (2), and on top of Z (3). In both the MSSe JLs, for sites 1 and 2, we observe a positive E$_{ad}$, and for site 3, we observe a negative E$_{ad}$ at both the doping concentrations. The E$_{ad}$ for WSSe is found to be slightly higher than that of MoSSe for H adsorption on sites 1 and 2, irrespective of the dopant and doping concentration. For site 3, i.e., on top of the dopant atom, E$_{ad}$ are found to be negative in all the cases. This indicates that the Z-H bonding is stronger than the S-H or Se-H bonding. The favorable E$_{ad}$ with less negative ($>$-0.45) are found for B doping at concentrations of 2.1\% and 3.8\%. 
This may be due to the locally broken structural symmetry around the dopant~\cite{Huibin-Appl.Surf.Sci2024}. In all the considered cases, the E$_{ad}$ is larger for site 2 (on top of the Se atom) and varies from 1.14 to 1.29 eV. If we compare the E$_{ad}$ on site-3 across all cases, it is found to be less negative for B compared to Si and Ge, which may be because B acts as a p-type dopant, whereas Si and Ge are n-type dopants.

%Additionally, the E$_{ad}$ value differs slightly for different DCs. For the dopant B, E$_{ad}$ decreases by an amount of 0.04 eV from 2.1\% DC to 3.8\%, whereas for Si and Ge doping, it increases by a maximum of 0.03 eV. A similar trend is seen for site 1 as well. Moreover, a strong attraction of H towards the JL is observed in the case of the Si@S dopant of WSSe.\textcolor{green}{report the change in the bond length here}

\begin{figure*}[htp]
     \centering
     \includegraphics[width=\textwidth]{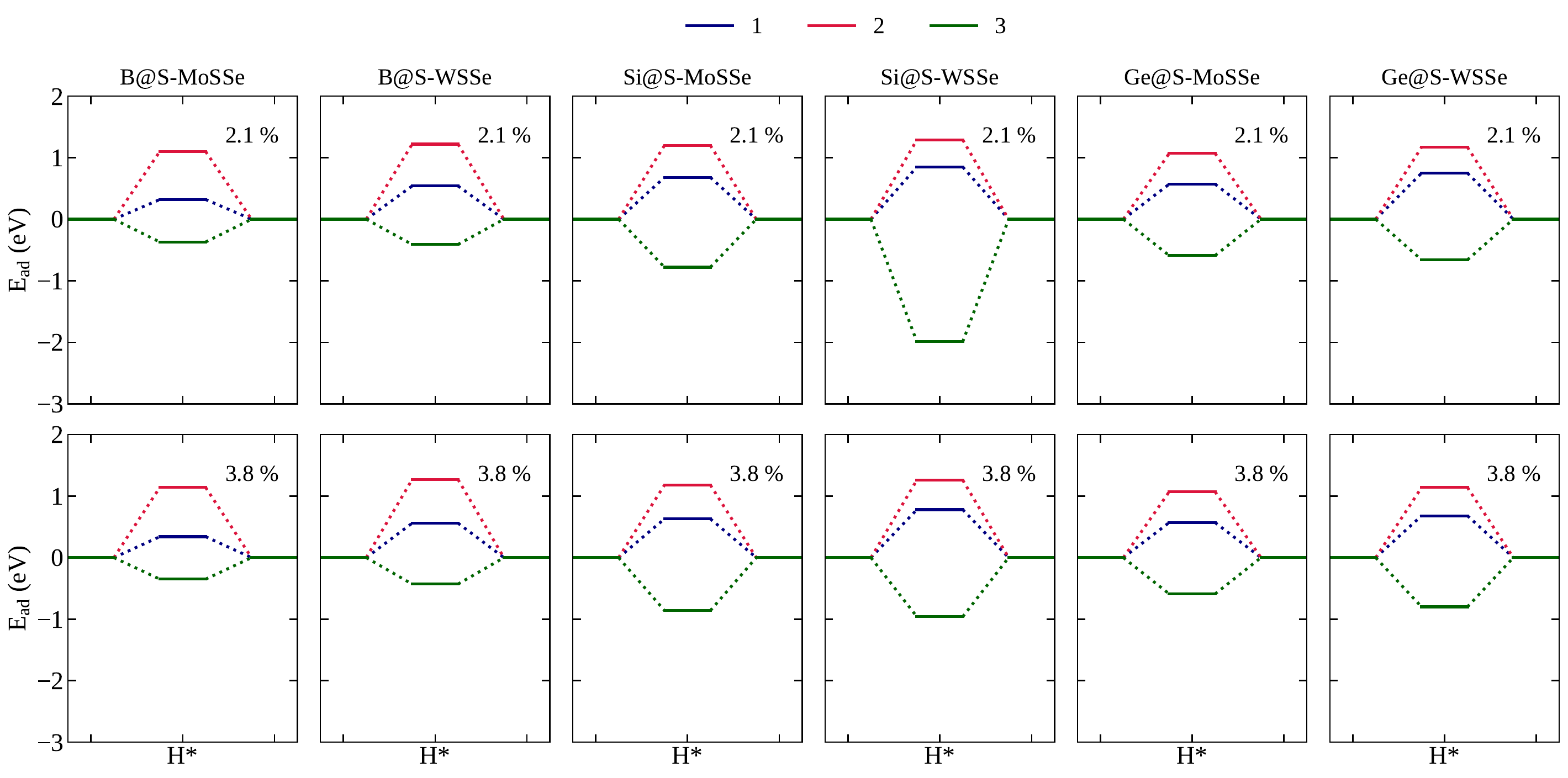} %16cm
     \caption{Atomic H adsorption energies (E$_{ad}$) when the metalloid elements B, Si, and Ge are doped in place of S of MSSe JLs under the doping concentrations (DCs) 2.1\% and 3.8\%.}
     \label{fig:xats}
\end{figure*}

\subsubsection{Dopant at Se site}
The trend for H adsorption observed for the Z@Se case is similar to that observed for the Z@S case. The H adsorption reaction is found to be endothermic for sites 1 and 2, and exothermic for site 3. The E$_{ad}$ on top of Se, i.e., for site 2, are found to be highly endothermic as compared to the other sites. For the H adsorption on site 1 the E$_{ad}$ is slightly positive and is negligible for the cases, Ge@Se-MoSSe (2.1\%), Si@Se-MoSSe(3.8\%) and Ge@Se-MoSSe (3.8\%).
For sites 1 and 2, the E$_{ad}$ on MoSSe are found to be lower than those on the WSSe. This trend is similar to the case seen in pristine MSSe JLs~\cite{Tejaswini-JMCA2025}. Due to the broken symmetry near the dopant, this trend is not observed for adsorption at site 3. For the HER reaction E$_{ad}$ should be only slightly negative. Since the Si and Ge-doped cases have larger negative adsorption energies, the B@Se cases will be more suitable for HER.
\begin{figure*}[htp]
     \centering
     \includegraphics[width=\textwidth]{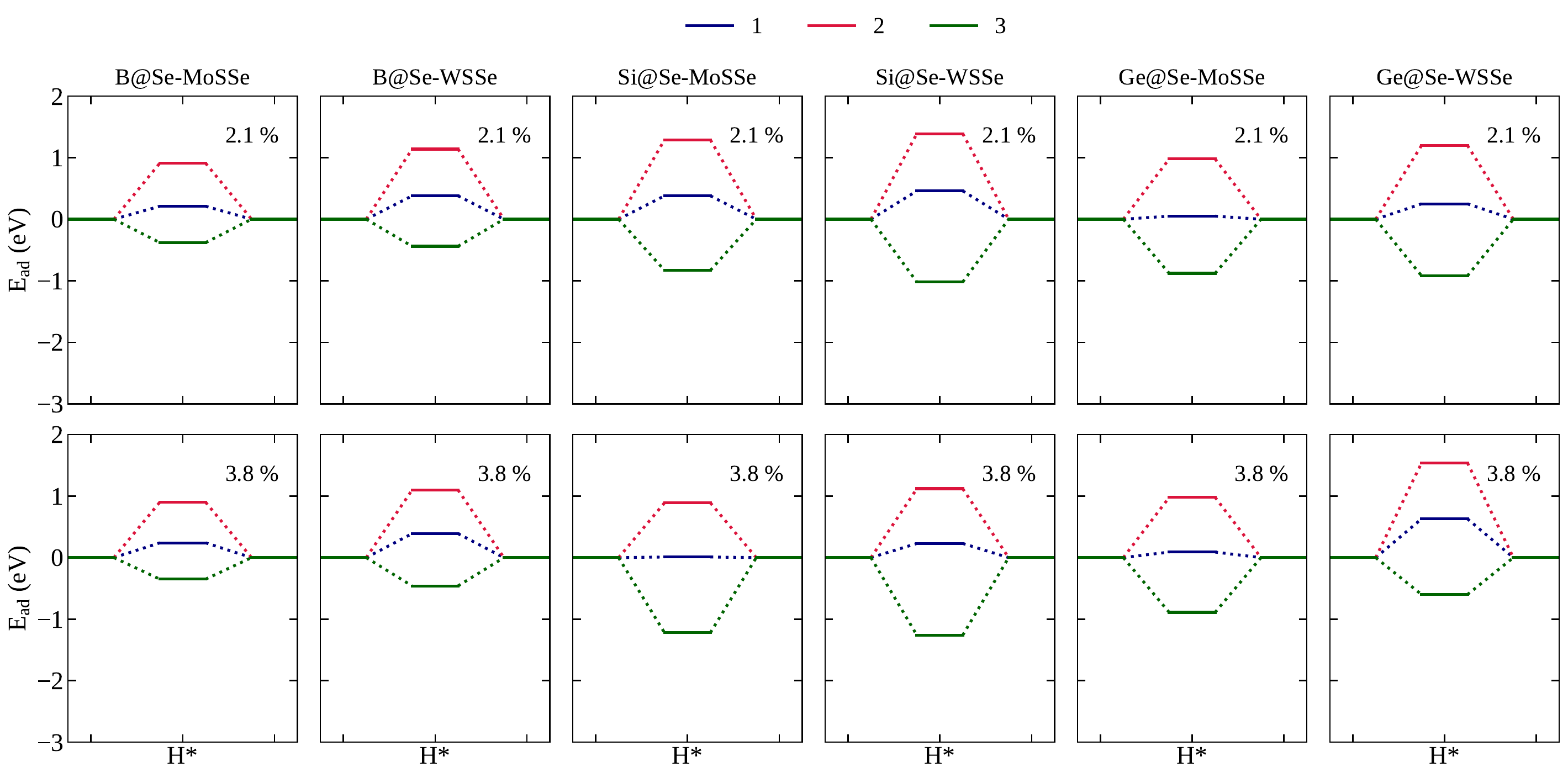}
     \caption{Atomic H adsorption energies (E$_{ad}$) when the metalloid elements B, Si, and Ge are doped on a Se site of MSSe JLs under the doping concentrations (DCs) 2.1\% and 3.8\%.}
     \label{fig:xatse}
\end{figure*}

\subsubsection{Dopant at the interstitial site}
\begin{figure*}[htp]
     \centering
     \includegraphics[width=\textwidth]{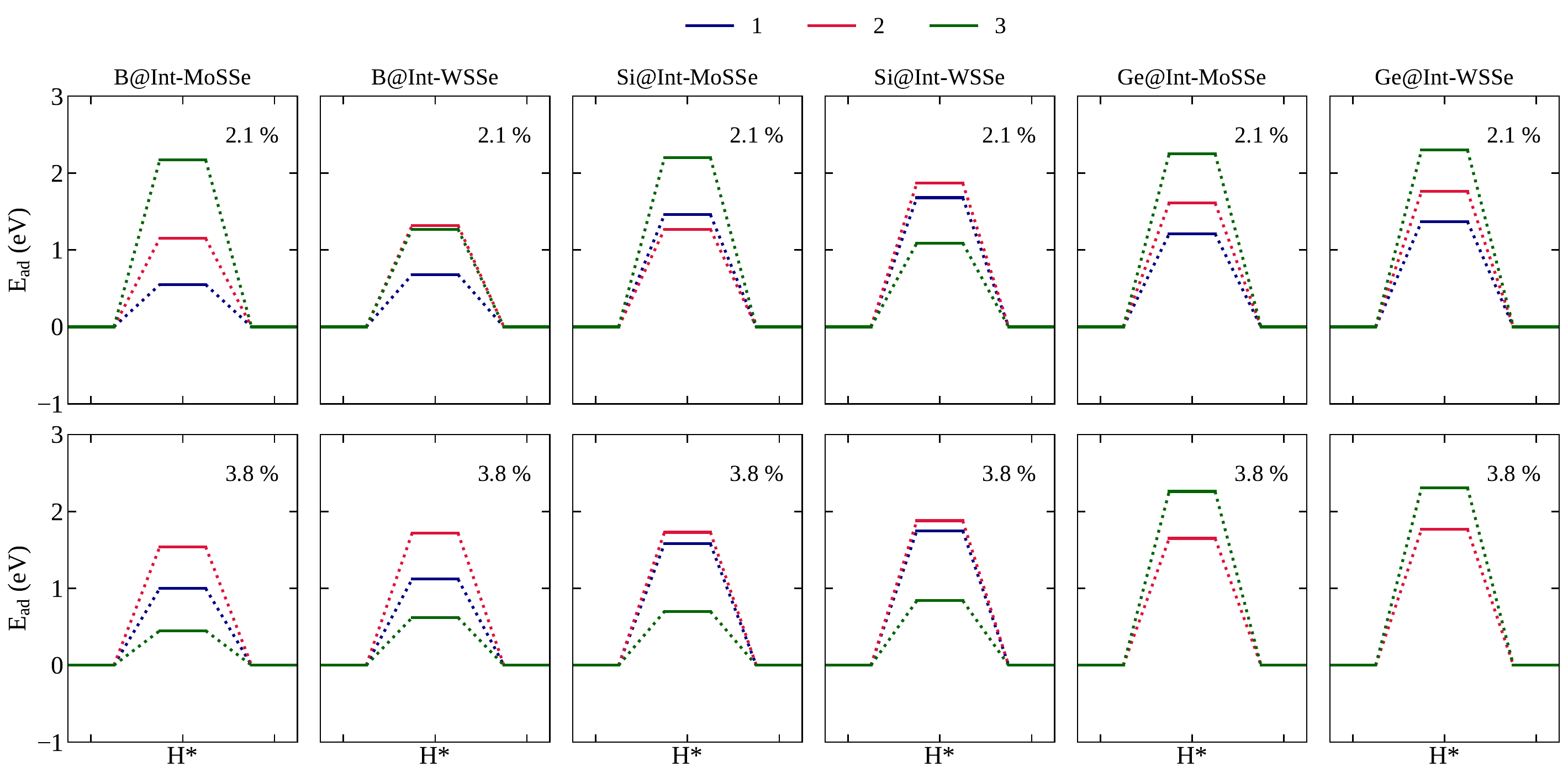}
     \caption{Atomic H adsorption energies (E$_{ad}$) when the metalloid elements B, Si, and Ge are doped at an interstitial of MSSe JLs under the doping concentrations (DCs) 2.1\% and 3.8\%.}
     \label{fig:xatint}
\end{figure*}

Finally, we study the H adsorption behavior when the dopant atom is present at an interstitial site of the MSSe JLs.
Similar to the earlier two cases, Z@S and Z@Se, the H adsorption is studied at three different sites on top of the S atom (1), on top of the Se atom (2), and on the dopant atom (3).
The calculated E$_{ad}$ is as shown in Fig.~\ref{fig:xatint}.
In the presence of the dopant at an interstitial site, the H adsorption is found to be highly endothermic for all the dopants, for all the H adsorption sites, and in both MSSe JLs at different doping concentrations. For the two cases, Ge@int of MSSe at a DC of 3.8\%, convergence is not achieved during relaxation. Hence, those two cases are not shown in Fig.~\ref{fig:xatint}. Though the doping of Z at the interstitial site leads to n-type doping in the MSSe layers, the depleted charge has not transferred to the S or Se atomic planes and is distributed in the interstitial region, which can be observed from Fig.~\ref{fig:cdd}.
At lower doping concentrations, similar H adsorption trends are observed for B and Ge doping.
The largest E$_{ad}$ is found for site 3, whereas the lowest E$_{ad}$ is found for site 1.
For the Si@Int, in MoSSe JL, the largest E$_{ad}$ is found for site 3 and the lowest E$_{ad}$ is found for site 2, with the E$_{ad}$ for site 1 lying in between. The trend is not the same for Si@Int in the WSSe JL.
In WSSe JL, the largest E$_{ad}$ is found for site 2 and the smallest is found for site 3, with the E$_{ad}$ of site 1 lying in between.
At 3.8\% DC, the B and Si doping at the interstitial site are showing a similar trend with the larger E$_{ad}$ corresponding to site 2, and the lowest E$_{ad}$ is observed for site 3, with site 2 lying in between. 

\subsection{Workflow for the ML predictions}
\begin{figure*}[htp]
     \centering
     \includegraphics[width=\textwidth]{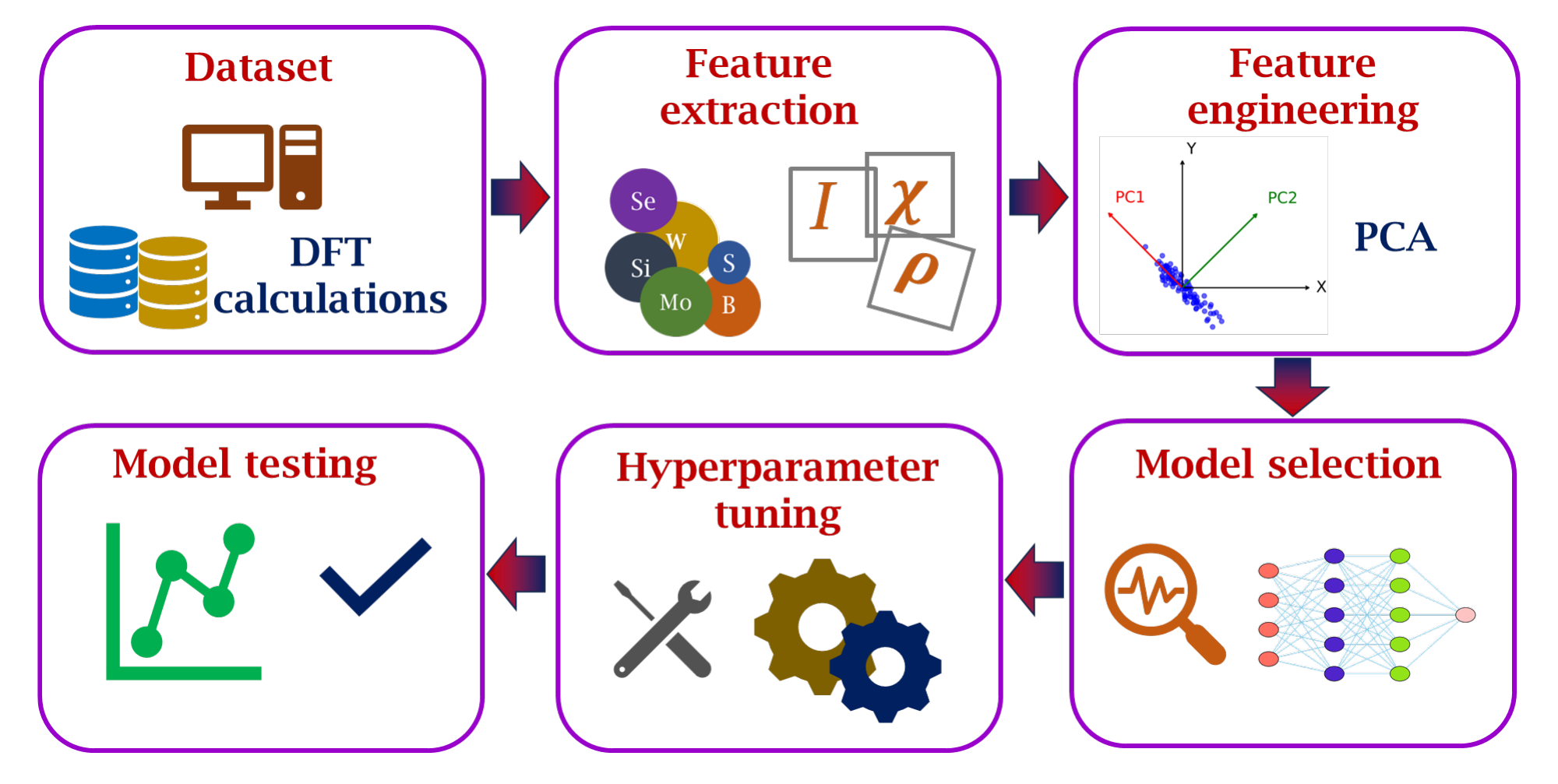}
     \caption{The schematic illustrating the workflow followed to predict the H atom adsorption energies in the presence of substitutional and interstitial dopants B, Si, and Ge in MSSe Janus layers. }
     \label{fig:workflow}
\end{figure*}

Due to the peculiar properties shown by the JLs, the number of studies on these JLs is increasing year by year. Due to the lack of available datasets containing H atom adsorption energies in the presence of dopants, this study will be beneficial for researchers exploring 2D photocatalysts for clean H$_2$ production. Using the DFT generated, we develop an ML model entirely based on the fundamental elemental properties of the atoms. The schematic illustrating the workflow is depicted in Fig.~\ref{fig:workflow}. The step-by-step procedure we follow for the implementation of the ML model is as follows:  

\begin{enumerate}
    \item In the first step, we extract the DFT calculated H atom adsorption energies in the presence of the metallolid dopants B, Si, and Ge under two distinct doping concentrations. All the above-discussed distinct sites for H atom adsorption are considered for each dopant. Thus, we ended up with a total of 178 data points.
    \item In the second step, we identify the elemental features of the atoms that characterize the dopant and its location. The chemical properties of the elements have already been successfully utilized in predicting various physical and chemical properties of materials. The details of the feature selection are discussed in ~\ref{sec:features}
    \item In the third step, we engineer the features into a better representable form for training the ML model. For this purpose, we utilize the dimensionality reduction technique of PCA, which not only reduces the dimension of the feature space but also yields independent features, known as principal components (PCs), that can be used as descriptors for the ML model. 
    \item In the fourth step, we provide the PCs resulting from PCA as input and select the ML model for training. In this scenario, we use the multilayer perceptron regression model.
    \item In the fifth step, we tune various hyperparameters such as the number of hidden layers, the number of neurons in each hidden layer, the type of activation function, the initial learning rate, and the number of epochs till the optimal performance on the training data has been achieved.
    \item In the final step, with the identified best hyperparameter set, we test the performance of the model by calculating the metrics RMSE and R$^2$. The robustness of the model is confirmed by calculating the metrics with 10 distinct random numbers.

\end{enumerate}
\subsection{Neural network model for predicting dopant adsorption energies}
\subsubsection{Dataset preparation}
Based on the discussions above regarding the DFT results, we have a total of 178 data points representing the E$_{ad}$ in the presence of three metalloid dopants, B, Si, and Ge, in the MSSe JLs. Two dopant concentrations are considered by the selection of the 3$\times$3$\times$1 and 4$\times$4$\times$1 supercells. Supercells, with dopant sites Z@Mo, Z@S, Z@Se, and Z@Int. For each dopant, we have 60 E$_{ad}$ data points for H adsorption on different sites as shown in Fig.~\ref{fig:ballstick}(b). Due to the structural distortion and convergence issues, we exclude two data points. The two excluded points are when H is adsorbed on the S atom of MSSe JLs in the presence of the Ge@Int at DC 3.8\%. Thus, finally, we have a total of 178 data points. The details of the E$_{ad}$ dataset are given in STable 1. However, the size of the available data is comparatively small to build a robust ML model. Thus, we adopt the data augmentation technique to expand the dataset by generating slightly deviated copies of the original data, which will be discussed in the \textit{Data augmentation} section. The use of data augmentation not only increases the size of the data, but it also makes the model more generalizable and addresses data imbalance. Using the augmented dataset, we develop an ML model to predict E$_{ad}$. In the following sections, we discuss the details of feature selection and the feature engineering techniques.

\begin{table*}[ht]
\centering
\caption{The notation and physical importance of identified elemental features.}
\label{tab:features}

\begin{adjustbox}{width=\textwidth}
\begin{ruledtabular}
\begin{tabular}{p{6cm} p{2cm} p{9.5cm}}

\textbf{Feature Name} & \textbf{Symbol} & \textbf{Physical Importance} \\
\midrule
Hydrogen concentration & $n_{\mathrm{H}}$ & Includes supercell dependent H coverage \\
Doping concentration & $n_{\mathrm{d}}$ & Represents the amount of doping introduced into the system. \\
Pauling electronegativity difference (eliminated atom vs. dopant) & $\delta\chi$ & Captures the charge transfer variation due to introduction of the dopant. \\
Valency difference (eliminated atom vs. dopant) & $\delta v$ & Describes the bonding character upon the introduction of the dopant \\
Ionization energy difference (eliminated atom vs. dopant) & $\delta I$ & Reflects the charge transfer from/to the dopant \\
Sum of atomic radii of H and adsorption atom & $\sum r^{a}$ & Indicates the H adsorption site and characterizes the bonding \\
Sum of covalent radii of H and adsorption atom & $\sum r^{c}$ & Characterizes the H bonding with the atom on top of which H is adsorbed \\
Atomic radius of first nearest neighbour & $r^{a}_{\mathrm{1nn}}$ & Provides information about the nearest atom influencing H adsorption. \\
Number of second nearest neighbours & $N_{\mathrm{2nn}}$ & Captures the local environment around the H-adsorbed site. \\
Number of metal atoms in SNNs & $N^{\mathrm{M}}_{\mathrm{2nn}}$ & Determines the role of metal atoms in the adsorption behavior. \\
Number of S atoms in SNNs & $N^{\mathrm{S}}_{\mathrm{2nn}}$ & Determines the role of S atoms in the adsorption behavior. \\
Number of Se atoms in SNNs & $N^{\mathrm{Se}}_{\mathrm{2nn}}$ & Determines the role of Se atoms in the adsorption behavior. \\
Number of dopant atoms in SNNs & $N^{\mathrm{Z}}_{\mathrm{2nn}}$ & Reflects the influence of dopant atoms in the SNNs on H adsorption. \\
Average atomic radius of SNNs & $\bar{r}^{a}_{\mathrm{2nn}}$ & Characterizes the local environment around the H-adsorption site \\
Average covalent radius of SNNs & $\bar{r}^{c}_{\mathrm{2nn}}$ & Characterizes the local environment around the H-adsorption site \\
Average electronegativity of SNNs & $\bar{\chi}_{\mathrm{2nn}}$ & Characterizes the local chemical environment around the H-adsorption site \\
Average ionization energy of SNNs & $\bar{I}_{\mathrm{2nn}}$ & Characterizes the local environment around the H-adsorption site \\
Sum of atomic radii of SNNs & $\sum r^{a}_{\mathrm{2nn}}$ & Describes the size-dependent nature of the SNN environment \\
Sum of covalent radii of SNNs & $\sum r^{c}_{\mathrm{2nn}}$ & Describes the bonding characteristics of the SNN environment \\
Sum of electronegativities of SNNs & $\sum\chi_{\mathrm{2nn}}$ & Reflects the collective electron-attracting ability of the SNN atoms. \\
Sum of ionization energies of SNNs & $\sum I_{\mathrm{2nn}}$ & Reflects the collective ionization energy of SNN atoms influencing H adsorption. \\
Chemical potential of M & $\mu^M$ & Distinguishes between MSSe monolayers. \\
Chemical potential of dopant & $\mu^X$ & Distinguishes the dopants B, Si, and Ge. \\

\end{tabular}
\end{ruledtabular}
\end{adjustbox}

\end{table*}

\subsubsection{\label{sec:features}Feature selection and physical importance}
From the DFT analysis, it is clear that E$_{ad}$ strongly depends on the surface of adsorption, charge distribution, dopant site, and the adsorption site. To train any ML model, the first step is to identify the most appropriate, accurate features that best describe the data. Herein, we design the descriptors based on the elemental features of the species and two less expensive features calculated from the DFT.
The identification of the features is done such that the chosen features should distinguish the JLs, dopant elements, and the H adsorption site. The 23 primary elemental features, along with their notation and physical importance, are listed in Table~\ref{tab:features}. 

The identified chemical properties of the elements $n_H$ and $n_d$ give information about the hydrogen coverage and doping concentration in the MSSe JLs. The features, $\delta\chi$, $\delta v$, and $\delta I$, characterizes the modified charge distribution, bonding and electron transfer with the substitution of the dopant. The features $\sum r^{a}$ and $\sum r^{c}$ pin the atom on top of which H adsorption occurs and also describes about the H bonding character to the adsorption site. To distinguish between distinct H adsorption sites on the MSSe JLs, we include the first and second nearest neighbours with respect to the adsorbed H atom. To distinguish between the first and second nearest neighbours, we use a tolerance factor of 0.05\AA. The $r^{a}_{\mathrm{1nn}}$ feature gives information about the atom on top of which the H atom is adsorbed. This means that it provides details on whether the H is adsorbed on S/Se or the dopant Z. Thus, based on the considered sites, one of S/Se/B/Si/Ge will always be the first nearest neighbour. The second-nearest neighbour features play a key role in distinguishing the H adsorption site more clearly. First nearest neighbour features provide only the information of the atom on which H is adsorbed, without distinguishing between all the considered adsorption sites. For example, if we consider the cases of Si@S and Ge@S in MoSSe with H adsorption on the S surface, the first nearest neighbours are the same for both cases. The second nearest neighbour features will give the change in the chemical environment for both cases. Thus, we include the second nearest neighbour features ($N_{\mathrm{2nn}}$, $N^{\mathrm{M}}_{\mathrm{2nn}}$, $N^{\mathrm{S}}_{\mathrm{2nn}}$, $N^{\mathrm{Se}}_{\mathrm{2nn}}$, $N^{\mathrm{Z}}_{\mathrm{2nn}}$) and their arithmetic combinations ($\bar{r}^{a}_{\mathrm{2nn}}$, $\bar{r}^{c}_{\mathrm{2nn}}$, $\bar{\chi}_{\mathrm{2nn}}$, $\bar{I}_{\mathrm{2nn}}$, $\sum r^{a}_{\mathrm{2nn}}$,$\sum r^{c}_{\mathrm{2nn}}$, $\sum \chi_{\mathrm{2nn}}$, $\sum I_{\mathrm{2nn}}$) as features for our dataset. The description of each feature is given in Table~\ref{tab:features}. To make a distinction between the MSSe JLs, we use the chemical potential of the Mo and W atom $\mu^M$ as a feature, which is calculated using DFT with a bulk system of Mo/W as a reference. To differentiate between the dopants, we also include the dopant chemical potential $\mu^{Z}$ as an additional feature. However, the use of too many features in training the ML model results in a significant increase in computational cost and also leads to overfitting. Thus, choosing the more appropriate and independent features is always preferred. For this purpose, we utilize the feature engineering and dimensionality reduction technique of PCA.

\begin{figure*}[htp]
     \centering
     \includegraphics[width=16cm]{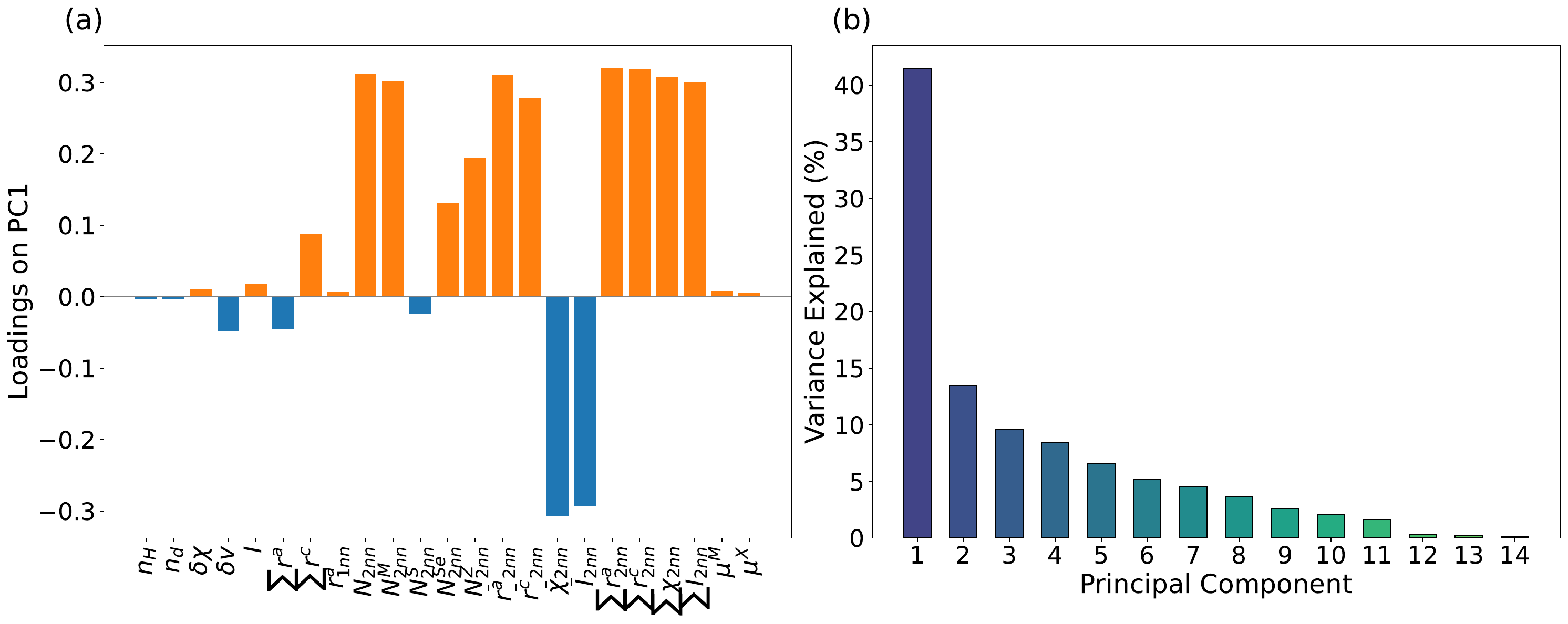}
     \caption{(a) Loadings plot representing the contribution of each feature in obtaining the PC1 and (b) The variance (\%) explained by each of the principal components.}
     \label{fig:pca}
\end{figure*}

\subsubsection{Principal component analysis}

Principal component analysis is a dimensionality reduction method that projects the $n$-dimensional feature vectors in the original feature space into principal components(PCs), which are the linear combination of the original features in a lower-dimensional space~\cite{Michael-Nat.Rev.Methods2022}. The eigenvectors of the covariance matrix of the original data produce PCs. Intuitively, they give the directions of maximum variance, and the corresponding eigenvalues give the variance values. The eigenvalues are sorted in decreasing order of their variances, and the top $k$ eigenvectors ($k$ is less than or equal to $n$) that accommodate most of the variance (like 95\%, 98\%, or 99\% as a parameter) in the data are chosen. The points in the original space are then projected onto the $k$-dimensional space spanned by the $k$ eigenvectors. As such, final predictions can be performed in a lower $k$-dimensional space instead of a higher $n$-dimensional space, thereby reducing computational complexity. Since the PCA-generated PCs are orthogonal to each other, they act as the best features for the ML model.

By performing PCA on the original 23 features with a variance recovery of 98\% we ended up with 11 PCs, which are independent of each other and hence each PC captures the unique information from the data. Due to the reduced dimensionality, the PCs hold the hidden correlations in the original data~\cite{Mohamed-Chem.Phys.Lett.2022}. The percentage of variance explained by each of the PC is given in Fig.~\ref{fig:pca}(b). The first PC (PC1) covers the highest amount of variance in the data, i.e., 41.46\%, and the second PC(PC2) covers an amount of 13.49\%. To understand the variance among all the features, we have analyzed the loading plot for PC1, which is given in Fig.~\ref{fig:pca}(a). Loadings are defined as the contribution of each variable to the PC. The loadings plot for PC2 is also analyzed and is given in SFig. 9. With only two PCs, nearly 55\% of the variance is retained. The contribution of each of the features can also be observed from Eq.~\ref{eq:pc1} and \ref{eq:pc2}. For PC1, the most important features are $N_{2nn}$, $N^M_{2nn}$, $\bar{r^a}_{2nn}$, $\bar{r^c}_{2nn}$, $\bar{\chi}_{2nn}$, $\bar{I}_{2nn}$, $\sum r^a_{2nn}$, $\sum r^c_{2nn}$, $\sum{\chi}_{2nn}$ and $\sum{I}_{2nn}$. This suggests that the SNN features play a crucial role in predicting E$_{ad}$, which characterizes the local bonding character and charge distribution. For PC2, the key features are $\sum r^a$, $\sum r^c$, $N^S_{2nn}$ and $N^{Se}_{2nn}$. The sum of atomic (covalent) radii of H and the atom on which H is being adsorbed well describes the H-host atom bonding, and $N^S_{2nn}$ and $N^{Se}_{2nn}$ represent the presence of S/Se in the neighbourhood of H, and hence are important features influencing E$_{ad}$.

\begin{equation}
\begin{split}
PC1 &= 
- 0.0028 \, n_H 
- 0.0028 \, n_d
+ 0.0105 \, \delta\chi
- 0.0480 \, \delta v\\
&\quad
+ 0.0184 \, I 
- 0.0457 \, \sum r^a
+ 0.0885 \, \sum r^c\\
&\quad
+ 0.0067 \, r^a_{1nn}
+ 0.3116 \, N_{2nn} 
+ 0.3023 \, N^M_{2nn}\\
&\quad 
- 0.0242 \, N^S_{2nn}
+ 0.1313 \, N^{Se}_{2nn}
+ 0.1941 \, N^Z_{2nn}\\
&\quad
+ 0.3105 \, \bar{r^a}_{2nn}
+ 0.2785 \, \bar{r^c}_{2nn} 
- 0.3062 \, \bar{\chi}_{2nn} \\
&\quad
- 0.2924 \, \bar{I}_{2nn}
+ 0.3205 \, \sum r^a_{2nn}
+ 0.3192 \, \sum r^c_{2nn}\\
&\quad
+ 0.3079 \, \sum{\chi}_{2nn}
+ 0.3009 \, \sum{I}_{2nn}
+ 0.0078 \, \mu^M \\
&\quad
+ 0.0061 \, \mu^X
\end{split}
\label{eq:pc1}
\end{equation}

\begin{equation}
\begin{split}
PC2 &= 
-0.0289\,n_H 
-0.0290\,n_d 
-0.0687\,\delta\chi 
+ 0.0119\,\delta v \\
&\quad
-0.0660\,I 
-0.4579\,\sum r^a 
-0.4237\,\sum r^c \\
&\quad
-0.0288\,r^a_{1nn} 
+0.1105\,N_{2nn} 
+0.0884\,N^M_{2nn}\\
&\quad
+0.4794\,N^S_{2nn} 
-0.4379\,N^{Se}_{2nn} 
+0.1114\,N^Z_{2nn} \\
&\quad
- 0.0747\,\bar{r^a}_{2nn} 
-0.2099\,\bar{r^c}_{2nn} 
+0.1022\,\bar{\chi}_{2nn} \\
&\quad
-0.0342\,\bar{I}_{2nn} 
+0.0507\,\sum r^a_{2nn} 
+0.0475\,\sum r^c_{2nn} \\
&\quad
+0.1135\,\sum\chi_{2nn} 
+0.1439\,\sum I_{2nn} 
+0.0024\,\mu^M \\
&\quad
-0.2131\,\mu^X
\end{split}
\label{eq:pc2}
\end{equation}

\subsubsection{Multilayer perceptron regression}
Using the 11 PCs as features, we train a neural network model based on multi-layer perceptrons~\cite{Duda2000, Goodfellow2016}. The training and testing data are divided into ratios of 70:30 and 80:20. We perform various experiments with the model to obtain the best prediction. For this, we examine the neural network model with the optimization techniques Adam~\cite{kingma2017}, Limited-memory Broyden–Fletcher–Goldfarb–Shanno (LBFGS)~\cite{Dong-MathematicalProgramming1989}, and stochastic gradient descent (SGD)~\cite{Bottou-2011} in combination with the activation functions rectifying linear unit (ReLU), tanh, and sigmoid function.
  
The key hyperparameters for the neural network model are the optimization method, the activation function, and the architecture of the hidden layers. Of the three optimization methods available in the \textit{sklearn} package~\cite{scikit-learn}, LBFGS and Adam, in combination with the ReLU activation function, yield good performance for our data. Thus, we test the performance of the model with this combination. We vary the number of hidden layers and the number of neurons in each of the hidden layers and test the performance of the model. The number of hidden layers is varied from 1 to 3, and their depth is varied in multiples of 10 from a minimum of 10 to 100. To assess the robustness of the neural network for a fixed set of all hyperparameters, we use 10 distinct random seeds and observe the performance of the model each time by calculating the root mean squared error (RMSE) and the coefficient of determination (R$^2$). The standard deviation of these metrics for 10 distinct random seeds quantifies the robustness of the model.

\begin{figure*}[htp]
     \centering
     \includegraphics[width=16cm]{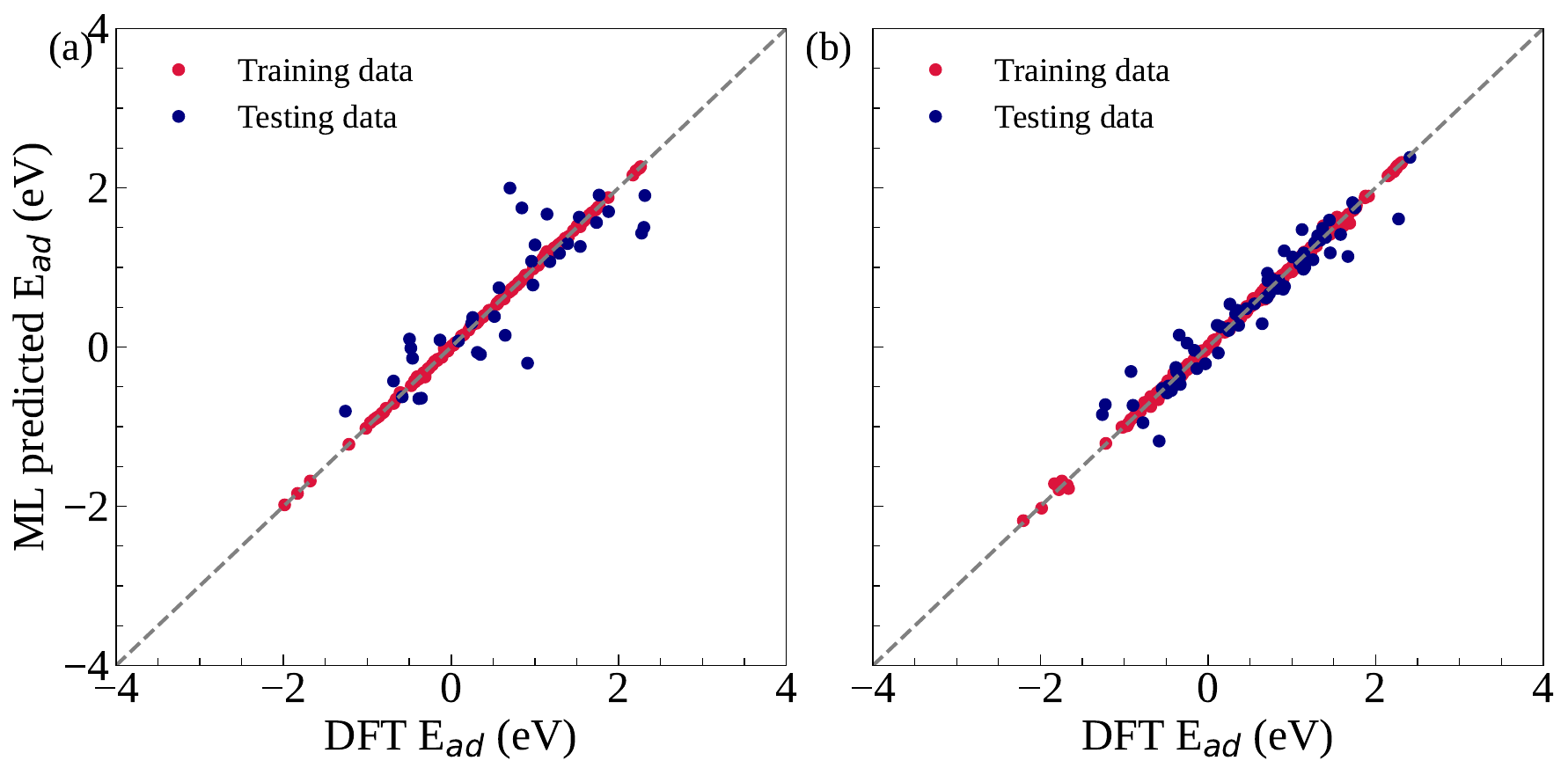}
     \caption{The DFT calculated versus ML predicted E$_{ad}$ with the principal components as the features (a) without and (d) with data augmentation, respectively. The performances reported here are the average performance on the test data achieved by using the 10 distinct random seeds.}
     \label{fig:MLPR}
\end{figure*}

When we analyze all 178 data points and evaluate the performance of the model with different hyperparameter sets, we find that data points related to 2.1\% B@Mo doping appear as outliers in the predicted data, which leads to reduced performance and stability of the model. The parity plot comparing DFT and ML predicted E${ad}$ for this case is shown in SFig. 10. The highest stability achieved for this case is 0.1412, which is quite high. The metrics RMSE and R$^2$ for the testing data are found to be 0.5928 eV and 0.57, respectively. This is due to the significant deviation of E${ad}$ values for 2.1\% B@Mo doping from all the other cases (see Fig.~\ref{fig:xatm}). Thus, we exclude the four data points corresponding to the B@Mo doping under 2.1\% doping concentration and consider the remaining 174 data points for our study. We train the neural network using 11 principal components as input, and the performance of the model is evaluated after hyperparameter tuning. From our experimentation, we found that the 80:20 training and testing data split ratio is the best compared to the 70:30 split, as the model is well-trained with a large number of data points in the 80:20 split case, resulting in the best performance. The optimization technique LBFGS, in combination with the ReLU activation function, yields the best prediction compared to other optimizer and activation function combinations. The calculated metrics RMSE and R$^2$ for the testing (training) data are 0.4322 eV (0.0110 eV) and 0.75 (0.99), respectively. The DFT versus ML predicted E$_{ad}$ parity plot is given in Fig~\ref{fig:MLPR}(a). The robustness of the model is confirmed by measuring the standard deviation of the testing data R$^2$ with 10 distinct random seeds. The observed standard deviation is 0.0881, which indicates the stability of the model. We also check the performance of the model with the Adam optimization technique. The Adam optimizer with the ReLU activation function yields R$^2$ of 0.72 (RMSE = 0.7244 eV) on the test data, which is lower compared to the LBFGS optimization technique. The standard deviation of 0.0995 is slightly higher compared to that with LBFGS. However, the performance of the model is limited to 75\% with either of the optimization methods due to the comparatively smaller size of the dataset. To improve the accuracy of the model and data imbalances, we employ the technique of data augmentation, which artificially generates data points in the neighbourhood of the original data points.

\subsubsection{Data augmentation}

We increase the dataset size by utilizing a data augmentation technique known as the Synthetic Minority Over-Sampling Technique for Regression (SMOTER)~\cite{Chawla_2002}. The SMOTER technique generates the synthetic data points using the K-Nearest Neighbours (KNN) algorithm. We double the size of the dataset to 348 and analyze the performance of the model on the combined dataset of original data points and synthetic data points. Similar to the earlier case, when PCA with 98\% variance coverage is implemented on the combined dataset, 11 principal components are obtained. We use these 11 principal components as input features and train the neural network. With data augmentation, its performance improves significantly, as indicated by the testing (training) data metrics: RMSE 0.2583 eV (0.0273 eV) and R$^2$ 0.90 (0.99), respectively. The DFT versus ML predicted scatter plot for the augmented data is shown in Fig.~\ref{fig:MLPR}(b). The optimized hyperparameters for the case of augmented data are an 80:20 training and testing data split, LBFGS optimizer with a ReLU activation function, and two hidden layers with 90 neurons each. We calculate the standard deviation of R$^2$ for the testing data, obtained using 10 distinct random seeds. The standard deviation is found to be 0.0387, indicating the robustness of the model. The parity plot of DFT versus ML predicted E$_{ad}$ for the MLPR model when all the 23 original features are used as descriptors is provided in SFig. 11 (a) and (b) for the original and augmented data, respectively. The performance of the neural network on the augmented data suggests that the proposed procedure of feature selection and feature engineering is beneficial for accurately predicting E$_{ad}$ using larger datasets. The performance of the model is also examined with the Adam optimizer, with the three hidden layers and 80 neurons in each layer. The metrics RMSE and R$^2$ for the testing data are found to be  0.3028 eV and 0.88, respectively. To understand the effectiveness of the training process with Adam, we calculate the training and validation losses for each epoch using the Adam optimizer, as shown in SFig. 12(a). The DFT versus ML predicted scatterplot is given in SFig. 12(b). For validation purposes, 10\% (28 data points) of the training data is being used. The given curves suggest that the model is neither overfitting nor underfitting, as indicated by the gradual decrease in both the training and validation losses. Thus, the developed model is robust and can predict the E$_{ad}$ using elemental information without requiring DFT calculations for feature extraction.

\section{Conclusion}
In summary, we systematically investigate the adsorption of H atoms on MSSe JLs in the presence of metalloid dopants B, Si, and Ge using DFT calculations. The doping concentration and hydrogen coverage are taken into account by considering the supercells of sizes 4$\times$4$\times$1 and 3$\times$3$\times$1 representing the concentration 2.1\% and 3.8\%, respectively. For the introduction of the dopants into the MSSe JLs, three atomic sites and one interstitial site have been selected. The calculated binding energies of the dopants suggested that the three metalloids are favorable for doping. The electronic structure calculations for the doped systems have revealed the spin-polarization and reduced band gaps. From DOS, it is found that the spin polarization mainly originated from the partially filled d orbitals of Mo/W atoms. It is observed that the MSSe layers undergo a transition from semiconducting to half-metallic and metallic phases, depending on the type of dopant and doping site. Due to the charge redistribution around the dopant, several sites have been activated for H atom adsorption. In certain cases, the H adsorption process on the MSSe JLs is found to be spontaneous with the introduction of dopants. In contrast, in the pristine form, it has resulted in a positive adsorption energy. However, interstitial site doping does not lead to spontaneous H adsorption, and the process remains exothermic. Overall, from the H adsorption energies, it is found that B doping at S/Se atomic planes is favourable for HER, as it results in less negative E$_{ad}$, satisfying the HER criteria. 

Using the total of 178 E$_{ad}$ data points, we have developed an ML model that predicts E$_{ad}$ using the 23 chemical properties of the atoms involved in the structure. By performing the PCA, the dimension of the feature space is reduced from 23 to 11. With the 11 principal components finally obtained from PCA, we trained a multi-layer perceptron regression model and achieved an R$^2$ of 0.75 on the testing data. The stability of the model is characterized by the calculation of the standard deviation, which is found to be 0.0881, obtained by varying the random seed. Due to the comparatively small size of the data set, the accuracy of the model is limited to 75\%. To increase the number of training data points and accuracy, we employed the data augmentation technique SMOTER, which doubled the number of data points. With the combined dataset of original and artificial data points, the accuracy of the model is increased to 90\% with a stability of 0.0387. Thus, the ML model based on fundamental atomic properties is the best model for predicting atomic H adsorption energies in monolayers with impurities.

\begin{acknowledgments}
G.T and D.M acknowledge funding by the DST-SERB within the project CRG/2022/006778. D.M thanks the Department of Information Services and Computing at Helmholtz-Zentrum Dresden-Rossendorf for providing extensive computational facilities. This work was in part supported by the Center for Advanced Systems Understanding (CASUS), which is financed by Germany’s Federal Ministry of Research, Technology and Space (BMFTR) and by the Saxon State government out of the State budget approved by the Saxon State Parliament.
\end{acknowledgments}

% The \nocite command causes all entries in a bibliography to be printed out
% whether or not they are actually referenced in the text. This is appropriate
% for the sample file to show the different styles of references, but authors
% most likely will not want to use it.
\nocite{*}

\bibliography{apssamp}% Produces the bibliography via BibTeX.

@Article{VM2020,
author ="Vallinayagam, M. and Posselt, M. and Chandra, S.",
title  ={{Electronic structure and thermoelectric properties of Mo-based dichalcogenide monolayers locally and randomly modified by substitutional atom}},
journal  ="RSC Adv.",
year  ="2020",
volume  ="10",
issue  ="70",
pages  ="43035-43044",
doi  ="10.1039/D0RA08463H",
url  ="http://dx.doi.org/10.1039/D0RA08463H"
}

@misc{VM2025,
year={2025},
eprint={2509.16992},
archivePrefix={arXiv},
primaryClass={cond-mat.mtrl-sci},
url={https://arxiv.org/abs/2509.16992}, 
author ="Vallinayagam, M. and Kumar, Amrendra and Sudheer, A. E. and Tejaswini, G. and Posselt, M. and Kamal, C. and Zschornak, M. and Murali, D.",
title  ={{Thermoelectric properties of Lead halide Janus layers - A theoretical investigatio}}
}

@article{kresse1996,
title = {{Efficiency of ab-initio total energy calculations for metals and semiconductors using a plane-wave basis set}},
journal = {Comput. Mater. Sci.},
volume = {6},
number = {1},
pages = {15-50},
year = {1996},
issn = {0927-0256},
doi = {https://doi.org/10.1016/0927-0256(96)00008-0},
url = {https://www.sciencedirect.com/science/article/pii/0927025696000080},
author = {G. Kresse and J. Furthmüller},
}

@article{pbe,
  title = {{Generalized Gradient Approximation Made Simple}},
  author = {Perdew, John P. and Burke, Kieron and Ernzerhof, Matthias},
  journal = {Phys. Rev. Lett.},
  volume = {77},
  issue = {18},
  pages = {3865-3868},
  numpages = {0},
  year = {1996},
  month = {Oct},
  publisher = {American Physical Society},
  doi = {10.1103/PhysRevLett.77.3865},
  url = {https://link.aps.org/doi/10.1103/PhysRevLett.77.3865}
}

@article{monkhorst1976,
  title = {{Special points for Brillouin-zone integrations}},
  author = {Monkhorst, Hendrik J. and Pack, James D.},
  journal = {Phys. Rev. B},
  volume = {13},
  issue = {12},
  pages = {5188-5192},
  numpages = {0},
  year = {1976},
  month = {Jun},
  publisher = {American Physical Society},
  doi = {10.1103/PhysRevB.13.5188},
  url = {https://link.aps.org/doi/10.1103/PhysRevB.13.5188}
}

@article{Blochl1994,
  title = {{Projector augmented-wave method}},
  author = {Bl\"ochl, P. E.},
  journal = {Phys. Rev. B},
  volume = {50},
  issue = {24},
  pages = {17953-17979},
  numpages = {0},
  year = {1994},
  month = {Dec},
  publisher = {American Physical Society},
  doi = {10.1103/PhysRevB.50.17953},
  url = {https://link.aps.org/doi/10.1103/PhysRevB.50.17953}
}

@article{brillouin,
  title = {{Improved tetrahedron method for Brillouin-zone integrations}},
  author = {Bl\"ochl, Peter E. and Jepsen, O. and Andersen, O. K.},
  journal = {Phys. Rev. B},
  volume = {49},
  issue = {23},
  pages = {16223-16233},
  numpages = {0},
  year = {1994},
  month = {Jun},
  publisher = {American Physical Society},
  doi = {10.1103/PhysRevB.49.16223},
  url = {https://link.aps.org/doi/10.1103/PhysRevB.49.16223}
}

@Article{Vallinayagam-JMCA2024,
author ="Vallinayagam, M. and Karthikeyan, J. and Posselt, M. and Murali, D. and Zschornak, M.",
title  ={{Metalloid-doping in SMoSe Janus layers: first-principles study on efficient catalysts for the hydrogen evolution reaction}},
journal  ="J. Mater. Chem. A",
year  ="2024",
volume  ="12",
issue  ="13",
pages  ="7742-7753",
publisher  ="The Royal Society of Chemistry",
doi  ="10.1039/D3TA07243F",
url  ="http://dx.doi.org/10.1039/D3TA07243F"}

@Article{Tejaswini-JMCA2025,
author ="Tejaswini, G. and Sudheer, Anjana E. and Kumar, Amrendra and Perepu, Pavan Kumar and Vallinayagam, M. and Kamal, C. and Prakash, S. Mani and M.Posselt and Zschornak, M. and Murali, D.",
title  ={{Hydrogen adsorption energy trends in Mo/WXY (X{,} Y = S{,} Se{,} Te) regular and Janus TMD monolayers: a first-principles and machine learning study}},                  
journal  ="J. Mater. Chem. A",
year  ="2025",
pages  ="-",
publisher  ="The Royal Society of Chemistry",
doi  ="10.1039/D5TA02028J",
url  ="http://dx.doi.org/10.1039/D5TA02028J",
}

@article{Ju-JPM2020,
doi = {10.1088/2515-7639/ab7c57},
url = {https://dx.doi.org/10.1088/2515-7639/ab7c57},
year = {2020},
month = {apr},
publisher = {IOP Publishing},
volume = {3},
number = {2},
pages = {022004},
author = {Ju, Lin and Bie, Mei and Shang, Jing and Tang, Xiao and Kou, Liangzhi},
title = {{Janus transition metal dichalcogenides: a superior platform for photocatalytic water splitting}},
journal = {J. Phys. Mater.},
}

@article{Huang-J.Phys.Chem.C2019,
author = {Huang, Aijian and Shi, Wenwu and Wang, Zhiguo},
title = {{Optical Properties and Photocatalytic Applications of Two-Dimensional Janus Group-III Monochalcogenides}},
journal = {J. Phys. Chem. C},
volume = {123},
number = {18},
pages = {11388-11396},
year = {2019},
doi = {10.1021/acs.jpcc.8b12450},
URL = {https://doi.org/10.1021/acs.jpcc.8b12450},
}

@article{Kumar-PRM2022,
  title = {{Predicting phase preferences of two-dimensional transition metal dichalcogenides using machine learning}},
  author = {Kumar, Pankaj and Sharma, Vinit and Shirodkar, Sharmila N. and Dev, Pratibha},
  journal = {Phys. Rev. Mater.},
  volume = {6},
  issue = {9},
  pages = {094007},
  numpages = {11},
  year = {2022},
  month = {Sep},
  publisher = {American Physical Society},
  doi = {10.1103/PhysRevMaterials.6.094007},
  url = {https://link.aps.org/doi/10.1103/PhysRevMaterials.6.094007}
}

@Article{Xiaofeng-RSC.adv2024,
author ="Cao, Xiaofeng and Luo, Wenjia and Liu, Huimin",
title  ={{A prediction model for CO2/CO adsorption performance on binary alloys based on machine learning"}},
journal  ="RSC Adv.",
year  ="2024",
volume  ="14",
issue  ="17",
pages  ="12235-12246",
publisher  ="The Royal Society of Chemistry",
doi  ="10.1039/D4RA00710G",
url  ="http://dx.doi.org/10.1039/D4RA00710G",
}

@article{Daniel-ACS2022,
author = {Willhelm, Daniel and Wilson, Nathan and Arroyave, Raymundo and Qian, Xiaoning and Cagin, Tahir and Pachter, Ruth and Qian, Xiaofeng},
title = {{Predicting Van der Waals Heterostructures by a Combined Machine Learning and Density Functional Theory Approach}},
journal = {ACS Appl. Mater. Interfaces},
volume = {14},
number = {22},
pages = {25907-25919},
year = {2022},
doi = {10.1021/acsami.2c04403},
URL = {https://doi.org/10.1021/acsami.2c04403}
}

@article{Peng-npj.Compt.Mater2024,
author = {Han, Peng and Zhang, Jingtong and Shi, Shengbin and Zhao, Yunhong and Zhang, Yajun and Wang, Jie},
title = {{Machine learning assisted screening of two dimensional chalcogenide ferromagnetic materials with Dzyaloshinskii Moriya interaction}},
journal = {npj Comput. Mater.},
volume = {10},
pages = {232},
year = {2024},
doi = {10.1038/s41524-024-01419-y},
URL = {https://doi.org/10.1038/s41524-024-01419-y}
}

@article{Minh-Sci.Rep.2023,
author = {Dau, Minh Tuan and Al Khalfioui, Mohamed and Michon, Adrien and Reserbat-Plantey, Antoine and Vézian, Stéphane and Boucaud, Philippe},
title = {{Descriptor engineering in machine learning regression of electronic structure properties for 2D materials}},
journal = {Sci. Rep.},
volume = {13},
pages = {5426},
year = {2023},
doi = {10.1038/s41598-023-31928-7},
URL = {https://doi.org/10.1038/s41598-023-31928-7}
}

@article{Jonathan-npjComput.Mater.2019,
author = {DSchmidt, Jonathan and Marques, Mário R. G. and Botti, Silvana and Marques, Miguel A. L.},
title = {{Recent advances and applications of machine learning in solid-state materials science}},
journal = {npj Comput. Mater.},
volume = {5},
pages = {83},
year = {2019},
doi = {10.1038/s41524-019-0221-0},
URL = {https://doi.org/10.1038/s41524-019-0221-0}
}

@article{Lu-Nature.Nanotech2017,
author = {Lu, Ang-Yu and Zhu, Hanyu and Xiao, Jun and Chuu, Chih-Piao and Han, Yimo and Chiu, Ming-Hui and Cheng, Chia-Chin and Yang, Chih-Wen and Wei, Kung-Hwa and Yang, Yiming and Wang, Yuan and Sokaras, Dimosthenis and Nordlund, Dennis and Yang, Peidong and Muller, David A. and Chou, Mei-Yin and Zhang, Xiang and Li, Lain-Jong
},
title = {{Janus monolayers of transition metal dichalcogenides}},
journal = {Nature. Nanotech. },
volume = {12},
pages = {744--749},
doi = {10.1038/nnano.2017.100},
url = {https://doi.org/10.1038/nnano.2017.100},
year = {2017},
}

@Article{Lin-ACSNano2022,
AUTHOR = {Lin, Yu-Chuan and Liu, Chenze and Yu, Yiling and Zarkadoula, Eva and Yoon, Mina and Puretzky, Alexander A. and Liang, Liangbo and Kong, Xiangru and Gu, Yiyi and Strasser, Alex and Meyer, Harry M. III and Lorenz, Matthias and Chisholm, Matthew F. and Ivanov, Ilia N. and Rouleau, Christopher M. and Duscher, Gerd and Xiao, Kai and Geohegan, David B.
},
TITLE = {{Low Energy Implantation into Transition-Metal Dichalcogenide Monolayers to Form Janus Structures}},
JOURNAL = {ACS Nano},
VOLUME = {14},
YEAR = {2020},
URL = {https://doi.org/10.1021/acsnano.9b10196},
ISSN = {2073-4352},
DOI = {10.1021/acsnano.9b10196},
PAGES = {3896--3906}
}

@article{Zhang-ACSNano2017,
author = {Zhang, Jing and Jia, Shuai and Kholmanov, Iskandar and Dong, Liang and Er, Dequan and Chen, Weibing and Guo, Hua and Jin, Zehua and Shenoy, Vivek B. and Shi, Li and Lou, Jun},
title = {{Janus Monolayer Transition-Metal Dichalcogenides}},
journal = {ACS Nano},
volume = {11},
number = {8},
pages = {8192-8198},
year = {2017},
doi = {10.1021/acsnano.7b03186},
URL = {https://doi.org/10.1021/acsnano.7b03186},
}

@article{Varjovi-PRB2021,
  title = {{Janus two-dimensional transition metal dichalcogenide oxides: First-principles investigation of $\mathrm{W}X\mathrm{O}$ monolayers with $X=\mathrm{S}$, Se, and Te}},
  author = {Varjovi, M. Jahangirzadeh and Yagmurcukardes, M. and Peeters, F. M. and Durgun, E.},
  journal = {Phys. Rev. B},
  volume = {103},
  issue = {19},
  pages = {195438},
  numpages = {11},
  year = {2021},
  month = {May},
  publisher = {American Physical Society},
  doi = {10.1103/PhysRevB.103.195438},
  url = {https://link.aps.org/doi/10.1103/PhysRevB.103.195438}
}

@article{Haman-IJHE2024,
title = {{Harnessing intrinsic electric fields in 2D Janus MoOX (X=S, Se, and Te) monolayers for enhanced photocatalytic hydrogen evolution}},
journal = {International Journal of Hydrogen Energy},
volume = {68},
pages = {566-574},
year = {2024},
issn = {0360-3199},
doi = {https://doi.org/10.1016/j.ijhydene.2024.04.257},
url = {https://www.sciencedirect.com/science/article/pii/S036031992401560X},
author = {Zakaryae Haman and Moussa Kibbou and Nabil Khossossi and Elhoussaine Ouabida and Poulumi Dey and Ismail Essaoudi and Abdelmajid Ainane}
}

@article{Bikerouin-Appl.Surf.Sci.2022,
title = {{Janus transition-metal dichalcogenides heterostructures for highly efficient excitonic solar cells}},
journal = {Applied Surface Science},
volume = {598},
pages = {153835},
year = {2022},
issn = {0169-4332},
doi = {https://doi.org/10.1016/j.apsusc.2022.153835},
url = {https://www.sciencedirect.com/science/article/pii/S0169433222013666},
author = {Mouad Bikerouin and Mohamed Balli},
}

@article{Rajneesh-Adv.Theory.Simul.2025,
author = {Chaurasiya, Rajneesh and Tyagi, Shubham and Kale, Abhijeet J. and Gupta, Goutam Kumar and Kumar, Rajesh and Dixit, Ambesh},
title = {{Advances in Physics and Chemistry of Transition Metal Dichalcogenide Janus Monolayers: Properties, Applications, and Future Prospects}},
journal = {Advanced Theory and Simulations},
volume = {8},
number = {4},
pages = {2400854},
doi = {https://doi.org/10.1002/adts.202400854},
url = {https://advanced.onlinelibrary.wiley.com/doi/abs/10.1002/adts.202400854},
year = {2025}
}

@Article{Yadong-PCCP2019,
author ="Wei, Yadong and Xu, Xiaodong and Wang, Songsong and Li, Weiqi and Jiang, Yongyuan",
title  ={{Second harmonic generation in Janus MoSSe a monolayer and stacked bulk with vertical asymmetry}},
journal  ="Phys. Chem. Chem. Phys.",
year  ="2019",
volume  ="21",
issue  ="37",
pages  ="21022-21029",
publisher  ="The Royal Society of Chemistry",
doi  ="10.1039/C9CP03395E",
url  ="http://dx.doi.org/10.1039/C9CP03395E"}

@Article{Sujata-Mater.Adv.2025,
author ="Sujata, KM and Verma, Nidhi and Solanki, Rekha Garg and Kumar, Ashok",
title  ={{Thermoelectric performance of Bi-based novel Janus monolayer structures}},
journal  ="Mater. Adv.",
year  ="2025",
volume  ="6",
issue  ="2",
pages  ="849-859",
publisher  ="RSC",
doi  ="10.1039/D4MA00924J",
url  ="http://dx.doi.org/10.1039/D4MA00924J",}

@article{Sant-npj.2D.Mater.Appl2020,
author = {Sant, Roberto and Gay, Maxime and Marty, Alain and Lisi, Simone and Harrabi, Rania and Vergnaud, Céline and Dau, Minh Tuan and Weng, Xiaorong and Coraux, Johann and Gauthier, Nicolas and Renault, Olivier and Renaud, Gilles and Jamet, Matthieu},
title = {{Synthesis of epitaxial monolayer Janus SPtSe}},
journal = {npj 2D Mater Appl},
volume = {4},
pages = {41},
doi = {10.1038/s41699-020-00175-z},
url = {https://doi.org/10.1038/s41699-020-00175-z},
year = {2020}
}

@online{webelem,
  title        = {{WebElements:}},
  year         = {1993},
  url          = {https://www.webelements.com},
  note         = {last accessed 15th July 2025}
}

@article{Michael-Nat.Rev.Methods2022,
author = {Greenacre, Michael and Groenen, Patrick J. F. and Hastie, Trevor and D’Enza, Alfonso Iodice and Markos, Angelos and Tuzhilina, Elena},
title = {{Principal component analysis}},
journal = {Nat. Rev. Methods. Primers.},
volume = {2},
pages = {100},
doi = {10.1038/s43586-022-00184-w},
url = {https://doi.org/10.1038/s43586-022-00184-w},
year = {2022}
}

@book{Duda2000,
author = {Duda, Richard O. and Hart, Peter E. and Stork, David G.},
title = {{Pattern Classification (2nd Edition)}},
year = {2000},
isbn = {0471056693},
publisher = {Wiley-Interscience},
address = {USA}
}

@book{Goodfellow2016,
  author = {Goodfellow, Ian and Bengio, Yoshua and Courville, Aaron},
  note = {Book in preparation for MIT Press},
  publisher = {MIT Press},
  title = {{Deep Learning}},
  url = {http://www.deeplearningbook.org},
  year = 2016
}

@article{Mohamed-Chem.Phys.Lett.2022,
title = {{The Principal Component Analysis as a tool for predicting the mechanical properties of Perovskites and Inverse Perovskites}},
journal = {Chemical Physics Letters},
volume = {798},
pages = {139615},
year = {2022},
issn = {0009-2614},
doi = {https://doi.org/10.1016/j.cplett.2022.139615},
url = {https://www.sciencedirect.com/science/article/pii/S0009261422002822},
author = {Mohamed Boubchir and Rachid Boubchir and Hafid Aourag}
}

@article{Yagmurcukardes-Appl.Phys.Rev.2020,
    author = {Yagmurcukardes, M. and Qin, Y. and Ozen, S. and Sayyad, M. and Peeters, F. M. and Tongay, S. and Sahin, H.},
    title = {{Quantum properties and applications of 2D Janus crystals and their superlattices}},
    journal = {Appl. Phys. Rev.},
    volume = {7},
    number = {1},
    pages = {011311},
    year = {2020},
    month = {02},
    issn = {1931-9401},
    doi = {10.1063/1.5135306},
    url = {https://doi.org/10.1063/1.5135306},
}

@article{Kandemir-PRB2018,
  title = {{Janus single layers of ${\mathrm{In}}_{2}\text{SSe}$: A first-principles study}},
  author = {Kandemir, A. and Sahin, H.},
  journal = {Phys. Rev. B},
  volume = {97},
  issue = {15},
  pages = {155410},
  numpages = {7},
  year = {2018},
  month = {Apr},
  publisher = {American Physical Society},
  doi = {10.1103/PhysRevB.97.155410},
  url = {https://link.aps.org/doi/10.1103/PhysRevB.97.155410}
}

@Article{Chauhan-JMCA2024,
author ="Chauhan, Poonam and Singh, Jaspreet and Kumar, Ashok",
title  ={{Two-dimensional Janus antimony chalcohalides for efficient energy conversion applications}},
journal  ="J. Mater. Chem. A",
year  ="2024",
volume  ="12",
issue  ="26",
pages  ="16129-16142",
publisher  ="The Royal Society of Chemistry",
doi  ="10.1039/D4TA02974G",
url  ="http://dx.doi.org/10.1039/D4TA02974G",}

@Article{Sudheer-PCCP2024,
author ="Sudheer, A. E. and Kumar, Amrendra and Tejaswini, G. and Vallinayagam, M. and Posselt, M. and Zschornak, M. and Kamal, C. and Murali, D.",
title  ={{A first principles study on the stability and electronic and optical properties of 2D SbXY (X = Se/Te and Y = I/Br) Janus layers}},
journal  ="Phys. Chem. Chem. Phys.",
year  ="2024",
volume  ="26",
issue  ="47",
pages  ="29371-29383",
publisher  ="The Royal Society of Chemistry",
doi  ="10.1039/D4CP04077E",
url  ="http://dx.doi.org/10.1039/D4CP04077E",}

@article{Johnson-Scilight2025,
    author = {Johnson-Groh, Mara},
    title = {{Breakthrough in green hydrogen energy production found in Janus materials}},
    journal = {Scilight},
    volume = {2025},
    number = {19},
    pages = {191101},
    year = {2025},
    month = {05},
    issn = {2572-7907},
    doi = {10.1063/10.0036743},
    url = {https://doi.org/10.1063/10.0036743},
}

@article{Maka-CleanEnergy2024,
    author = {Maka, Ali O M and Mehmood, Mubbashar},
    title = {{Green hydrogen energy production: current status and potential}},
    journal = {Clean Energy},
    volume = {8},
    number = {2},
    pages = {1-7},
    year = {2024},
    month = {03},
    issn = {2515-4230},
    doi = {10.1093/ce/zkae012},
    url = {https://doi.org/10.1093/ce/zkae012},
}

@article{Bidattul-RSER2024,
title = {{Recent advancement and assessment of green hydrogen production technologies}},
journal = {Renewable and Sustainable Energy Reviews},
volume = {189},
pages = {113941},
year = {2024},
issn = {1364-0321},
doi = {https://doi.org/10.1016/j.rser.2023.113941},
url = {https://www.sciencedirect.com/science/article/pii/S1364032123007992},
author = {Bidattul Syirat Zainal and Pin Jern Ker and Hassan Mohamed and Hwai Chyuan Ong and I.M.R. Fattah and S.M. Ashrafur Rahman and Long D. Nghiem and T M Indra Mahlia}
}

@article{Rahman-ACSOmega2022,
title = {{Molybdenum Disulfide-Based Nanomaterials for Visible-Light-Induced Photocatalysis}},
journal = {ACS Omega},
volume = {7},
pages = {22089-22110},
year = {2022},
doi = {10.1021/acsomega.2c01314},
url = {https://doi.org/10.1021/acsomega.2c01314},
author = {Rahman, Ashmalina and Jennings, James Robert and Tan, Ai Ling and Khan, Mohammad Mansoob
}
}

@Article{Ma-JMCA2025,
author ="Ma, Jiancheng and Pang, Jiafei and Yang, Jinni and Xie, Wanying and Kuang, Xiaoyu and Mao, Aijie",
title  ={{Prediction of high photoconversion efficiency and photocatalytic water splitting in vertically stacked TMD heterojunctions MX2/WS2 and MX2/MoSe2 (M = Cr{,} Mo{,} W; X = S{,} Se{,} Te)}},
journal  ="J. Mater. Chem. A",
year  ="2025",
volume  ="13",
issue  ="13",
pages  ="9312-9322",
publisher  ="The Royal Society of Chemistry",
doi  ="10.1039/D4TA08513B",
url  ="http://dx.doi.org/10.1039/D4TA08513B",
}

@article{Mukesh-ATAS2025,
author = {Choudhary, Mukesh K. and V, Amal Raj and S, Gowri Sankar and Ravindran, P.},
title = {{Composition and Structure Based GGA Bandgap Prediction Using Machine Learning Approach}},
journal = {Advanced Theory and Simulations},
volume = {n/a},
year = {2025},
number = {n/a},
pages = {e00771},
doi = {https://doi.org/10.1002/adts.202500771},
url = {https://advanced.onlinelibrary.wiley.com/doi/abs/10.1002/adts.202500771}
}

@article{Zhiheng-CATC2024,
title = {{Machine learning prediction of hydrogen adsorption energy on platinum nanoclusters: A comparative study of SOAP descriptors}},
journal = {Computational and Theoretical Chemistry},
volume = {1241},
pages = {114923},
year = {2024},
issn = {2210-271X},
doi = {https://doi.org/10.1016/j.comptc.2024.114923},
url = {https://www.sciencedirect.com/science/article/pii/S2210271X24004626},
author = {Zhiheng Yu and Yanli Li and Yanwei Wen and Bin Shan and Jiaqiang Yang}
}

@article{Victor-Nat.Commun2021,
title = {{Machine learned features from density of states for accurate adsorption energy prediction}},
journal = {Nat Commun},
volume = {12},
pages = {88},
year = {2021},
doi = {10.1038/s41467-020-20342-6},
url = {https://doi.org/10.1038/s41467-020-20342-6},
author = {Fung, Victor and Hu, Guoxiang and Ganesh, P. and Sumpter, Bobby G.
}
}

@article{Nokubonga-JES-2025,
title = {{Application of machine learning in adsorption energy storage using metal organic frameworks: A review}},
journal = {Journal of Energy Storage},
volume = {111},
pages = {115363},
year = {2025},
issn = {2352-152X},
doi = {https://doi.org/10.1016/j.est.2025.115363},
url = {https://www.sciencedirect.com/science/article/pii/S2352152X25000763},
author = {Nokubonga P. Makhanya and Michael Kumi and Charles Mbohwa and Bilainu Oboirien},
}

@article{Brian-npj2DMaterAppl2025,
title = {{Understanding and predicting trends in adsorption energetics on monolayer transition metal dichalcogenides}},
journal = {npj 2D Mater Appl},
volume = {9},
pages = {61},
year = {2025},
doi = {10.1038/s41699-025-00579-9v},
url = {https://doi.org/10.1038/s41699-025-00579-9},
author = {Lee, Brian H. and Fatheema, Jameela and Akinwande, Deji and Wang, Wennie},
}

@article{Wen-PRB2025,
  title = {{Advancing the understanding and prediction accuracy of molecular adsorption energy with artificial intelligence}},
  author = {Liu, Wen and Xu, Ning and Li, Zheng and Ma, Meiliang and Hu, Xiaojuan and Han, Zhong-Kang and Jiang, Ying and Yuan, Wentao and Yang, Hangsheng and Levchenko, Sergey V. and Wang, Yong},
  journal = {Phys. Rev. B},
  volume = {111},
  issue = {19},
  pages = {195413},
  numpages = {8},
  year = {2025},
  month = {May},
  publisher = {American Physical Society},
  doi = {10.1103/PhysRevB.111.195413},
  url = {https://link.aps.org/doi/10.1103/PhysRevB.111.195413}
}

@article{Ghanekar-NatCommun2022,
  title = {{Adsorbate chemical environment-based machine learning framework for heterogeneous catalysis}},
  author = {Ghanekar, Pushkar G. and Deshpande, Siddharth and Greeley, Jeffrey},
  journal = {Nat Commun },
  volume = {13},
  pages = {5788},
  year = {2022},
  doi = {10.1038/s41467-022-33256-2},
  url = {https://doi.org/10.1038/s41467-022-33256-2}
}

@article{Wang-IJHE2024,
title = {{Machine learning-assisted design of transition metal-doped 2D WSn$_2$N$_4$ electrocatalysts for enhanced hydrogen evolution reaction}},
journal = {International Journal of Hydrogen Energy},
volume = {90},
pages = {599-606},
year = {2024},
issn = {0360-3199},
doi = {https://doi.org/10.1016/j.ijhydene.2024.10.011},
url = {https://www.sciencedirect.com/science/article/pii/S0360319924042022},
author = {Guang Wang and Yi Wang and YingChao Wang and Tengteng Chen and Lei Li and Zhengli Zhang and Zhao Ding and Xiang Guo and Zijiang Luo and Xuefei Liu},
}

@article{Tao-PRB2018,
  title = {{Intrinsic and anisotropic Rashba spin splitting in Janus transition-metal dichalcogenide monolayers}},
  author = {Hu, Tao and Jia, Fanhao and Zhao, Guodong and Wu, Jiongyao and Stroppa, Alessandro and Ren, Wei},
  journal = {Phys. Rev. B},
  volume = {97},
  issue = {23},
  pages = {235404},
  numpages = {6},
  year = {2018},
  month = {Jun},
  publisher = {American Physical Society},
  doi = {10.1103/PhysRevB.97.235404},
  url = {https://link.aps.org/doi/10.1103/PhysRevB.97.235404}
}

@article{Ayushi-JAP2024,
    author = {Jain, Ayushi and Bera, Chandan},
    title = {{Rashba spin-splitting and spin Hall effect in Janus monolayers Sb2XSX’ (X, X’= S, Se, or Te; X $\neq$ X’)}},
    journal = {Journal of Applied Physics},
    volume = {135},
    number = {11},
    pages = {114302},
    year = {2024},
    month = {03},
    doi = {10.1063/5.0192623},
    url = {https://doi.org/10.1063/5.0192623},
}

@article{Shi-Qi-Adv.Optical.Mater.2022,
author = {LI, Shi-Qi and He, Chuan and Liu, Hongsheng and Zhao, Luneng and Xu, Xinlong and Chen, Maodu and Wang, Lu and Zhao, Jijun and Gao, Junfeng},
title = {{Dramatically Enhanced Second Harmonic Generation in Janus Group-III Chalcogenide Monolayers}},
journal = {Advanced Optical Materials},
volume = {10},
number = {15},
pages = {2200076},
doi = {https://doi.org/10.1002/adom.202200076},
url = {https://advanced.onlinelibrary.wiley.com/doi/abs/10.1002/adom.202200076},
year = {2022},
}

@article{Gorkan-PRM2023,
  title = {{Skyrmion formation in Ni-based Janus dihalide monolayers: Interplay between magnetic frustration and Dzyaloshinskii-Moriya interaction}},
  author = {Gorkan, Taylan and Das, Jyotirish and Kapeghian, Jesse and Akram, Muhammad and Barth, Johannes V. and Tongay, Sefaattin and Akturk, Ethem and Erten, Onur and Botana, Antia S.},
  journal = {Phys. Rev. Mater.},
  volume = {7},
  issue = {5},
  pages = {054006},
  numpages = {13},
  year = {2023},
  month = {May},
  publisher = {American Physical Society},
  doi = {10.1103/PhysRevMaterials.7.054006},
  url = {https://link.aps.org/doi/10.1103/PhysRevMaterials.7.054006}
}

@article{Weiy-PRB2024,
  title = {{Strain-induced intrinsic antiferromagnetic skyrmions in two-dimensional Janus magnets}},
  author = {Pan, Weiyi and Ji, Shilei and Xu, Zhiming},
  journal = {Phys. Rev. B},
  volume = {110},
  issue = {14},
  pages = {144408},
  numpages = {9},
  year = {2024},
  month = {Oct},
  publisher = {American Physical Society},
  doi = {10.1103/PhysRevB.110.144408},
  url = {https://link.aps.org/doi/10.1103/PhysRevB.110.144408}
}

@article{Ting-NanoLetters2021,
author = {Zheng, Ting and Lin, Yu-Chuan and Yu, Yiling and Valencia-Acuna, Pavel and Puretzky, Alexander A. and Torsi, Riccardo and Liu, Chenze and Ivanov, Ilia N. and Duscher, Gerd and Geohegan, David B. and Ni, Zhenhua and Xiao, Kai and Zhao, Hui},
title = {{Excitonic Dynamics in Janus MoSSe and WSSe Monolayers}},
journal = {Nano Letters},
volume = {21},
number = {2},
pages = {931-937},
year = {2021},
doi = {10.1021/acs.nanolett.0c03412},
URL = {https://doi.org/10.1021/acs.nanolett.0c03412},
}

@article{Zhang-2DMaterials2023,
doi = {10.1088/2053-1583/ace5bb},
url = {https://dx.doi.org/10.1088/2053-1583/ace5bb},
year = {2023},
month = {jul},
publisher = {IOP Publishing},
volume = {10},
number = {4},
pages = {045005},
author = {Zhang, Long and Liu, Yuqi and Xu, Zhiyuan and Gao, Guoying},
title = {{Electronic phase transition, perpendicular magnetic anisotropy and high Curie temperature in Janus FeClF}},
journal = {2D Materials},
}

@article{Manzeli-Nat.Rev.Mater2017,
doi = {10.1038/natrevmats.2017.33},
url = {https://doi.org/10.1038/natrevmats.2017.33},
year = {2017},
volume = {2},
pages = {17033},
author = {Manzeli, Sajedeh and Ovchinnikov, Dmitry and Pasquier, Diego and Yazyev, Oleg V. and Kis, Andras},
title = {{2D transition metal dichalcogenides}},
journal = {Nat Rev Mater},
}

@article{Xiangfeng-Chem.Reviews2024,
author = {Duan, Xiangfeng and Zhang, Hua},
title = {{Introduction: Two-Dimensional Layered Transition Metal Dichalcogenides}},
journal = {Chemical Reviews},
volume = {124},
number = {19},
pages = {10619-10622},
year = {2024},
doi = {10.1021/acs.chemrev.4c00586},
URL = {https://doi.org/10.1021/acs.chemrev.4c00586},
}

@article{Ruijie-Angew.Chem.Int.Ed.2023,
author = {Yang, Ruijie and Fan, Yingying and Zhang, Yuefeng and Mei, Liang and Zhu, Rongshu and Qin, Jiaqian and Hu, Jinguang and Chen, Zhangxing and Hau Ng, Yun and Voiry, Damien and Li, Shuang and Lu, Qingye and Wang, Qian and Yu, Jimmy C. and Zeng, Zhiyuan},
title = {{2D Transition Metal Dichalcogenides for Photocatalysis}},
journal = {Angewandte Chemie International Edition},
volume = {62},
number = {13},
pages = {e202218016},
doi = {https://doi.org/10.1002/anie.202218016},
url = {https://onlinelibrary.wiley.com/doi/abs/10.1002/anie.202218016},
year = {2023}
}

@Article{Eftekahri-JMCA2017,
author ="Eftekhari, Ali",
title  ={{Tungsten dichalcogenides (WS2{,} WSe2{,} and WTe2): materials chemistry and applications}},
journal  ="J. Mater. Chem. A",
year  ="2017",
volume  ="5",
issue  ="35",
pages  ="18299-18325",
publisher  ="The Royal Society of Chemistry",
doi  ="10.1039/C7TA04268J",
url  ="http://dx.doi.org/10.1039/C7TA04268J"}

@article{Haoyang-MaterialsTodayChemistry2023,
title = {{A review of strategies to improve the performance of photocatalysts for CO2 reduction}},
journal = {Materials Today Chemistry},
volume = {34},
pages = {101802},
year = {2023},
issn = {2468-5194},
doi = {https://doi.org/10.1016/j.mtchem.2023.101802},
url = {https://www.sciencedirect.com/science/article/pii/S2468519423004299},
author = {Haoyang Xu and Yue Shen and Xuen Guo and Liang Zhang},
}

@Article{Tejaswini-PCCP2024,
author ="Tejaswini, G. and Sudheer, Anjana E. and Vallinayagam, M. and Posselt, M. and Zschornak, M. and Maniprakash, S. and Murali, D.",
title  ={{Band alignment in CdS–$\alpha$-Te van der Waals heterostructures for photocatalytic applications: influence of biaxial strain and electric field}},
journal  ="Phys. Chem. Chem. Phys.",
year  ="2024",
volume  ="26",
issue  ="47",
pages  ="29339-29350",
publisher  ="The Royal Society of Chemistry",
doi  ="10.1039/D4CP03368J",
url  ="http://dx.doi.org/10.1039/D4CP03368J"}

@Article{Karthikeyan-JMCC2025,
author ="Karthikeyan, C. and Tejaswini, G. and Sudheer, Anjana E. and Vallinayagam, M. and Posselt, M. and Zschornak, M. and Murali, D.",
title  ={{Theoretical insights into PtSSe–SnSSe heterostructures for renewable energy applications}},
journal  ="J. Mater. Chem. C",
year  ="2025",
volume  ="13",
issue  ="23",
pages  ="11904-11916",
publisher  ="The Royal Society of Chemistry",
doi  ="10.1039/D5TC00301F",
url  ="http://dx.doi.org/10.1039/D5TC00301F"}

@article{Chen-ChemPhotoChem2023,
author = {Chen, Zhengnan and Chi, YuHua and Ma, Hao and Yuan, Saifei and Hao, Chunlian and Ren, Hao and Zhao, Wen and Zhu, HouYu and Guo, Wenyue},
title = {{Strain-Tunable Photocatalytic Performance in Two-Dimensional Janus In2S2X (X=Se,Te) Monolayer as Photocatalyst for CO2 Reduction}},
journal = {ChemPhotoChem},
volume = {7},
number = {8},
pages = {e202300037},
doi = {https://doi.org/10.1002/cptc.202300037},
url = {https://chemistry-europe.onlinelibrary.wiley.com/doi/abs/10.1002/cptc.202300037},
year = {2023}
}

@article{Hou-PRM2021,
  title = {{Hybrid density functional study of band gap engineering of $\mathrm{SrTi}{\text{O}}_{3}$ photocatalyst via doping for water splitting}},
  author = {Hou, Y. S. and Ardo, S. and Wu, R. Q.},
  journal = {Phys. Rev. Mater.},
  volume = {5},
  issue = {6},
  pages = {065801},
  numpages = {9},
  year = {2021},
  month = {Jun},
  publisher = {American Physical Society},
  doi = {10.1103/PhysRevMaterials.5.065801},
  url = {https://link.aps.org/doi/10.1103/PhysRevMaterials.5.065801}
}

@article{Yu-Small2024,
author = {Yu, Xiangxiang and Peng, Zhuiri and Xu, Langlang and Shi, Wenhao and Li, Zheng and Meng, Xiaohan and He, Xiao and Wang, Zhen and Duan, Shikun and Tong, Lei and Huang, Xinyu and Miao, Xiangshui and Hu, Weida and Ye, Lei},
title = {{Manipulating 2D Materials through Strain Engineering}},
journal = {Small},
volume = {20},
number = {38},
pages = {2402561},
doi = {https://doi.org/10.1002/smll.202402561},
url = {https://onlinelibrary.wiley.com/doi/abs/10.1002/smll.202402561},
year = {2024}
}

@Article{Maarisetty-JMCA2020,
author ="Maarisetty, Dileep and Baral, Saroj Sundar",
title  ={{Defect engineering in photocatalysis: formation{,} chemistry{,} optoelectronics{,} and interface studies}},
journal  ="J. Mater. Chem. A",
year  ="2020",
volume  ="8",
issue  ="36",
pages  ="18560-18604",
publisher  ="The Royal Society of Chemistry",
doi  ="10.1039/D0TA04297H",
url  ="http://dx.doi.org/10.1039/D0TA04297H"}

@article{Mehdipour-PRB2022,
  title = {{Structural defects in a Janus MoSSe monolayer: A density functional theory study}},
  author = {Mehdipour, Hamid and Kratzer, Peter},
  journal = {Phys. Rev. B},
  volume = {106},
  issue = {23},
  pages = {235414},
  numpages = {17},
  year = {2022},
  month = {Dec},
  publisher = {American Physical Society},
  doi = {10.1103/PhysRevB.106.235414},
  url = {https://link.aps.org/doi/10.1103/PhysRevB.106.235414}
}

@article{Ouahrani-PRM2023,
  title = {{Effect of intrinsic point defects on the catalytic and electronic properties of ${\mathrm{Cu}}_{2}{\mathrm{WS}}_{4}$ single layer: Ab initio calculations}},
  author = {Ouahrani, Tarik and Boufatah, Reda M. and Benaissa, Mohammed and Morales-Garc\'{\i}a, \'Angel and Badawi, Michael and Errandonea, Daniel},
  journal = {Phys. Rev. Mater.},
  volume = {7},
  issue = {2},
  pages = {025403},
  numpages = {11},
  year = {2023},
  month = {Feb},
  publisher = {American Physical Society},
  doi = {10.1103/PhysRevMaterials.7.025403},
  url = {https://link.aps.org/doi/10.1103/PhysRevMaterials.7.025403}
}

@Article{Obeid-PCCP2020,
author ="Obeid, Mohammed M. and Stampfl, C. and Bafekry, A. and Guan, Z. and Jappor, H. R. and Nguyen, C. V. and Naseri, M. and Hoat, D. M. and Hieu, N. N. and Krauklis, A. E. and Vu, Tuan V. and Gogova, D.",
title  ={{First-principles investigation of nonmetal doped single-layer BiOBr as a potential photocatalyst with a low recombination rate}},
journal  ="Phys. Chem. Chem. Phys.",
year  ="2020",
volume  ="22",
issue  ="27",
pages  ="15354-15364",
publisher  ="The Royal Society of Chemistry",
doi  ="10.1039/D0CP02007A",
url  ="http://dx.doi.org/10.1039/D0CP02007A"}

@article{Vallinayagam-JPCC2023,
author = {Vallinayagam, M. and Sudheer, A. E. and Aravindh, S. Assa and Murali, D. and Raja, N. and Katta, R. and Posselt, M. and Zschornak, M.},
title = {{Novel Metalless Chalcogen-Based Janus Layers: A Density Functional Theory Study}},
journal = {The Journal of Physical Chemistry C},
volume = {127},
number = {34},
pages = {17029-17038},
year = {2023},
doi = {10.1021/acs.jpcc.3c02248},
URL = {https://doi.org/10.1021/acs.jpcc.3c02248}
}

@article{Sudheer-Comput.Mater.Sci2024,
title = {{First principles investigation on structural and optoelectronic properties of newly designed Janus lead halides PbXY (X, Y = F, Cl, Br, I )}},
journal = {Computational Materials Science},
volume = {243},
pages = {113123},
year = {2024},
issn = {0927-0256},
doi = {https://doi.org/10.1016/j.commatsci.2024.113123},
url = {https://www.sciencedirect.com/science/article/pii/S0927025624003446},
author = {Anjana E. Sudheer and Golla Tejaswini and Matthias Posselt and D. Murali}
}

@article{Abhishek-ACS2020,
author = {{Patel, Abhishek and Singh, Deobrat and Sonvane, Yogesh and Thakor, P. B. and Ahuja, Rajeev}},
title = {High Thermoelectric Performance in Two-Dimensional Janus Monolayer Material WS-X (X = Se and Te)},
journal = {ACS Applied Materials \& Interfaces},
volume = {12},
number = {41},
pages = {46212-46219},
year = {2020},
doi = {10.1021/acsami.0c13960},
URL = {https://doi.org/10.1021/acsami.0c13960}
}

@article{Ozbey-PRB2024,
  title = {{Structural, electronic, vibrational, and thermoelectric properties of Janus ${\mathrm{Ge}}_{2}\mathrm{P}\mathit{X}(\mathit{X}=\mathrm{N},\mathrm{As},\mathrm{Sb}, \text{and} \mathrm{Bi})$ monolayers}},
  author = {Ozbey, Dogukan Hazar and Varjovi, Mirali Jahangirzadeh and Sarg\ifmmode \imath \else \i \fi{}n, G\"ozde \"Ozbal and Sevin\ifmmode \mbox{\c{c}}\else \c{c}\fi{}li, H\^aldun and Durgun, Engin},
  journal = {Phys. Rev. B},
  volume = {110},
  issue = {3},
  pages = {035411},
  numpages = {13},
  year = {2024},
  month = {Jul},
  publisher = {American Physical Society},
  doi = {10.1103/PhysRevB.110.035411},
  url = {https://link.aps.org/doi/10.1103/PhysRevB.110.035411}
}

@article{Ju-ACS2020,
author = {Ju, Lin and Bie, Mei and Tang, Xiao and Shang, Jing and Kou, Liangzhi},
title = {{Janus WSSe Monolayer: An Excellent Photocatalyst for Overall Water Splitting}},
journal = {ACS Applied Materials \& Interfaces},
volume = {12},
number = {26},
pages = {29335-29343},
year = {2020},
doi = {10.1021/acsami.0c06149},
URL = {https://doi.org/10.1021/acsami.0c06149}
}

@Article{Jamdagni-PCCP2022,
author ="Jamdagni, Pooja and Kumar, Ashok and Srivastava, Sunita and Pandey, Ravindra and Tankeshwar, K.",
title  ={{Photocatalytic properties of anisotropic $\beta$-PtX2 (X = S{,} Se) and Janus $\beta$-PtSSe monolayers}},
journal  ="Phys. Chem. Chem. Phys.",
year  ="2022",
volume  ="24",
issue  ="36",
pages  ="22289-22297",
publisher  ="The Royal Society of Chemistry",
doi  ="10.1039/D2CP02549C",
url  ="http://dx.doi.org/10.1039/D2CP02549C"}

@article{Zhou-PRM2021,
  title = {{Enhancement effects of interlayer orbital hybridization in Janus MoSSe and tellurene heterostructures for photovoltaic applications}},
  author = {Zhou, Bin and Cui, Anyang and Gao, Lichen and Jiang, Kai and Shang, Liyan and Zhang, Jinzhong and Li, Yawei and Gong, Shi-Jing and Hu, Zhigao and Chu, Junhao},
  journal = {Phys. Rev. Mater.},
  volume = {5},
  issue = {12},
  pages = {125404},
  numpages = {9},
  year = {2021},
  month = {Dec},
  publisher = {American Physical Society},
  doi = {10.1103/PhysRevMaterials.5.125404},
  url = {https://link.aps.org/doi/10.1103/PhysRevMaterials.5.125404}
}

@article{Huiqin-Chem.Phys2022,
title = {{Janus In2SeTe for photovoltaic device applications from first-principles study}},
journal = {Chemical Physics},
volume = {553},
pages = {111384},
year = {2022},
issn = {0301-0104},
doi = {https://doi.org/10.1016/j.chemphys.2021.111384},
url = {https://www.sciencedirect.com/science/article/pii/S0301010421002950},
author = {Huiqin Zhao and Yan Gu and Naiyan Lu and Yushen Liu and Yu Ding and Bingjie Ye and Xinxia Huo and Baoan Bian and Chunlei Wei and Xiumei Zhang and Guofeng Yang}
}

@article{Norskov-JPCB2004,
author = {Nørskov, J. K. and Rossmeisl, J. and Logadottir, A. and Lindqvist, L. and Kitchin, J. R. and Bligaard, T. and Jónsson, H.},
title = {{Origin of the Overpotential for Oxygen Reduction at a Fuel-Cell Cathode}},
journal = {The Journal of Physical Chemistry B},
volume = {108},
number = {46},
pages = {17886-17892},
year = {2004},
doi = {10.1021/jp047349j},
URL = {https://doi.org/10.1021/jp047349j}
}

@article{Novoselov-Science2004,
author = {K. S. Novoselov  and A. K. Geim  and S. V. Morozov  and D. Jiang  and Y. Zhang  and S. V. Dubonos  and I. V. Grigorieva  and A. A. Firsov },
title = {{Electric Field Effect in Atomically Thin Carbon Films}},
journal = {Science},
volume = {306},
number = {5696},
pages = {666-669},
year = {2004},
doi = {10.1126/science.1102896},
URL = {https://www.science.org/doi/abs/10.1126/science.1102896},
}

@article{Eugene-IJHE2012,
title = {{Hydrogen adsorption on and diffusion through MoS2 monolayer: First-principles study}},
journal = {International Journal of Hydrogen Energy},
volume = {37},
number = {19},
pages = {14323-14328},
year = {2012},
note = {HYFUSEN},
issn = {0360-3199},
doi = {https://doi.org/10.1016/j.ijhydene.2012.07.069},
url = {https://www.sciencedirect.com/science/article/pii/S0360319912016461},
author = {Eugene Wai {Keong Koh} and Cheng Hsin Chiu and Yao Kun Lim and Yong-Wei Zhang and Hui Pan}
}

@Article{Wang-Nanoscale.Adv.2020,
author ="Wang, Caiyun and Yang, Fuchao and Gao, Yihua",
title  ={{The highly-efficient light-emitting diodes based on transition metal dichalcogenides: from architecture to performance}},
journal  ="Nanoscale Adv.",
year  ="2020",
volume  ="2",
issue  ="10",
pages  ="4323-4340",
publisher  ="RSC",
doi  ="10.1039/D0NA00501K",
url  ="http://dx.doi.org/10.1039/D0NA00501K"
}

@article{Zhou-Adv.Funct.Mater.2024,
author = {Zhou, Zhanren and Lv, Junling and Tan, Chao and Yang, Lei and Wang, Zegao},
title = {{Emerging Frontiers of 2D Transition Metal Dichalcogenides in Photovoltaics Solar Cell}},
journal = {Adv. Funct. Mater.},
volume = {34},
number = {29},
pages = {2316175},
doi = {https://doi.org/10.1002/adfm.202316175},
url = {https://advanced.onlinelibrary.wiley.com/doi/abs/10.1002/adfm.202316175},
year = {2024}
}

@article{Priyakshi-Mater.Today.Sus2024,
title = {{2D transition metal dichalcogenides for efficient hydrogen generation}},
journal = {Materials Today Sustainability},
volume = {27},
pages = {100914},
year = {2024},
issn = {2589-2347},
doi = {https://doi.org/10.1016/j.mtsust.2024.100914},
url = {https://www.sciencedirect.com/science/article/pii/S2589234724002501},
author = {Priyakshi Bora and Suraj Kumar and Dipak Sinha}
}

@article{Ju-Appl.Mater.Interfaces2020,
author = {Ju, Lin and Bie, Mei and Tang, Xiao and Shang, Jing and Kou, Liangzhi},
title = {{Janus WSSe Monolayer: An Excellent Photocatalyst for Overall Water Splitting}},
journal = {ACS Applied Materials \& Interfaces},
volume = {12},
number = {26},
pages = {29335-29343},
year = {2020},
doi = {10.1021/acsami.0c06149},
URL = {https://doi.org/10.1021/acsami.0c06149}
}

@article{Komsa-PRL2013,
  title = {{Finite-Size Supercell Correction for Charged Defects at Surfaces and Interfaces}},
  author = {Komsa, Hannu-Pekka and Pasquarello, Alfredo},
  journal = {Phys. Rev. Lett.},
  volume = {110},
  issue = {9},
  pages = {095505},
  numpages = {5},
  year = {2013},
  month = {Feb},
  publisher = {American Physical Society},
  doi = {10.1103/PhysRevLett.110.095505},
  url = {https://link.aps.org/doi/10.1103/PhysRevLett.110.095505}
}

@misc{kingma2017,
      title={Adam: A Method for Stochastic Optimization}, 
      author={Diederik P. Kingma and Jimmy Ba},
      year={2017},
      eprint={1412.6980},
      archivePrefix={arXiv},
      primaryClass={cs.LG},
      url={https://arxiv.org/abs/1412.6980}, 
}

@article{Dong-MathematicalProgramming1989,
  title = {{On the limited memory BFGS method for large scale optimization}},
  author = {Liu, Dong C. and Nocedal, Jorge},
  journal = {Mathematical Programming},
  volume = {45},
  pages = {503-528},
  year = {1989},
  doi = {10.1007/BF01589116},
  url = {https://doi.org/10.1007/BF01589116}
}

@incollection{Bottou-2011,
    author = {Bottou, Léon and Bousquet, Olivier},
    isbn = {9780262298773},
    title = {{The Tradeoffs of Large-Scale Learning}},
    booktitle = {{Optimization for Machine Learning}},
    publisher = {The MIT Press},
    year = {2011},
    month = {09},
    doi = {10.7551/mitpress/8996.003.0015},
    url = {https://doi.org/10.7551/mitpress/8996.003.0015},
}

@article{Huibin-Appl.Surf.Sci2024,
title = {{Symmetry breaking induced local electron rearrangement to enhance photocatalytic tetracycline hydrochloride oxidation and hydrogen evolution of carbon nitride}},
journal = {Applied Surface Science},
volume = {649},
pages = {158964},
year = {2024},
issn = {0169-4332},
doi = {https://doi.org/10.1016/j.apsusc.2023.158964},
url = {https://www.sciencedirect.com/science/article/pii/S0169433223026442},
author = {Huibin Zong and Guixin Zeng and Honghai Miao and Zhao Mo and Xianglin Zhu and Hui Xu}
}

@article{Thajitr_2022,
doi = {10.1088/1402-4896/ac850c},
url = {https://doi.org/10.1088/1402-4896/ac850c},
year = {2022},
month = {aug},
publisher = {IOP Publishing},
volume = {97},
number = {9},
pages = {095805},
author = {Thajitr, W and Busayaporn, W and Rai, D P and Sukkabot, W},
title = {{Modulation of electronic and magnetic properties of MoX2 (X=S and Se) monolayer via mono- and co-transition metal dopants: Spin density functional theory}},
journal = {Physica Scripta},
}

@article{Zhang_2020,
doi = {10.1088/2053-1591/ab5b43},
url = {https://doi.org/10.1088/2053-1591/ab5b43},
year = {2019},
month = {dec},
publisher = {IOP Publishing},
volume = {7},
number = {1},
pages = {015501},
author = {Zhang, Zheng and Chen, Kai and Zhao, Qiang and Huang, Mei and Ouyang, Xiaoping},
title = {Effects of noble metal doping on hydrogen sensing performances of monolayer MoS2},
journal = {Materials Research Express},
}

@article{scikit-learn,
  title={Scikit-learn: Machine Learning in {P}ython},
  author={Pedregosa, F. and Varoquaux, G. and Gramfort, A. and Michel, V.
          and Thirion, B. and Grisel, O. and Blondel, M. and Prettenhofer, P.
          and Weiss, R. and Dubourg, V. and Vanderplas, J. and Passos, A. and
          Cournapeau, D. and Brucher, M. and Perrot, M. and Duchesnay, E.},
  journal={Journal of Machine Learning Research},
  volume={12},
  pages={2825--2830},
  year={2011}
}

@article{Chawla_2002,
   title={SMOTE: Synthetic Minority Over-sampling Technique},
   volume={16},
   ISSN={1076-9757},
   url={http://dx.doi.org/10.1613/jair.953},
   DOI={10.1613/jair.953},
   journal={Journal of Artificial Intelligence Research},
   publisher={AI Access Foundation},
   author={Chawla, N. V. and Bowyer, K. W. and Hall, L. O. and Kegelmeyer, W. P.},
   year={2002},
   month=jun, pages={321–357} }

\end{document}

\maketitle
\thispagestyle{empty}
\clearpage
\pagenumbering{arabic}

%-------------------------------------------------
% EPA in P2 
%-------------------------------------------------

\begin{figure*}[htp]
     \centering
     \includegraphics[width=\textwidth]{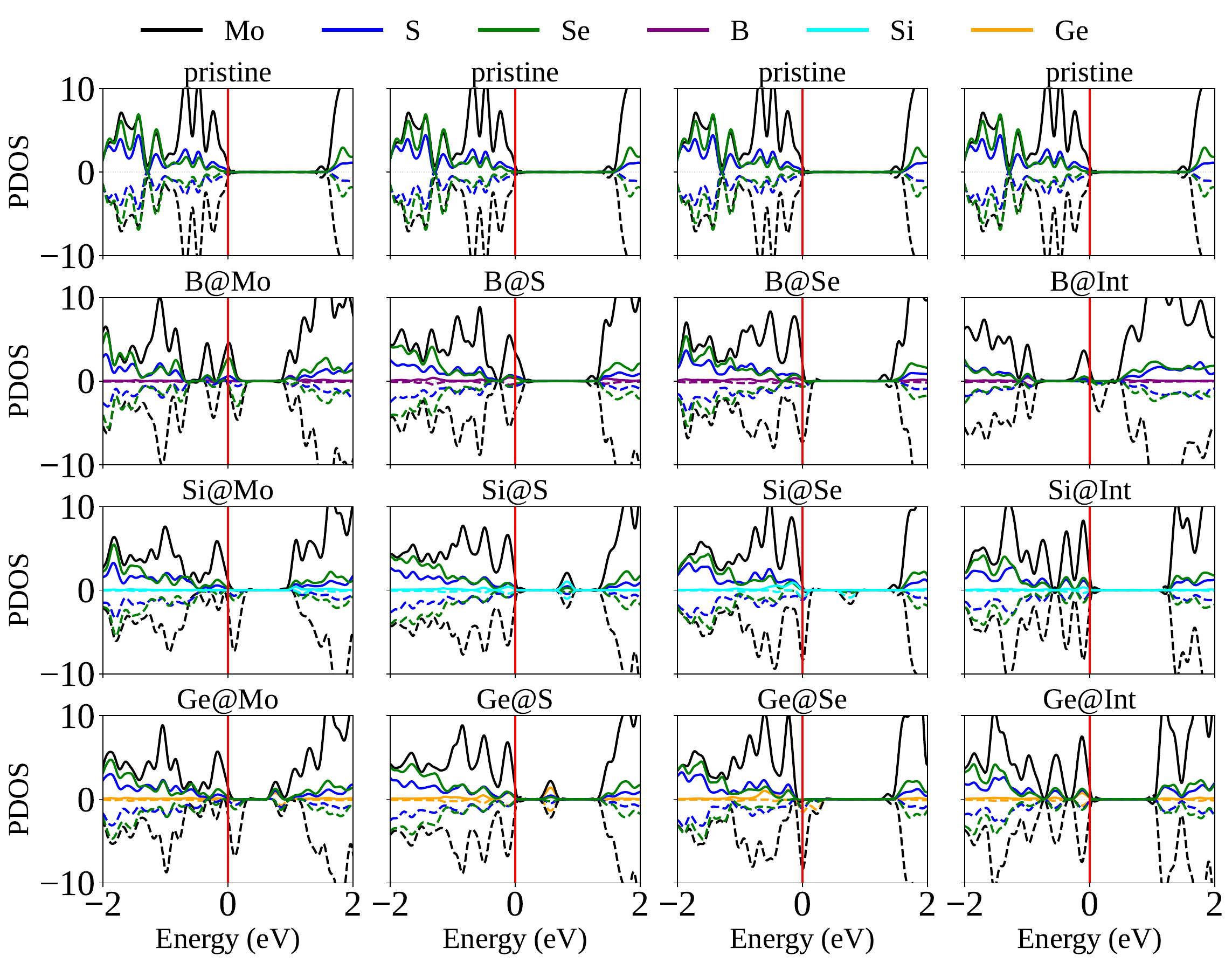} %16cm
     \caption{Atom resolved density of states of pristine and B, Si, and Ge doped MoSSe layer at a doping concentration of 3.8\% (3$\times$3$\times$1 supercell). The solid red line represents the Fermi level. }
     \label{fig:331-MoSSe-PDOS}
\end{figure*}

\begin{figure*}[htp]
     \centering
     \includegraphics[width=\textwidth]{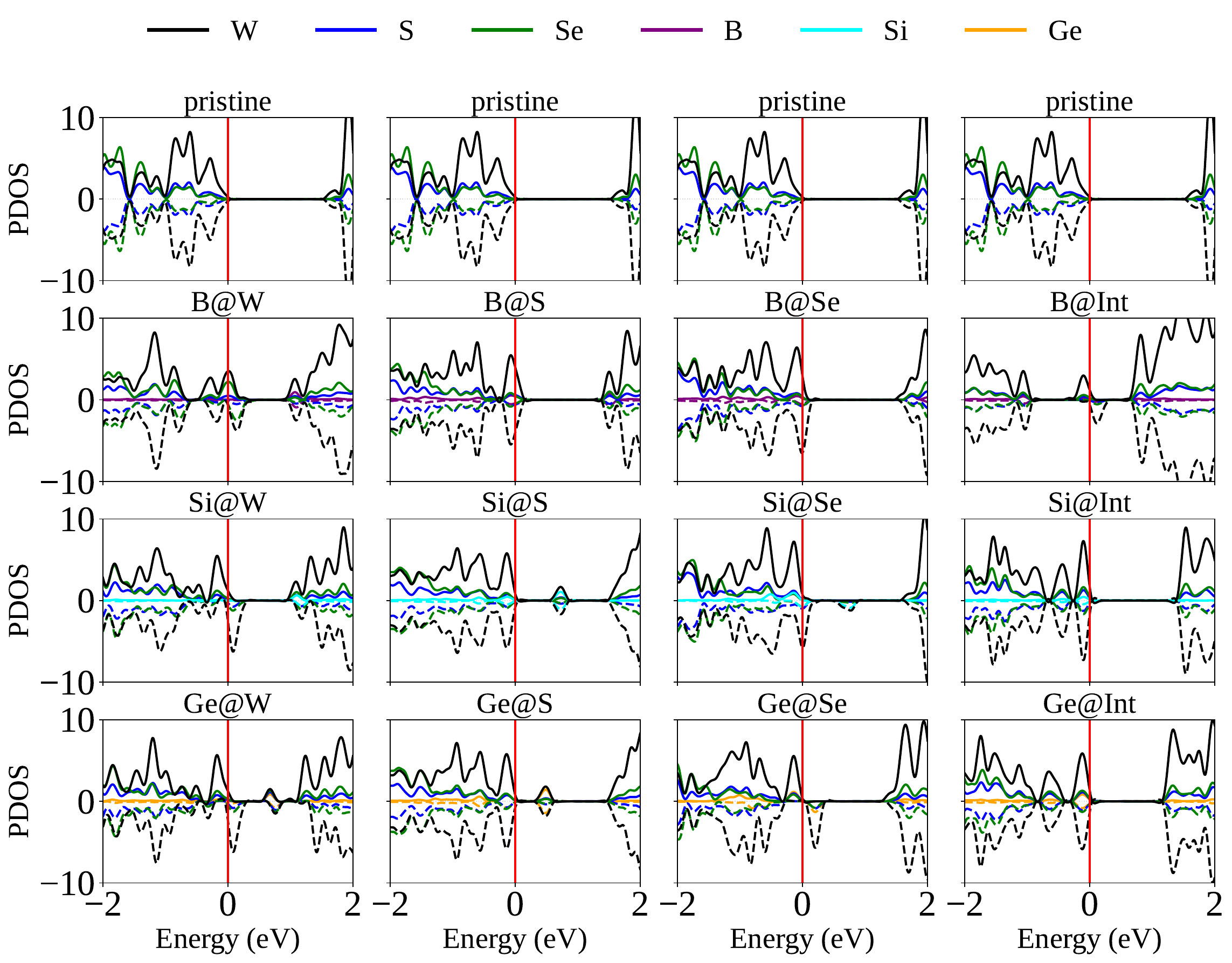} %16cm
     \caption{Atom resolved density of states of pristine and B, Si, and Ge doped WSSe layer at a doping concentration of 3.8\% (3$\times$3$\times$1 supercell). The solid red line represents the Fermi level. }
     \label{fig:331-WSSe-PDOS}
\end{figure*}

\begin{figure*}[htp]
     \centering
     \includegraphics[width=\textwidth]{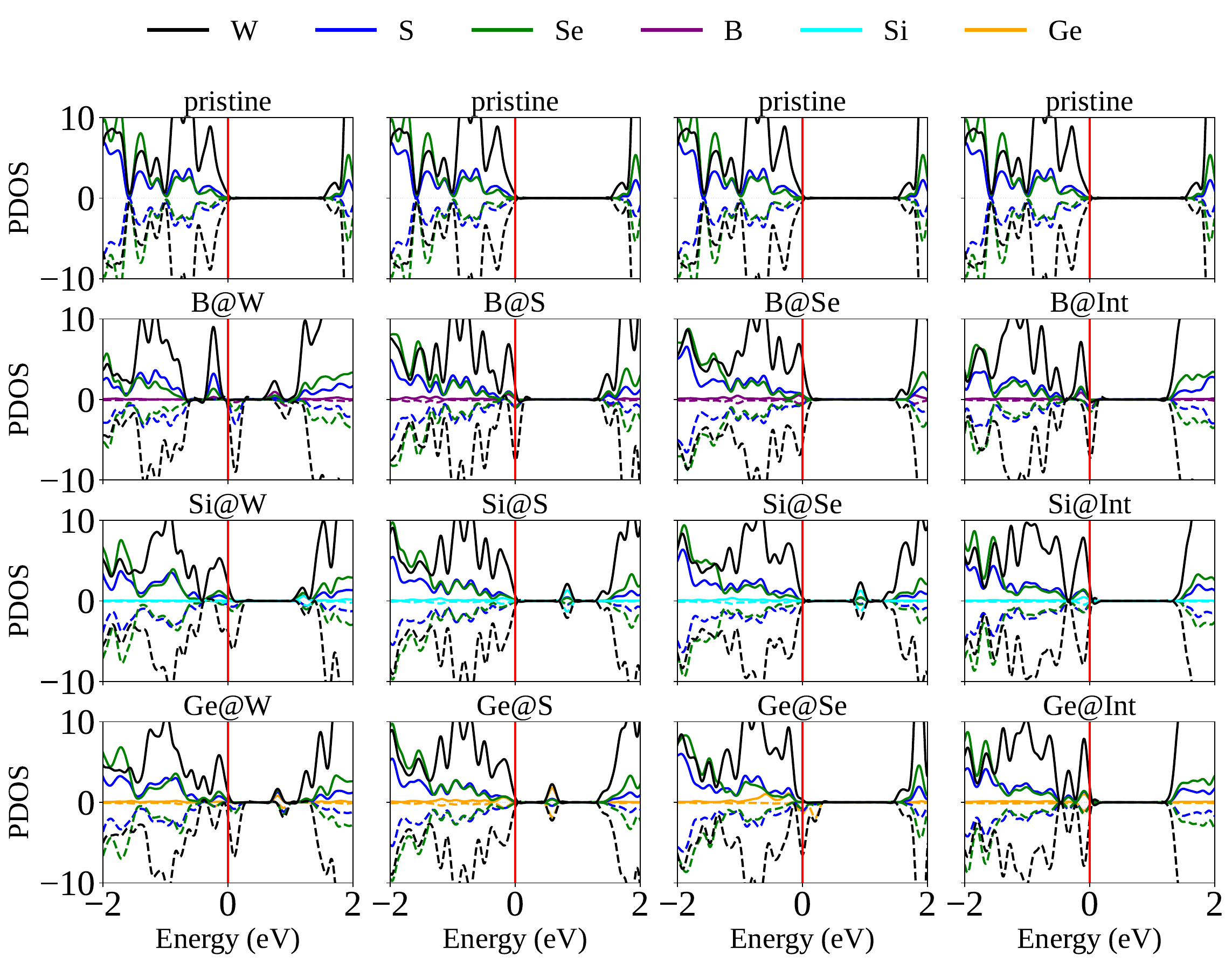} %16cm
     \caption{Atom resolved density of states of pristine and B, Si, and Ge doped WSSe layer at a doping concentration of 2.1\% (4$\times$4$\times$1 supercell). The solid red line represents the Fermi level. }
     \label{fig:441-WSSe-PDOS}
\end{figure*}

\begin{figure*}[htp]
     \centering
     \includegraphics[width=\textwidth]{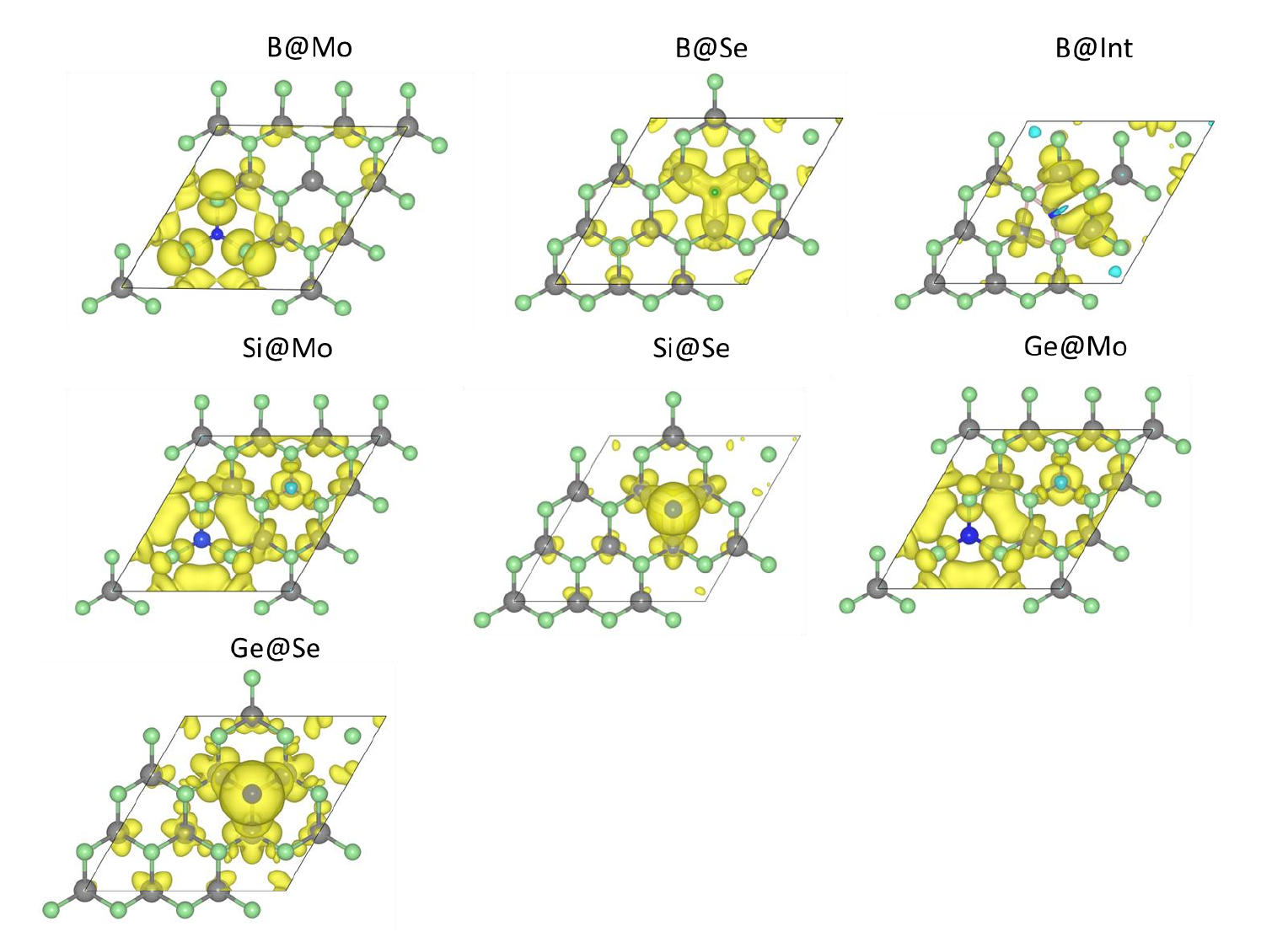} %16cm
     \caption{Spin density isosurfaces of the 3×3×1 supercell of MoSSe in the presence of the dopants B, Si and Ge. The grey, pink, green and blue spheres represent the W, S, Se and dopant atoms, respectively. The yellow and cyan regions represent spin-up and spin-down electron densities, respectively. The isosurface value was taken at 0.001 e/$\AA^3$. }
     \label{fig:331-MoSSe-spin}
\end{figure*}

\begin{figure*}[htp]
     \centering
     \includegraphics[width=\textwidth]{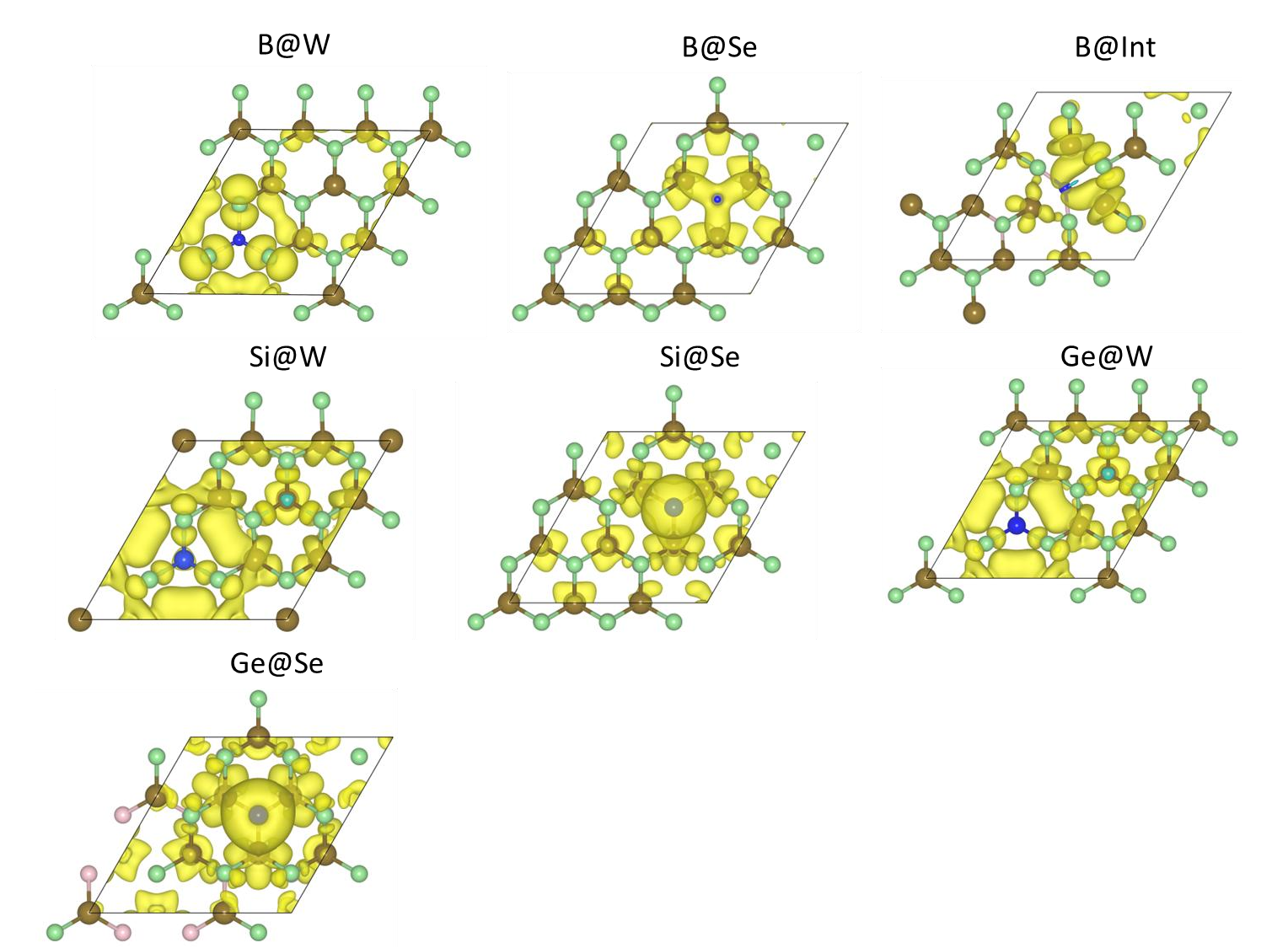} %16cm
     \caption{Spin density isosurfaces of the 3×3×1 supercell of WSSe in the presence of the dopants B, Si and Ge. The olive, pink, green and blue spheres represent the W, S, Se and dopant atoms, respectively. The yellow and cyan regions represent spin-up and spin-down electron densities, respectively. The isosurface value was taken at 0.001 e/$\AA^3$. }
     \label{fig:331-WSSe-spin}
\end{figure*}

\begin{figure*}[htp]
     \centering
     \includegraphics[width=\textwidth]{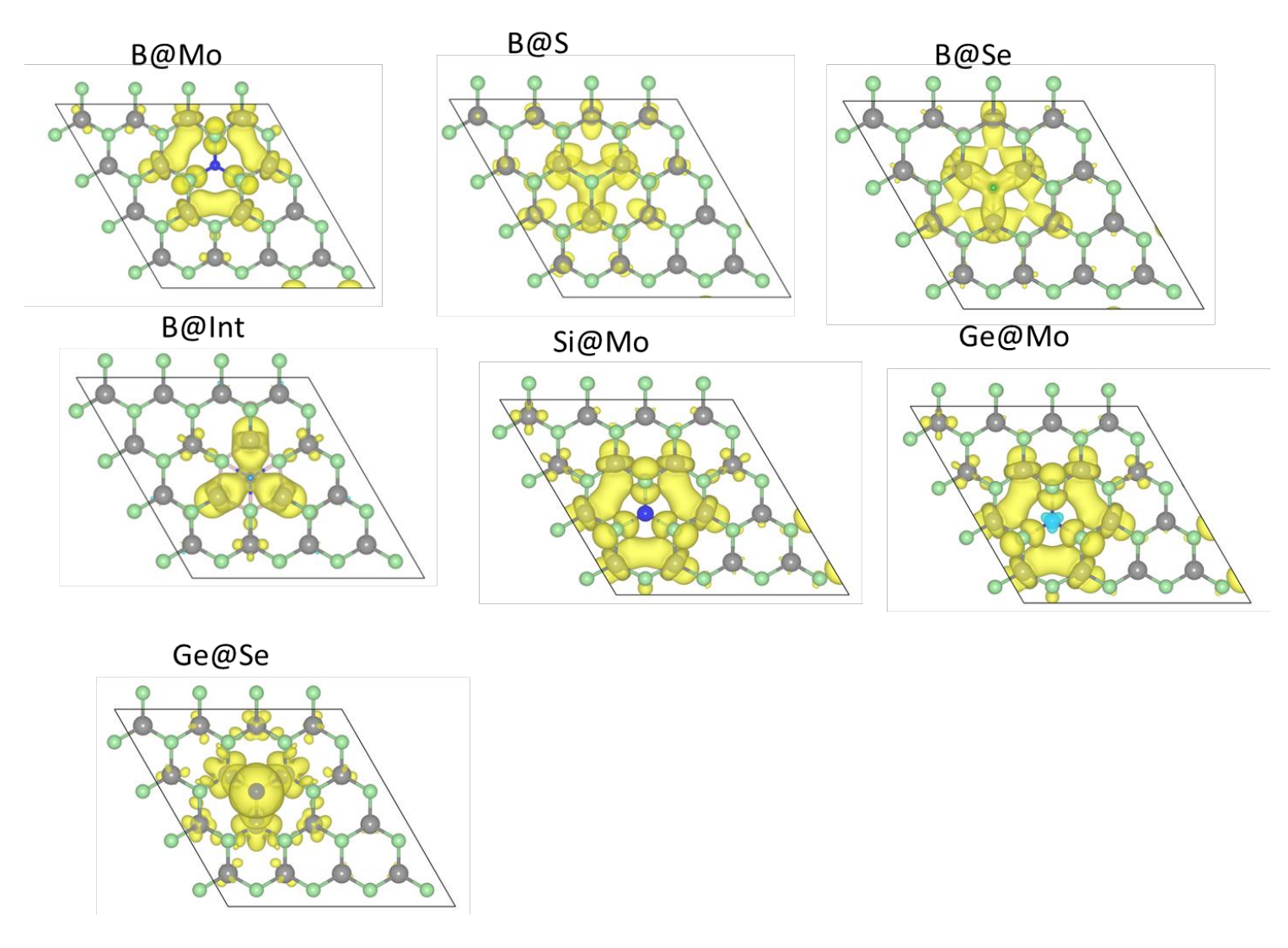} %16cm
     \caption{Spin density isosurfaces of the 4×4×1 supercell of MoSSe in the presence of the dopants B, Si and Ge. The grey, pink, green and blue spheres represent the W, S, Se and dopant atoms, respectively. The yellow and cyan regions represent spin-up and spin-down electron densities, respectively. The isosurface value was taken at 0.001 e/$\AA^3$. }
     \label{fig:441-MoSSe-spin}
\end{figure*}

\begin{figure*}[htp]
     \centering
     \includegraphics[width=\textwidth]{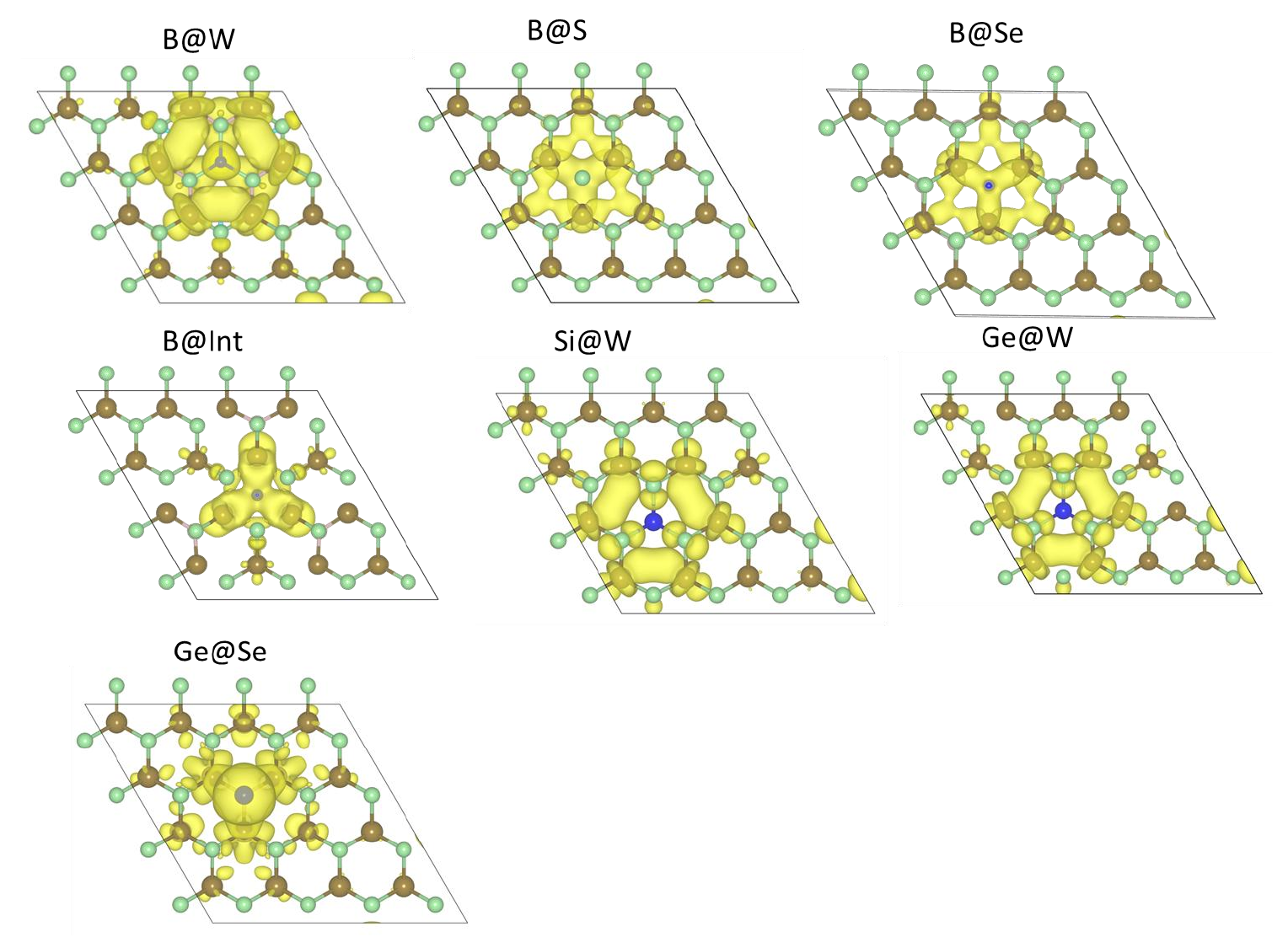} %16cm
     \caption{Spin density isosurfaces of the 4×4×1 supercell of WSSe in the presence of the dopants B, Si and Ge. The olive, pink, green and blue spheres represent the W, S, Se and dopant atoms, respectively. The yellow and cyan regions represent spin-up and spin-down electron densities, respectively. The isosurface value was taken at 0.001 e/$\AA^3$. }
     \label{fig:441-WSSe-spin}
\end{figure*}
\newpage
%\begin{longtable}[h!]
\centering
\begin{longtable}{c c c c c c c}
\caption{The DFT calculated H adsorption energies (E$_{ad}$) in the presence of the dopants B, Si, and Ge doped at distinct sites. }\\
\hline
S.No. & Layer & Dopant & Site of Doping & H adsorption site & $E_{\text{ad}}$ ($n_d$  =  2.1\%) &$E_{\text{ad}}$ ($n_d$ = 3.8\%) \\
\hline
1 & MoSSe & B & B@Mo & 1 & -2.5635 & 0.6460 \\
2 & & & & 2 & -2.5635 & 0.8157 \\
3 & & & & 3 & -2.8179 & -0.1325 \\
4 & & & & 4 & -2.8179 & 1.3206 \\
5 & & & & 5 & -1.8327 & 0.3030 \\
6 & & & & 6 & -1.6805 & 0.3501 \\
7 & & & B@S & 1 & 0.3223 & 0.3443 \\
8 & & & & 2 & 1.0959 & 1.1401 \\
9 & & & & 3 & -0.3734 & -0.3541 \\
10 & & & B@Se & 1 & 0.2110 & 0.2455 \\
11 & & & & 2 & 0.9090 & 0.8964 \\
12 & & & & 3 & -0.3846 & -0.3402 \\
13 & & & B@Int & 1 & 0.5522 & 1.0015 \\
14 & & & & 2 & 1.1540 & 1.5415 \\
15 & & & & 3 & 2.1697 & 0.4502 \\
16 & MoSSe & Si & Si@Mo & 1 & -0.6844 & -0.5939 \\
17 & & & & 2 & 0.4508 & 0.5529 \\
18 & & & & 3 & -0.4805 & -0.4074 \\
19 & & & & 4 & 1.1487 & 1.1384 \\
20 & & & & 5 & 1.2328 & -0.4754 \\
21 & & & & 6 & 0.7128 & 0.9123 \\
22 & & & Si@S & 1 & 0.6790 & 0.6325 \\
23 & & & & 2 & 1.2045 & 1.1783 \\
24 & & & & 3 & -0.7765 & -0.8604 \\
25 & & & Si@Se & 1 & 0.3791 & 0.0096 \\
26 & & & & 2 & 1.2914 & 0.8871 \\
27 & & & & 3 & -0.8258 & -1.2197 \\
28 & & & Si@Int & 1 & 1.4581 & 1.5815 \\
29 & & & & 2 & 2.2737 & 1.7351 \\
30 & & & & 3 & 2.2057 & 0.7022 \\
31 & MoSSe & Ge & Ge@Mo & 1 & -0.4284 & -0.3848 \\
32 & & & & 2 & -0.4284 & 0.5783 \\
33 & & & & 3 & -0.2248 & -0.1970 \\
34 & & & & 4 & -0.2248 & 1.2288 \\
35 & & & & 5 & 0.6725 & -0.2764 \\
36 & & & & 6 & 0.8843 & -0.0734 \\
37 & & & Ge@S & 1 & 0.5710 & 0.5173 \\
38 & & & & 2 & 1.0722 & 1.0405 \\
39 & & & & 3 & -0.5940 & -0.6870 \\
40 & & & Ge@Se & 1 & 0.0499 & 0.0879 \\
41 & & & & 2 & 0.9764 & 0.9799 \\
42 & & & & 3 & -0.8818 & -0.8959 \\
43 & & & Ge@Int & 1 & 1.2110 & 1.6462 \\
44 & & & & 2 & 1.6149 & 2.2610 \\
45 & & & & 3 & 2.2484 & - \\
46 & WSSe & B & B@W & 1 & -0.3305 & 0.8046 \\
47 & & & & 2 & 0.9108 & 0.9603 \\
48 & & & & 3 & 1.2010 & -0.0382 \\
49 & & & & 4 & 1.5390 & 1.5236 \\
50 & & & & 5 & -0.5835 & 0.1225 \\
51 & & & & 6 & -0.4065 & 0.1507 \\
52 & & & B@S & 1 & 0.5421 & 0.5565 \\
53 & & & & 2 & 1.2285 & 1.2696 \\
54 & & & & 3 & -0.4090 & -0.4333 \\
55 & & & B@Se & 1 & 0.3828 & 0.3929 \\
56 & & & & 2 & 1.1447 & 1.1051 \\
57 & & & & 3 & -0.4363 & -0.4600 \\
58 & & & B@Int & 1 & 0.6842 & 1.1190 \\
59 & & & & 2 & 1.3251 & 1.7248 \\
60 & & & & 3 & 1.2715 & 0.6242 \\
61 & WSSe & Si & Si@W & 1 & -0.6096 & -0.4962 \\
62 & & & & 2 & 0.7325 & 0.8379 \\
63 & & & & 3 & -0.4053 & -0.3107 \\
64 & & & & 4 & 1.4025 & 1.5291 \\
65 & & & & 5 & 0.6986 & -0.3841 \\
66 & & & & 6 & -0.1623 & 0.3133 \\
67 & & & Si@S & 1 & 0.8522 & 0.7859 \\
68 & & & & 2 & 1.2929 & 1.2626 \\
69 & & & & 3 & -1.9864 & -0.9638 \\
70 & & & Si@Se & 1 & 0.4638 & 0.2310 \\
71 & & & & 2 & 1.3912 & 1.1218 \\
72 & & & & 3 & -1.0165 & -1.2615 \\
73 & & & Si@Int & 1 & 1.6772 & 1.7547 \\
74 & & & & 2 & 1.8745 & 1.8794 \\
75 & & & & 3 & 1.0930 & 0.8447 \\
76 & WSSe & Ge & W@Mo & 1 & -0.3109 & -0.2630 \\
77 & & & & 2 & 0.7179 & 0.8708 \\
78 & & & & 3 & -0.1088 & -0.0829 \\
79 & & & & 4 & 1.3610 & 1.5060 \\
80 & & & & 5 & 0.8261 & -0.1654 \\
81 & & & & 6 & 0.4591 & 0.0322 \\
82 & & & Ge@S & 1 & 0.7542 & 0.6858 \\
83 & & & & 2 & 1.1723 & 1.1438 \\
84 & & & & 3 & -0.6565 & -0.8056 \\
85 & & & Ge@Se & 1 & 0.2575 & 0.6329 \\
86 & & & & 2 & 1.2003 & 1.5405 \\
87 & & & & 3 & -0.9195 & -0.5976 \\
88 & & & Ge@Int & 1 & 1.3668 & 1.7703 \\
89 & & & & 2 & 1.7657 & 2.3123 \\
90 & & & & 3 & 2.3005 & - \\

\hline
%\end{tabular}
\end{longtable}

\newpage
\section*{Principal component analysis (PCA)}
\justifying
Whenever we are dealing with a large dataset, one has to look at the “curse of dimensionality” which leads to large computational time. Principal component analysis is a dimensionality reduction technique, which reduces the dimension of the feature space by excluding a minimal information from the data. PCA results in a novel set of features which explains the maximum variance hidden in the original feature space. Let us suppose, we have two set of features X and Y for which we have a data distribution as shown in SFig. 8. The first principal component PC1 is a linear combination of the two features X and Y such that it retains the maximum variance from the data. The second principal component PC2 covers the second largest direction of variance and lies orthogonal to the PC1. In this way, each of the principal component (PC) will be independent of each other and acts a best predictor for the training of the machine learning model.

\begin{figure*}[htp]
     \centering
     \includegraphics[width=14cm]{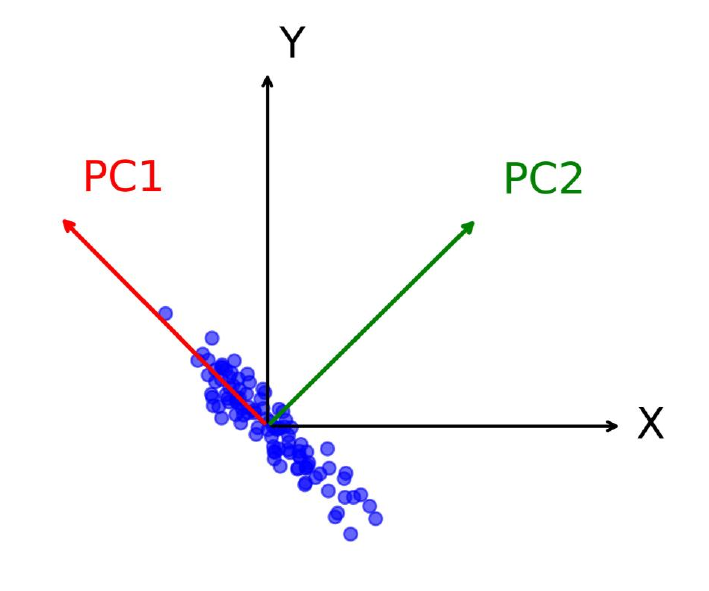} %16cm
     \caption{The schematic representation of the technique of principal component analysis (PCA) with two features X and Y.}
     \label{fig:pca}
\end{figure*}
\newpage

\begin{figure*}[htp]
     \centering
     \includegraphics[width=14cm]{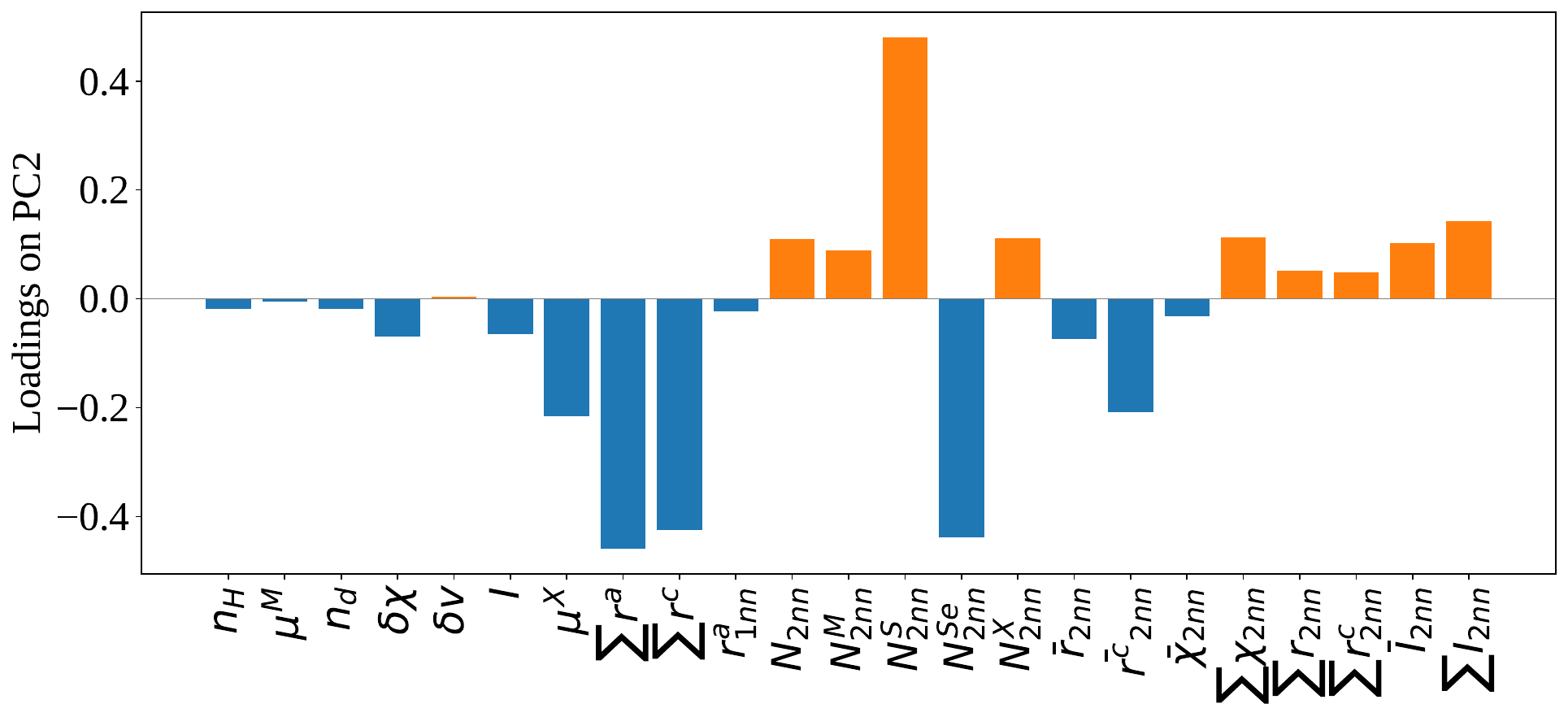} %16cm
     \caption{Loadings plot representing the contribution of each feature in obtaining the PC2}
     \label{fig:PC2}
\end{figure*}

\begin{figure*}[htp]
     \centering
     \includegraphics[width=12cm]{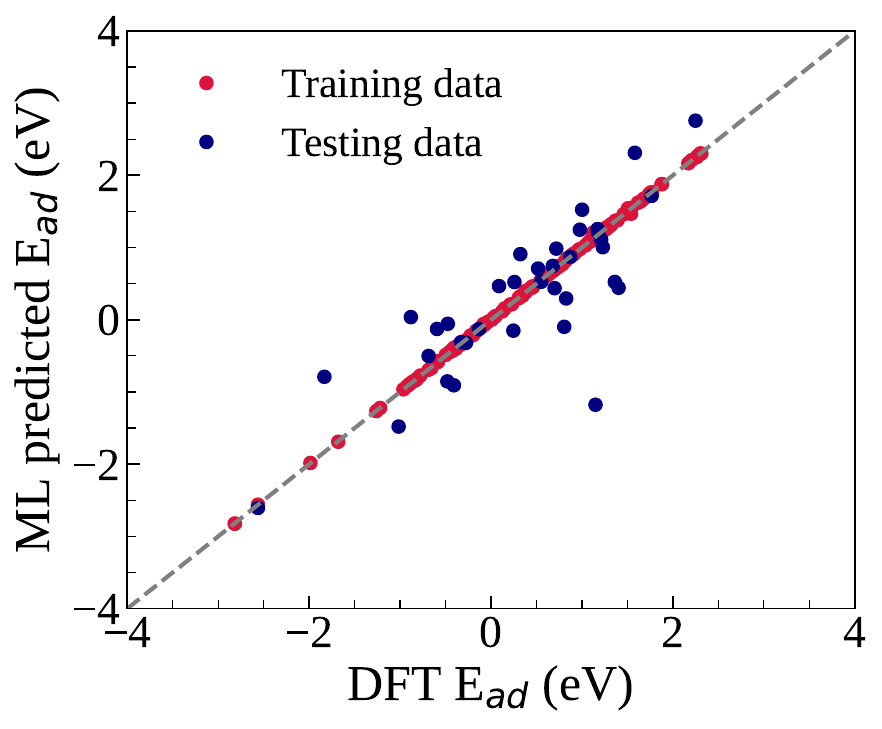} %16cm
     \caption{ The DFT-calculated versus ML-predicted E$_{ad}$, including all 178 data points and 11 PCs as features in the MLPR model.}
     \label{fig:178-data}
\end{figure*}

\begin{figure*}[htp]
     \centering
     \includegraphics[width=\textwidth]{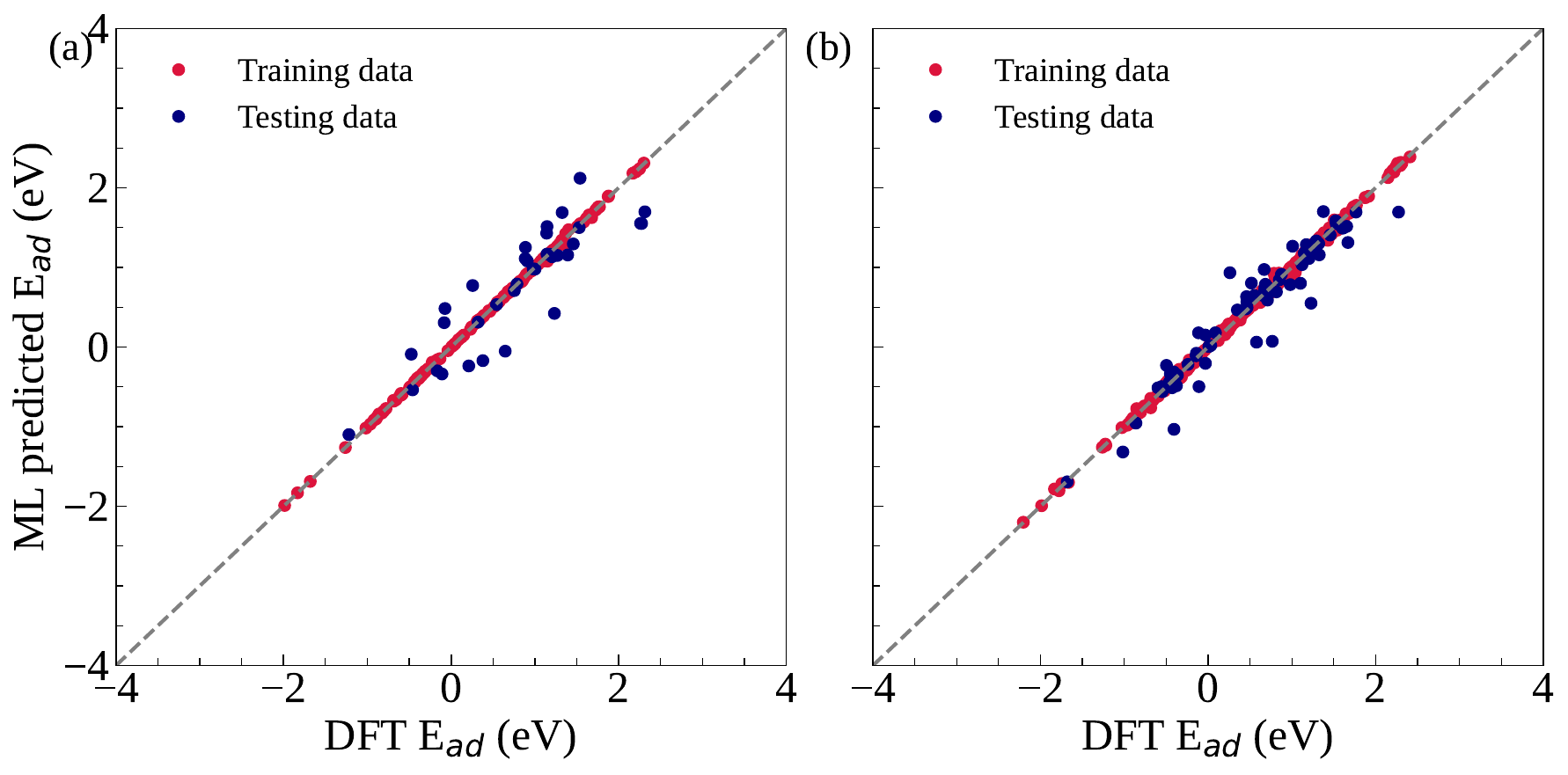} %16cm
     \caption{The DFT calculated versus ML predicted E$_{ad}$ with all 23 features as input to the MLPR model (a) without and (d) with data augmentation, respectively.}
     \label{fig:WOPCA}
\end{figure*}

\begin{figure*}[htp]
     \centering
     \includegraphics[width=\textwidth]{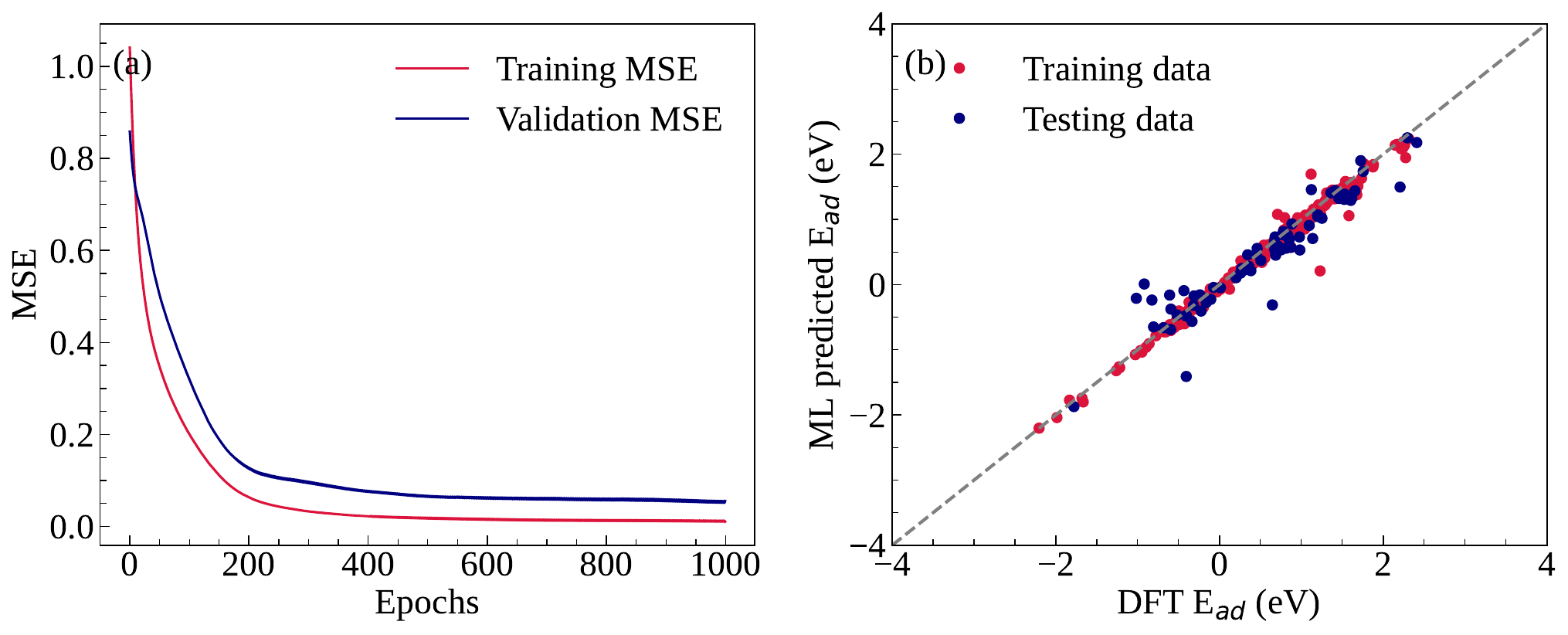} %16cm
     \caption{(a) The curves indicating the training and validation mean squared errors (MSE) variation with the increase in the number of epochs with the Adam optimization technique. (b) The DFT versus ML predicted E$_{ad}$ with the multi-layer perceptron regression model with the Adam optimization technique.}
     \label{fig:Adam}
\end{figure*}